# Proceedings of the International Workshop on Methodologies for Translating Legal Norms into Formal Representations (LN2FR 2022)

*in association with*
*35th International Conference on Legal Knowledge and Information Systems (JURIX 2022)*

LN2FR 2022 Co-Chairs

Georg Borges, Saarland University, Germany
Ken Satoh, Institute of Informatics, Japan
Erich Schweighofer , University of Vienna, Austria

December 14, 2022

# Preface

This volume contains the papers presented at LN2FR 2022: The International Workshop on Methodologies for Translating Legal Norms into Formal Representations, held on December 14, 2022 in a hybrid form (in person workshop was held in Saarland University, Saarbrucken) in association with 35th International Conference on Legal Knowledge and Information Systems (JURIX 2022).

Using symbolic logic or similar methods of knowledge representation to formalise legal norms is one of the most traditional goals of legal informatics as a scientific discipline. More than mere theoretical value, this approach is also connected to promising real-world applications involving, e.g., the observance of legal norms by highly automated machines or even the (partial) automatisation of legal reasoning, leading to new automated legal services. Albeit the long research tradition on the use of logic to formalise legal norms – be it by using classic logic systems (e.g., first-order logic), be it by attempting to construct a specific system of logic of norms (e.g., deontic logic) –, many challenges involved in the development of an adequate methodology for the formalisation of concrete legal regulations remain unsolved. This includes not only the choice of a sufficiently expressive formal language or model, but also the concrete way through which a legal text formulated in natural language is to be translated into the formal representation. The workshop LN2FR sought to explore the various challenges connected with the task of using formal languages and models to represent legal norms in a machine-readable manner. We had 13 submissions, which were reviewed by 2 or 3 reviewers. Among these, we selected 11 papers (seven long papers, three short papers, one published paper) for presentation and discussion.

Last but not least, we would like to thank all the authors who submitted papers, as well as the members of the PC and additional reviewers for reviewing the submitted papers.

December 14, 2022                                                                                  Georg Borges
Ken Satoh
Erich Schweighofer

## Table of Contents

# Program Committee

| | |
|---|---|
| Georg Borges | Saarland University, Germany |
| Mireille Hildebrandt | Vrije Universiteit Brussel (VUB), Belgium |
| Sarah Lawsky | Northwestern University, USA |
| Denis Merigoux | Centre Inria de Paris, France |
| Adrian Paschke | Freie Universität Berlin |
| Livio Robaldo | Swansea University, UK |
| Diogo Sasdelli | Saarland University, Germany |
| Ken Satoh | National Institute of Informatics, Japan |
| Burkhard Schafer | University of Edinburgh, UK |
| Erich Schweighofer | University of Vienna, Austria |
| Christoph Sorge | Saarland University, Germany |
| Tom van Engers | University of Amsterdam, The Netherlands |

# Additional Reviewers

**F**

Fungwacharakorn, Wachara

**L**

Le Minh, Nguyen

**T**

Tojo, Satoshi
Tsushima, Kanae

# Law to Binary Tree - An Formal Interpretation of Legal Natural Language

Ha-Thanh Nguyen[1,*], Vu Tran[2], Ngoc-Cam Le[3], Thi-Thuy Le[4], Quang-Huy Nguyen[5], Le-Minh Nguyen[6], and Ken Satoh[1]

[1] National Institute of Informatics, Tokyo, Japan
[2] The Institute of Statistical Mathematics (ISM), Tokyo, Japan
[3] Vietnam Judicial Academy, Hanoi, Vietnam
[4] Hanoi Law University, Hanoi, Vietnam
[5] VINASECO JSC, Hanoi, Vietnam
[6] Japan Advanced Institute of Science and Technology, Ishikawa, Japan

**Abstract.** Knowledge representation and reasoning in law are essential to facilitate the automation of legal analysis and decision-making tasks. In this paper, we propose a new approach based on legal science, specifically legal taxonomy, for representing and reasoning with legal documents. Our approach interprets the regulations in legal documents as binary trees, which facilitates legal reasoning systems to make decisions and resolve logical contradictions. The advantages of this approach are twofold. First, legal reasoning can be performed on the basis of the binary tree representation of the regulations. Second, the binary tree representation of the regulations is more understandable than the existing sentence-based representations. We provide an example of how our approach can be used to interpret the regulations in a legal document.

## 1 Introduction

In recent years, artificial intelligence technology has been widely used in the legal field to facilitate legal analysis and decision-making tasks [1, 15, 9]. Although legal documents usually describe the regulations in natural language, there is an ordered and logical system based on legal doctrines and theoretical legal issues behind the words. If it is only based on natural language processing without focusing on the systematic structure of legal documents, an artificial intelligence system is likely to give inaccurate or meaningless outcomes for the tasks which require logical analysis, or systematic legal knowledge. Therefore, using a formal logical representation to represent the regulations in legal documents is a promising approach. In more detail, jurists can express systematic legal knowledge through formal logical representation, while computers can easily process formal logical representations and "understand" the legal logic made by jurists. Thereby, artificial intelligence can simulate and learn how lawyers analyze certain legal documents. In other words, through formal logical representation, artificial intelligence and jurists can communicate logical problems with each other.

---

[*] Corresponding: nguyenhathanh@nii.ac.jp.

Many approaches have been proposed to represent the regulations in legal documents as logic formulas. The legal rules can be represented in the form of a structured representation [8], a rule-based representation [10] or a graph-based representation [12]. These approaches share the same general idea of representing the regulations as formal rules. We find that the main challenges for bringing the work to practical applications are not only in the technology of representing legal norms but also in the method of developing the technology with the involvement of legal experts. It costs a huge amount of time for lawyers to understand the logical expression and then to produce the logical formulas from the legal documents.

In this paper, we propose a new approach for representing and reasoning with legal documents. Our approach interprets the regulations in legal documents as binary trees and employs a legal reasoning system to resolve logical contradictions. We prove that our approach can be used to perform legal reasoning on the basis of the binary tree representation of the regulations. The biggest advantage of our approach is that the binary tree representation is easy to understand and to produce by jurists. As a result, our approach can be used to develop an automatic legal reasoning system that is understandable by jurists.

## 2   Preliminaries

The legal system of a country is composed of vast legal norms. Taxonomy plays an essential role in the arrangement of these legal regulations. Thanks to taxonomy, jurists think logically about legal problems by sorting legal rules, deciding them into categories, and generalizing them into fields of law [11]. Therefore, although the legal system consists of massive legal norms, it is still orderly, logical, and coherent. That is why jurists often use a logic diagram that we call a latent tree structure to analyze legal regulations and deal with legal documents. In this way, it is easier for them to see the relationship and hierarchy between the elements of the legal system. The word "latent" means that there is a hidden tree structure behind the legal system, and the tree structure is generated by the relationship between the legal rules and the fields of law.

A latent tree structure often consists of two parts: (i) **The root** presents legal taxonomies which are widely analyzed and acknowledged in legal science. This part may have similarities between the laws of countries in the same legal traditions, i.e., civil law, common law, religious law, etc.; and (ii) **The rest** shows the contents of current legal norms stuck with each legal taxonomy mentioned above. Depending on the policies of each state or nation, the part may be differences between the laws of each country, even though they share the same legal traditions.

# 3 Proposed Approach

As mentioned in the previous section, the latent tree structure is an effective tool for jurists to deal with legal documents. The latent tree can be drafted manually on paper or in jurists' minds without any standard. If there is a method to build a latent tree structure from a legal document automatically, it will make jurists' work more efficient and accurate. In addition, we can build an AI system that could automatically interpret legal rules in a manner similar to how lawyers and judges would.

Humans can use their common sense and legal experience to prioritize the rules. However, this is difficult for the machine or even lay people. For example, when checking a contract, if the agreement contains both particular rules and general rules, the particular rules take priority over the general rules. Another example is that before checking the contract, we need to check the parties to determine if they are qualified. Are they eighteens or older? Do they have the mental capacity to understand the contract? Which qualifications should take priority when there is a conflict between them? When the contract contains many qualifications, it increases the complexity of priority checking.

We propose a new way to resolve the problem of priority relationships between rules in latent trees. The idea is to construct a binary tree structure, in other words, convert an ordinary latent tree into a binary tree. In a binary tree, there are only two child nodes for each parent node. The position of nodes can tell which rule needs to be considered first. For example, the root is always the first node, and based on the result of the root we can go to the left or right node. When we get to the leaf, we can get the final result. This representation can be used by the machine to automatically resolve the conflict between the rules and determine the priority.

# 4 Case Study

In this section, we use a part of legal norms on inheritance under wills with a focus on the regulations related to testators to demonstrate how to convert the latent tree into a binary tree.

**Step 1: Drawing the latent tree**
Please note that, in practice, jurists will draw this tree on an AI system based on (i) technically fixed standards; and (ii) their systematic legal knowledge of relevant legal norms.

First of all, the root of the latent tree structure can be drawn based on legal taxonomies of laws on inheritance under wills.

From a legal point of view, the legal norm is created to govern the respective legal relations. Jurists can classify a legal relation into three elements: **a subject**, **an object**, and **contents** of legal relation[1, 6]. Besides, when analyzing and assessing a legal relation, it is essential to consider its creation, change, and termination. Thus, first of all, we draw the root of the latent tree structure includes four nodes: **subjects** in the inheritance relation under wills, the **object** of the inheritance relation under wills, **contents** of the inheritance relation under wills, and **the creation, change, and termination** of the inheritance relation under wills. Then, we break down each element separately in detail as follows (Figure 1):

(i) **Subjects**: In legal science, a subject (a legal person) includes two kinds: **a natural person** (human person, sometimes also a physical person), and **a legal entity** (non-human person) which is a body that can function legally, sue or be sued, and make decisions through agents such as association, corporation, partnership [4, 7]. Thus, main subjects in the inheritance under wills may include the testator, heirs, administrators of estates, and witnesses.

(ii) **The object**: In legal science, object is what the subjects want to achieve when entering into a legal relation, or the aim or purpose of legal norms which govern the respective legal relation [4, 13]. So, estates are the object of the inheritance under wills relation.

(iii) **Contents**: In legal science, contents of a legal relation are legal rights, and obligations of the relevant subject(s)[1, 6, 13]. Hence, the contents of the inheritance relation under wills are rights as well as obligations of each relevant subject, i.e., the testator, heirs, administrators of estates, and witnesses.

(iv) **the creation, change and termination** of the inheritance relation under wills.

The laws of inheritance under the wills of many countries such as Vietnam[14], China[2], Germany[5], France[3] etc. should be classified the same taxonomy mentioned above. Figure 1 shows the root of the latent tree structure based on legal taxonomies of laws on inheritance under wills.

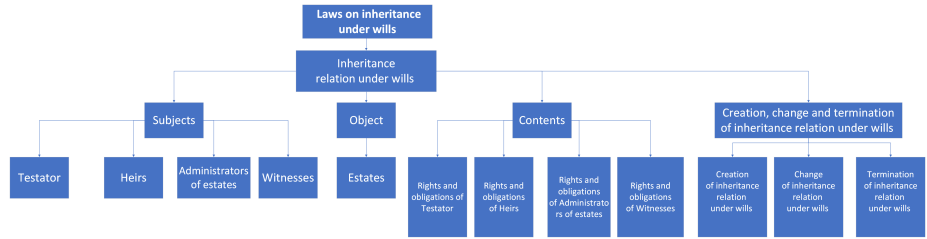

**Fig. 1.** The root of the latent tree structure based on legal taxonomies of laws on inheritance under wills

**Next**, we draw the rest of the latent tree structure which shows the contents of current legal norms on the inheritance relation under wills stuck with each legal taxonomy mentioned above. As mentioned above, this part of the tree structure is different when representing the legal norms of different countries due to differences in the practical laws of each country. After the task, we have a complete latent tree structure on the inheritance relation under wills. For illustration purposes, we only demonstrate legal norms on testators in two countries: Vietnam, and China.

Figure 2 and 3 show latent trees of Vietnamese and Chinese legal norms. We can see that both trees have a similar structure. In general, legal norms for this case are categorized into four groups: subjects, objects, contents, and the problem of creation, change, and termination of inheritance relation under wills. The trees are slightly different in terms of gestation and subject capacity. These differences may lead to big differences in legal consequences. For example, in Vietnam law, the legal capacity of the testator is divided into three cases: under 15 years of age, from 15 to 18 years of age, and over 18 years of age. The age of the testator is one of the important elements that determine the status of inheritance relation under wills.

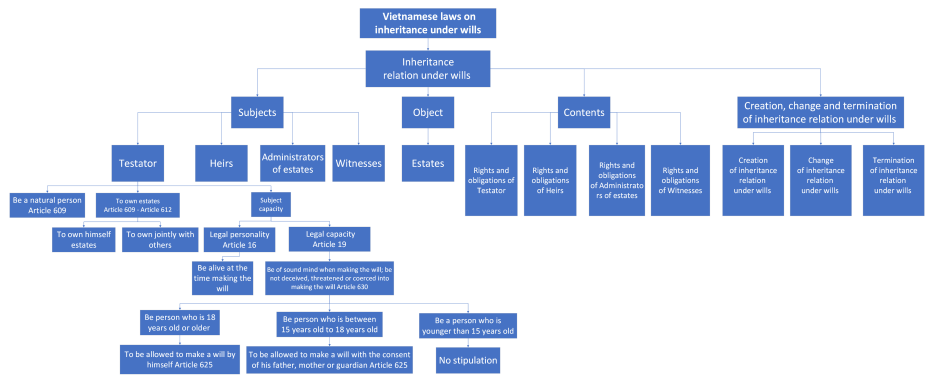

**Fig. 2.** The latent tree structure based on Vietnamese laws on inheritance under wills

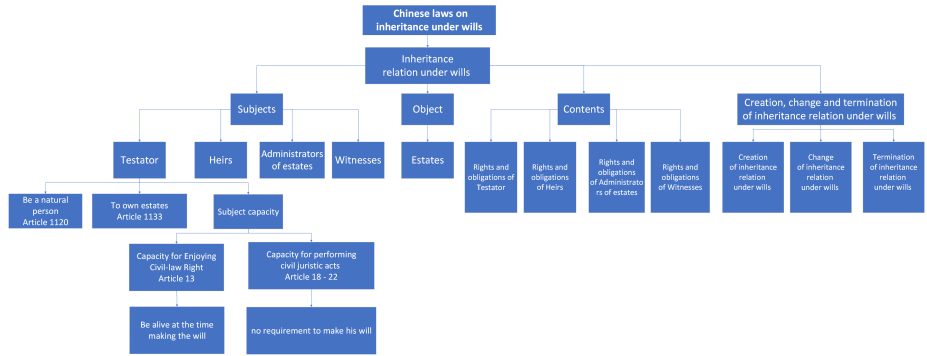

**Fig. 3.** The latent tree structure based on Chinese laws on inheritance under wills

**Step 2: Converting the latent tree into the binary tree**

An example of the Vietnamese binary tree is shown in Figure 4. This tree has a structure similar to a decision tree, with two branching arrows standing for Yes and No. The leaf nodes are the legal consequences of the interpretation. For example, in the case of Vietnamese law, if the testator is not a natural person, then there is no right to make a will. This approach allows us to reason about legal conclusions automatically. Besides, we can also utilize the game theory or tree-based algorithm to make efficient and effective legal decision-making applications with this structure.

With this binary tree structure, we create a deterministic and complete reasoning system, in which the legal consequences are determined by the input features. This reasoning system is not only suitable for legal decision-making applications but also an effective tool for analyzing legal regulations. For example, if there is no logical way to arrange the nodes, this can be considered a contradiction in the legal regulation. In addition, this structure can be used to frame the reasoning process of black-box models when learning from data.

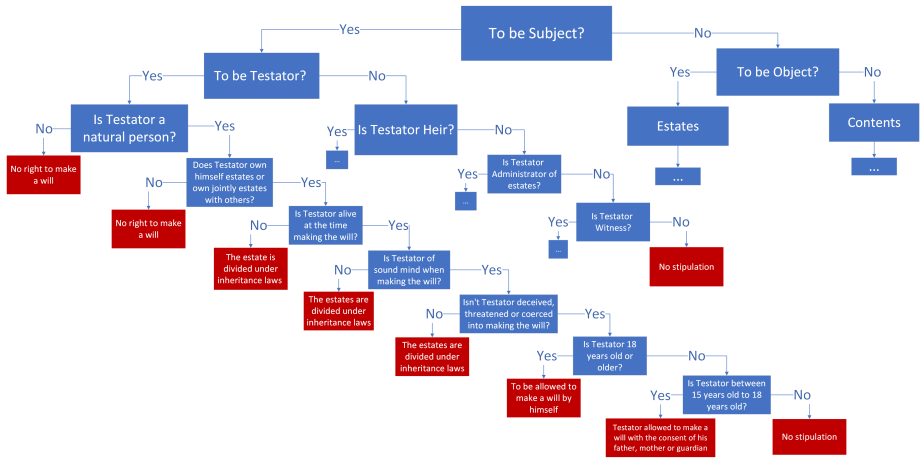

**Fig. 4.** The binary tree structure based on Vietnamese laws on inheritance under wills

## 5  Discussion and Conclusion

The proposed method for building the binary tree from the legal document is a way to understand the structure of legal concepts and regulations and resolve conflicts between rules automatically. This is important because it can make the machine automatically interpret legal rules in a manner that is similar to how lawyers and judges would. It will also make lawyers' work more efficient and accurate. In this paper, we also introduce a case study in which we can see the application of the proposed method to inheritance analysis to demonstrate our idea better. Note that the example tree presented in this paper is only one possible way to construct the binary tree. The tree can be customized to fit the different needs of the particular legal problem. This representation can be used to validate the outcome and reasoning ability of black-box machine learning models. In future works, we will involve this proposal in our systems, research activities, legal activities, and competitions.

## References

1. Araszkiewicz, M., Francesconi, E., Zurek, T.: Identification of contradictions in regulation. In: Legal Knowledge and Information Systems, pp. 151–160. IOS Press (2021)
2. China civil code (2020), issuance date: 20 May 2020, Issued by China Thirteenth National People's Congress
3. France civil code (2020), issuance date: 04 November 2022, Issued by France National Assembly
4. Garner, B.A.: Black's Law Dictionary,Abridged, 9th. Thomson Reuters (2014)
5. German civil code (2020), issuance date: 02 January 2002, Issued by German Federal Law Gazette
6. Khoshimov, S., Nikulina, V.: Legal framework for providing/obtaning financing in blockchain. russian perspective. In: Proceedings of the 2nd International Scientific and Practical Conference on Digital Economy (ISCDE 2020). pp. 589–594. Atlantis Press (2020). https://doi.org/https://doi.org/10.2991/aebmr.k.201205.098, https://doi.org/10.2991/aebmr.k.201205.098
7. Legal Information Institute: Legal person, https://www.law.cornell.edu/wex/legal_person, last accessed 03 November 2022
8. Nguyen, H.T., Nguyen, V.H., Vu, V.A.: A knowledge representation for vietnamese legal document system. In: 2017 9th International Conference on Knowledge and Systems Engineering (KSE). pp. 30–35. IEEE (2017)
9. Nguyen, H.T., Shirai, K., Nguyen, L.M.: Few-shot tuning framework for automated terms of service generation. In: Legal Knowledge and Information Systems, pp. 113–118. IOS Press (2021)
10. Satoh, K., Asai, K., Kogawa, T., Kubota, M., Nakamura, M., Nishigai, Y., Shirakawa, K., Takano, C.: Proleg: an implementation of the presupposed ultimate fact theory of japanese civil code by prolog technology. In: JSAI international symposium on artificial intelligence. pp. 153–164. Springer (2010)
11. Sherwin, E.: Legal taxonomy. Cornell Law Faculty Publications. (2019)
12. Sovrano, F., Palmirani, M., Vitali, F.: Legal knowledge extraction for knowledge graph based question-answering. In: Legal Knowledge and Information Systems, pp. 143–153. IOS Press (2020)
13. Uc, D.T., Que, H.T.K.: Textbook on State and Law. Vietnam National University (2017)
14. Vietnamese civil code (2015), issuance date: 24 November 2015, Issued by Vietnamese National Assembly
15. Xu, H., Savelka, J., Ashley, K.D.: Accounting for sentence position and legal domain sentence embedding in learning to classify case sentences. In: Legal Knowledge and Information Systems, pp. 33–42. IOS Press (2021)

# Towards a Congruent Interpretation of Traffic Rules for Automated Driving – Experiences and Challenges

Lukas Westhofen[1][0000−0003−1065−4182], Ingo Stierand[1], Jan Steffen Becker[1], Eike Möhlmann[1], and Willem Hagemann[1]

German Aerospace Center (DLR) e.V., Institute of Systems Engineering for Future Mobility, Oldenburg, Germany
{lukas.westhofen,ingo.stierand,jan.becker,
eike.moehlmann,willem.hagemann}@dlr.de

**Abstract.** The homologation of automated driving systems for public roads requires a rigorous safety case. Regulations of the United Nations demand to demonstrate the compliance of the developed system with local traffic rules. Hence, evidences for this have to be delivered by means of formal proofs, online monitoring, and other verification techniques in the safety case. In order for such methods to be applicable traffic rules have to be made machine-interpretable. However, that pursuit is highly challenging. This work reports on our practical experiences regarding the formalization of a non-trivial part of the German road traffic act. We identify a central issue when formalizing traffic rules within a development process, coined as the congruence problem, which is concerned with the semantic equality of the legal and system interpretation of traffic rules. As our main contribution, we delineate potential challenges arising from the congruence problem, hence impeding a congruent yet formal interpretation of traffic rules. Finally, we aim to initiate discussions by highlighting steps to partially address these challenges.

**Keywords:** Automated Driving · Traffic Rules · Formalization.

## 1 Introduction to Traffic Rules for Automated Driving

Automated driving systems (ADSs) are anticipated to take over a large part of the tasks that are currently performed by human drivers. Vehicles shall recognize their environment and perform maneuvers without the need of human intervention. This holds especially for critical situations within the operational design domain. The task of recognizing a dynamic environment, interpreting this in the context of traffic rules, and in particular reacting accordingly, is performed by human beings with a high fault tolerance, as long as they are attentively following what is happening. Due to ever extending operational domains, new risk potentials arise, which must be analyzed before the release of the ADS.

Already in the early stages of the development of an ADS, the question arises as to which requirements must be met to be approved for participation in road traffic. This includes technical requirements (resources, real-time, ...), control requirements (longitudinal and lateral control of vehicle dynamics, ...), and behavioral requirements (appropriate indication of actions, ...). At least for the top-level requirement of traffic rule compliance, this question can be answered with certainty: the United Nations regulation number 157, which is concerned with the approval of automated lane keeping systems, states that '*the activated system shall comply with traffic rules relating to the dynamic driving task in the country of operation*' [22, clause 5.1.2]. This raises two important questions:

1. How ADSs can be designed to comply with the traffic rules, and
2. how it can be proven that they do so?

Traffic rules are written by legal experts to be interpreted again by humans. Therefore, developing solutions for the first question is hindered by their inherently and intentionally vague character. Scenario-based approaches are anticipated to approach the second question [14]. However, due to the high number of test cases, even with scenario-based testing not every test run can be presented to a legal expert for evaluation. Thus, an automated assessment is necessary not only for a correct implementation but also for verification. This directly results in the need for a machine-interpretable understanding of these rules.

A promising way to make knowledge machine-interpretable is the use of formal methods for specification and verification. These allow to analyze the consistency of requirements, enable formal proofs by means of model checking, and evaluate runs of a system during operation. The question therefore arises as to how a formal understanding of traffic rules and case law can be realized. Specifically, the following issue must be clarified: *How can we ensure and demonstrate that the implemented semantics correctly reflects the traffic rule interpretation by the legal expert?* The work at hand contributes to this core issue by

1. an *experience report* on formalizing a part of the German road traffic act,
2. a *formal framework* and *key challenges* distilled from our practical experience, inhibiting a sound and complete rule interpretation, and,
3. potential directions of *future work* for the resolution of these challenges.

We highlight the second point as our main contribution. Our experience indicates solutions therefor to be a prerequisite for the homologation of ADSs.

## 2 Related Work

The formalization of legal texts, such as the British nationality act [20], is not a new endeavor. These early approaches already included applications on traffic rules in 1991 [6]. Since ADS technology was still in its infancy, the focus was on the support of legal experts and teaching aspects.

Recent advances in ADSs gave updraft to the formalization of traffic rules. We are largely concerned with methodical issues. Related work includes a method for formalizing traffic rules into a defeasible deontic logic [3]. Here, a special focus lies on the identification of terms, the definition of atoms based on these terms, the type of the sentence and the eventual premise-consequence structure of the formalized rule. Similarly, Costescu delineates the necessity of having a legal analysis prior to the actual traffic rule formalization process as to achieve a common understanding between engineers and legal experts [5]. The key observation is that the natural language character of traffic rules is preventing a straightforward formalization. Moreover, the German research project VVMethods investigates, among others, methods for the derivation of behavioral requirements from traffic rules [18]. It highlights that a thorough legal analysis is required, adopting methods for the creation of legal opinions and its associated mindset.

Apart from methodical considerations, most work focuses on formalization techniques. Notably, work has been driven by the Technical University of Munich and partners. First steps were performed by Rizaldi et al. [17], where monitors were derived from linear temporal logic (LTL). Buechel et al. use an ontology-based approach with a joint description logic and rule reasoner [4]. Work since then focused on metric temporal logic, both for highways [13] and intersections [12]. Traffic rule formalization within a development process was examined by Esterle [7]. The thesis analyzes exemplary rules, albeit without apparent legal support, understating the interplay of different expert domains. Formalization is done by defeasible LTL rules, and observer automata are evaluated on realistic data. Besides (derivatives of) LTL, related work relies on a multi-lane spatial logic [19], a defeasible deontic logic [23], and answer set programming [11].

Technical advances have been pursued as well, such as LegalRuleML [1], which partially addresses the challenges – e.g. exceptions – later presented. Due to the large amount of prior work on the technicalities of traffic rule formalization, we refrain from detailing formalisms and focus on the open methodical gaps.

## 3 An Experience Report

In joint work together with an Original Equipment Manufacturer, a law firm, and BTC Embedded Systems AG (www.btc-embedded.com), we developed an approach to formalize German traffic act (StVO) rules relevant for highway driving. The approach was realized, resulting in 111 formalized StVO rules. An implementation of observers for the formalized rules has been applied to simulated traffic scenarios. The challenges raised in this paper originate in this work.

The taken approach is depicted in Fig. 1. In a first step, the terms and rules of the relevant part of the StVO were assessed and, where ever possible, clarified from a legal perspective. This provided the basis for the formalization, which consists of two layers. The base layer is established by a description logic ontology containing the basic terms (concepts and roles) used in the StVO, like 'vehicle' and 'in front of' [2]. We also call them 'observable entities', as these are assumed to be entities recognized by a suitable perception chain. The second layer is established by a definition of predicates and function terms, which take the ontology as their domain. Along this, a light-weight fragment of duration calculus has been defined in order to allow specifying temporal terms. Finally, the traffic rules were formalized based on those *atoms* using the same language.

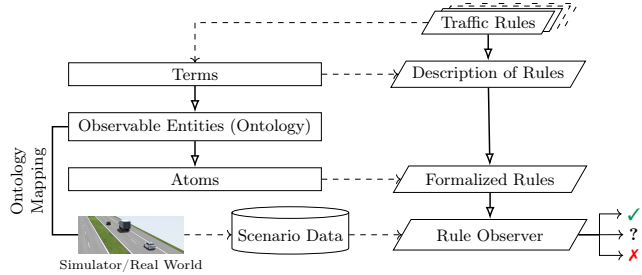

**Fig. 1.** Approach for the formalization and observer implementation of traffic rules.

As the subsequent section illustrates, traffic rules contain vague terms (e.g. 'to endanger'). Many terms have therefore been defined in a sufficient and necessary version. Hence, each rule has been formalized in two versions: One to safely detect the satisfaction of the traffic rule, and one for the safe detection of violations.

Finally, a selected subset of the rules was manually implemented in the BTC Embedded Platform (EP) [21]. For this, a set of traffic scenarios has been created, simulated, and logged. A mapping of the relevant observable entities within the legal ontology to the parameters observable in the ontology of the simulator has been defined. The mapped simulation logs have been fed to the generated observers in order to detect rule satisfactions and violations, respectively.

### 3.1 Example

In order to illustrate our approach, let us examine the exemplary rule of section four, paragraph one, sentence one of the StVO: *'The distance to a preceding vehicle shall be, generally, large enough to allow stopping behind this vehicle even if it is suddenly braked.'* We also refer to this rule as the *distance rule*.

We identified the following relevant terms: *distance*, *preceding*, *vehicle*, *generally*, *large enough*, *stopping behind*. For those, legal experts delivered a (recursive) definition. For example, explicating a *large enough* distance involves the *stopping distance*, which in turn relies on the *braking distance*, *speed*, and the driver's *reaction time*. Moreover, it relies on the *minimal stopping distance*, which is dependent on various complex factors such as road surfaces, tire types, weather conditions, and temperatures, allowing only a partial definition. The set of relevant atoms and their observability assumptions are depicted in Fig. 2.

Omitting details due to conciseness, we can assemble the safe satisfaction as

$$\texttt{sat}_{4.1.1} \coloneqq \forall f_1, f_2 \in \texttt{Vehicle} : \texttt{preceding}(f_1, f_2) \land \texttt{generally}_n(f_1, f_2) \implies \texttt{enough\_distance}(f_1, f_2).$$

We introduced the atom `generally` to model exceptions, both in a sufficient and a necessary condition. This concerns e.g. situations of driving in a convoy or during starting after waiting in traffic. This leads to the aforementioned two versions of the rule; one including the necessary condition of `generally` for a safe satisfaction and one including the sufficient condition of `generally` for a safe violation of the rule. In the example of a convoy, a partially formalized necessary condition is $\texttt{generally}_n(f_1, f_2) \coloneqq f_1, f_2 \notin \texttt{Convoy\_Vehicle}$. A partial formalization of a sufficient condition is, e.g., $\texttt{generally}_s(f_1, f_2) \coloneqq f_1, f_2 \notin \texttt{Convoy\_Vehicle} \wedge f_1.\texttt{speed} > 10\text{m/s}$. This can be used to safely detect violations of the safe distance, where $\texttt{vio}_{4.1.1}$ evaluates to true only if it is violated:

$$\texttt{vio}_{4.1.1} \coloneqq \exists f_1, f_2 \in \texttt{Vehicle} : \texttt{preceding}(f_1, f_2) \wedge \texttt{generally}_s(f_1, f_2) \wedge \neg \texttt{enough\_distance}(f_1, f_2).$$

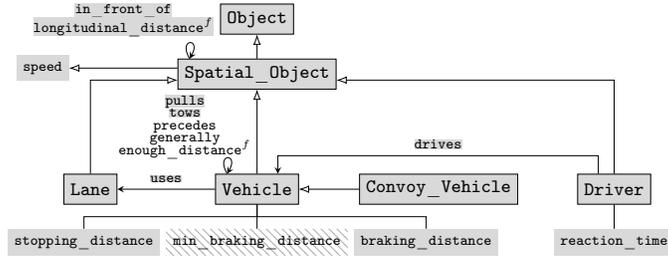

**Fig. 2.** Ontology and atoms including the relevant terms for the exemplary distance rule. Observable entities are denoted by a gray background, partially formalized atoms are hatched. A directed edge is a binary relation where an open tip denotes the subsumption relation. Properties are indicated by solid lines and functions by superscript $^f$.

Observers are created by translation to EP Universal Patterns, where observable quantities, e.g. `speed`, and atoms, e.g. `generally`, are represented as EP macros.

### 3.2 Results

In summary, over 160 traffic rules from the StVO have been identified as relevant for highway driving. 69% of these rules were formalized as described above, while 26% defines particular terms and thus are covered at one of the two layers. 4% of the rules are rather explanations and covered otherwise by the formalization. The remaining rules define priorities. For rule formalization, more than 600 terms were needed, from which half have been defined as observable entities, and the other half as atoms. Less than 60 atoms could not be (completely) formalized, where half of them were identified as vague legal concepts. This led to the overall result that 62% out of the formalized rules were *completely* formalized. The remainder depends on not or only partially formalized terms. Simulation scenarios for 26 rules have been created, where both satisfaction and violation of the corresponding rule have been checked by handcrafted observers.

## 4 Lessons Learned and Future Research Directions

The described practical experience was accompanied by various challenges. We aim to extrapolate our lessons learned to enable, in the end, elaborations on future research directions. We structure the discussion of our report as follows:

- Firstly, we introduce the central problem – called *congruence problem* – whose resolution we distilled as an abstract key goal during our activities;
- secondly, we present challenges that arise from the congruence problem;
- and finally, we sketch potential research directions for these challenges.

### 4.1 The Congruence Problem

In a nutshell, the *congruence problem* is concerned with the equivalence of semantics between a legal interpretation and the system's implementation. Before in-depth elaborations, we introduce the framework in which we place its definition.

**The Semiotic Triangle as a Foundational Model**  As we are concerned with the equivalence in understanding of stakeholders, the *semiotic triangle* as introduced by Ogden and Richards [15] is a natural fit. It models how humans express conceptualizations of their world and is concerned with

1. symbols, which are identifiers in a given language,[1]
2. concepts, which are semantic interpretations of the symbols, and
3. referents, which are the entities in some world referred to by a concept.

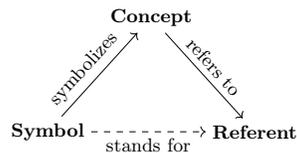

**Fig. 3.** The semiotic triangle, relating symbols, concepts, and referents.

Their relation is depicted in Fig. 3. Here, a symbol, e.g. 'vehicle', symbolizes an (intended) concept, e.g. 'a machine destined for transportation'. Communicating the symbol 'vehicle' invokes in the receiver the concept of a vehicle. This concept in turn maps to a set of things, called referents. This set can include, e.g., the bicycle in front of your house. Note that the mapping is not necessarily explicit. Because concepts are learned, humans with different background may use the same symbol for different concepts, or use different symbols for the same concepts. We now formally introduce the notion of a semiotic triangle.

---
[1] In our experience report, we referred to terms, which are defined as symbols that occur in the natural language specification of the rule set.

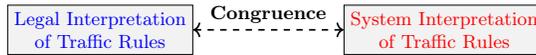

**Fig. 4.** Congruence between legal and system interpretation of rules

**Definition 1 (Semiotic Triangle).** *A semiotic triangle is a triple* $T = (S, R, C)$, *where* $S$ *is the set of all symbols,* $R$ *the set of referents, and the conceptualization* $C$ *a function* $C : S \to 2^R$, *assigning every symbol a set of referents.*

We emphasize that referents can be anything that can be referred to; this includes imaginary objects as well as traffic rules and scenarios. Furthermore, we highlight that $C$ may not be perfectly explicated. The interpretation of traffic rules is especially affected by this, as traffic rules are designed by and for humans. They exploit, e.g., inherent vagueness of concepts in order to reduce the degree of required explicitness, keeping the rule set concise yet comprehensible.

**The Congruence Problem** The eventual goal of the system's development is the implementation of some valid interpretation of the relevant legal concepts, a circumstance we call *congruence*. This relation is depicted in Fig. 4. Of course, this view is simplifying as there may be dissensions also among legal experts. Though this work assumes a harmonized legal interpretation, our general idea may be applied to consensus-building as well. The goal is to maximize congruence of legal and system interpretations of traffic rules, i.e. implement a system that judges a scenario (1) as conform to a traffic rule if and only if a legal expert does, and conversely (2) as violating a traffic rule if and only if a legal expert does.

Our practical experience shows that perfect congruence will not be achievable for traffic rules. Rather, we are interested to maximize the cases where the system judges a scenario as conform implies that the legal expert would have done so as well. This is due to an automated legal judgement of all scenarios being unrealistic; despite, we are interested in maximizing legal compliance of the system. Based on Definition 1, we are now ready to formally define congruence.

**Definition 2 (Congruence of Semiotic Triangles).** *The semiotic triangles* $T_1$ *and* $T_2$ *over symbol sets* $S_1$ *and* $S_2$ *with* $S_1 \cap S_2 \neq \emptyset$, *and concepts* $C_1$ *of* $T_1$, $C_2$ *of* $T_2$, *are* congruent *if for every* $s \in S_1 \cap S_2$ *it holds that* $C_1(s) = C_2(s)$.

Congruence hence requires that each shared symbol stands for the same set of referents. A visual interpretation is depicted in Fig. 5, assuming that both parties refer to a shared set of referents, i.e., to the same universe.

The definition is motivated by the implication that, if the semiotic triangles of the involved parties are congruent, their interpretation of the traffic rules and all involved symbols has to be congruent as well, therefore achieving the goal sketched in Fig. 4. In our practical experience, we identified congruence as a key step required for correctly formalizing, implementing, and verifying traffic rules. It is only due to congruence that subsequent discussions around technical means, e.g. suitable logics as presented in Sect. 2, become substantiated.

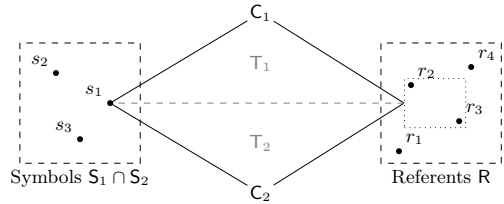

**Fig. 5.** Visualization of the definition of congruence between two semiotic triangles.

Up to this point, we have introduced both semiotic triangles – a formal framework on comprehension and communication – as well as congruence – a central problem in constructing a system that obeys traffic rules. The problem of establishing congruence is non-trivial. We emphasize this by understanding its emergence in the development process using the framework of semiotic triangles.

**Emergence of the Congruence Problem** Our report relates to the development of a safety-critical ADS by means of a state-of-the-art development process, i.e. compliant to ISO 26262 [9] and ISO 21448 [10]. A significant portion of the safety case will be concerned with establishing congruence between the legal and system interpretation of traffic rules. Obviously, a system development process involves a variety of relevant parties apart from the legal expert and the system. This includes requirement engineers, software programmers, test designers, certification authorities, and even those interacting with the product during operation. Each is equipped with a semiotic triangle, bidirectionally trying to achieve congruence with other stakeholders' triangles, such as those up- and downstream but also on the opposite development phase. This harmonization chain along a representative development scheme is depicted in Fig. 6.

A prototypical development process is displayed as a bridge between the legal expert and the system. Initially, the legal expert interprets laws and acts s.t. there arises a set of requirements on the system. They are written using the symbols and conceptualization of the legal expert and systematically transformed into a verified and validated system along a complex, iterative sequence of steps. For example, initially, legal requirements are decomposed by a requirements engineer into item-level and technical requirements which can then be used in the design phase. In the end, the system thus implements neither the traffic rules nor the interpretation of the legal expert directly but rather concepts that originated through the alignment of various semiotic triangles. This leads to the core issue:

> *How can the congruence of the semiotic triangles of the system and the legal expert be maximized in a development process?*

A solution is the inductive definition of concepts based on other concepts.

*A Decompositional Approach for Congruence:* A natural way of aligning semiotic triangles of expert groups is to ask the experts for definitions of the relevant legal symbols. This is based on the *compositionality principle* of symbols [16].

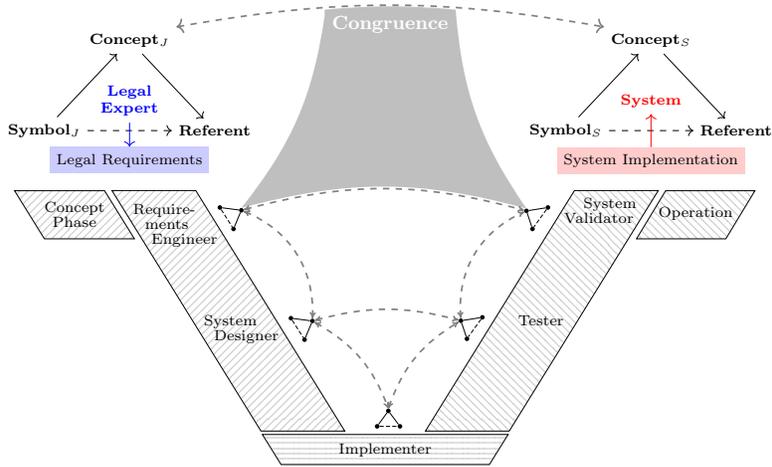

**Fig. 6.** The emergence of congruence along a representative development process.

For example, we can use the symbols *distance to*, *preceding*, and *vehicles* to construct a new symbol *distance to preceding vehicles*. In general, symbols can have different contextual semantics, so it is not possible to reconstruct the semantics of a sentence from the non-contextual semantics of its atoms as obtained by syntactic decomposition. However, we assume a clear context with no syntactic ambiguity, justified by the observation that traffic rules are intentionally written this way. Under this assumption, the compositionality principle states that the meaning of a sentence is unambiguously inferred from the contextual meaning and the ordering of contained symbols. We can inductively build definitions from an initial symbol set $S_0$ symbolizing the same concepts for the first (C) and the second stakeholder (C'). As Fig. 7 shows, by having congruent semiotic triangles $T_0$ and $T_0'$, and $S_0$ for definitions of new symbols $S \setminus S_0$, the semiotic triangles T and T' are also congruent. The two-level approach of Sect. 3 exploits this idea.

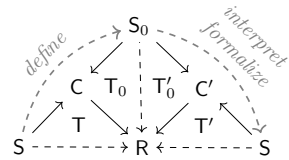

**Fig. 7.** An approach for congruence by compositionality.

### 4.2 Key Challenges in Addressing the Congruence Problem

We introduced the congruence problem as a central issue. This sections substantiates its hardness by challenges that need to be solved as to mitigate its effect. We inferred them from our practical insights, but we do not claim completeness.

**Alignment** A development process involves various cross-domain experts, including lawyers, logicians, and engineers (cf. Fig. 6). Alignment is the establishment of congruence between these parties as to achieve an overall congruence of the legal and system's interpretation. This challenge was already experienced during our work, where comparatively small teams collaborated. It becomes even more pressing when involving entire companies along automotive supply chains.

**Example 1** *For the distance rule, we identified the symbol* vehicle. *An alignment has to be reached as to what a vehicle entails. Whereas a machine learning engineer may learn a perception component on a data set with only wheeled vehicles, a legal expert may assume that a vehicle may not need to have wheels. Thus, the system may exhibit illegitimate behavior near rail vehicles.*

In the framework of semiotic triangles, alignment includes explicating the set vocabulary of relevant legal symbols $\mathsf{S}$, and then agreeing on a definition of $\mathsf{C}(s)$ for each $s \in \mathsf{S}$ and each involved stakeholder in the development process.

**Observability** In our experience report, we relied on decomposing terms into sub-terms until one assumes observability by the perception chain. The question arises: how do we agree on a justification for this observability assumption? This requires in-depth stakeholder discussions during the design and implementation phase, e.g. when selecting suitable perception technologies from suppliers.

**Example 2** *We assumed vehicles to be observable based on a given natural language explanation for the concept of a* vehicle. *We now have to justify this, or, alternatively, need to ask the legal experts for a detailed definition for the concepts used in the definition of a vehicle, such as* machine *and* transportation.

Formally, the challenge of observability asks for a justification of the assumption that it is sufficient to give an informal explication of $\mathsf{C}(s)$ for an $s \in \mathsf{S}$.

**Vagueness** Vagueness can be understood as the circumstance that concepts necessary for situation assessment do not have a clear truth condition. This implies that their set of referents is not *crisp* but rather *fuzzy*. For traffic rules, a common source of fuzziness are vague legal concepts. This can lead to the use of non-exhaustive enumerations of examples in definitions, which, during the development process, hinders the alignment. Thus, this challenge is located even prior to system development, namely within the traffic rules themselves.

**Example 3** `generally` *is a vague legal term as exceptions can not be exhaustively explicated, which, legally, indicates the necessity of case-by-case decisions. However, there are special cases such as platooning that can be safely excluded.*

In the framework of semiotic triangles, a concept $\mathsf{C}(s)$ of some $s \in \mathsf{S}$ is vague if a referent $r \in \mathsf{R}$ exists for which the truth value of the statement $r \in \mathsf{C}(s)$ can not be exactly identified even if all necessary information is perfectly present.

**Uncertainty** If we assume a crisp membership function $\mathsf{C}(s)$, we could, in theory, clearly label referents by their symbols, yielding an evaluation of a scenario with absolute certainty. Although, for this to hold, we have to assume that we

1. considered all parts of the vehicle's environment needed for concept evaluation during its design and built a perception system able to recognize them,
2. knew every legal detail and the perception is perfectly certain, and
3. created a non-probabilistic environment model based on a certain perception.

Those assumptions correspond to *ontological*, *epistemic*, and *aleatoric* uncertainties [8]. In the development process of Fig. 6, we find ontological and aleatoric uncertainty arising mostly during system design and implementation. Epistemic uncertainty can be located in both the early design phase – a lack of knowledge of the designer – or the operation phase – a lack of knowledge of the system. In practice, these assumptions are unrealistic due to physical and economical limitations; thus, to achieve congruence, we have to consider such uncertainties.

**Example 4** *For the distance rule, ontological uncertainty is the neglect of weather conditions when formalizing the stopping distance. Subsequently, epistemic uncertainty is the reception of uncertain information about the weather condition from a rain sensor of the ADS. Aleatoric uncertainty exists if we estimate a probability distribution of the current coefficient of friction based on this data.*

Ontological uncertainty requires a relevant symbol $s$ to be not considered when collecting the legal symbols $\mathsf{S}$, and subsequently developing the perception system. Epistemic uncertainty exists if there is some $r \in \mathsf{R}$ for which the truth value of the statement $r \in \mathsf{C}(s)$ can not be exactly identified due to some theoretically crisp information not being known at design- or runtime. Aleatoric uncertainty can be seen as a concept function $\mathsf{C}(s)$ that answers $r \in \mathsf{C}(s)$ only with a certainty $p \in [0, 1]$ due to unmeasurable random factors influencing the evaluation.

**Interrelations** Terminologically, we encounter *synonyms*, different symbols with the same meaning, and *polysemy*, similar symbols with different meanings.

**Example 5** *Note that* distance *is a homograph. It may refer to a spatial, temporal, or even social distance. A legal analysis clarifies that it is to be interpreted spatially. Here, a synonym for spatial distance, e.g.* gap*, can be helpful.*

For some symbols $s \neq s'$, $\mathsf{C}(s) = \mathsf{C}(s')$ indicates a synonym. Polysemy can be seen as $\mathsf{C}(s) \neq \mathsf{C}(s')$ for some $s \approx s'$, where $\approx$ denotes linguistic similarity.

Interrelations of concepts – including rules – are often present in traffic rules. Traffic rules are especially affected by *priorities and exceptions* as well as *inconsistencies*. Even different rule sets can interact, e.g. by lex specialis. First of all, the challenge arises how to identify those interrelations inherent to traffic rules. The subsequent handling of priorities is then necessary to congruently reflect their semantics. Assume, e.g., the rules $\varphi_1 \coloneqq a \implies b$ and $\varphi_2 \coloneqq c \implies d$, not necessarily taken from the same rule set. A thorough legal analysis delivers us with the fact that these rules interrelate by $a \wedge c$ being satisfiable but $b \wedge d$ being unsatisfiable. If $a \wedge c$ holds, the implied behavior is unclear, as the system can not be compliant with both $b$ and $d$. Prioritization can resolve this ambiguity by e.g. assigning $\varphi_1$ a higher precedence, therefore implying that $b$ should follow in case $a \wedge c$ holds. Exceptions arise when the constraint of the higher prioritized rule (e.g. $\varphi_1$) is implied by the general case, i.e. if $a \implies c$.

Inconsistencies arise due to incompatible concepts. For the above sketch, without prioritization, it would have followed that $\varphi_1$ and $\varphi_2$ are contradictory and thus unsatisfiable. This is undesired during formalization and implementation – it indicates formalization errors and can lead to unspecified behavior.

**Example 6** *We gave a necessary and sufficient version for the distance rule using* $\mathtt{generally}_n(f_1, f_2) \coloneqq f_1, f_2 \notin \mathtt{Convoy\_Vehicle}$ *and* $\mathtt{generally}_s(f_1, f_2) \coloneqq f_1, f_2 \notin \mathtt{Convoy\_Vehicle} \wedge f_1.\mathtt{speed} > 10m/s$. *For these to be valid,* $\mathtt{generally}_s \implies \mathtt{generally}_n$ *has to hold. Imagine we made a modeling error by means of* $\mathtt{generally}_s(f_1, f_2) \coloneqq f_1.\mathtt{speed} > 10m/s$. *Then,* $\mathtt{generally}_s \wedge \neg \mathtt{generally}_n$ *is satisfiable but inconsistent with* $\mathtt{generally}_s \implies \mathtt{generally}_n$.

Generally, a priority is an arbitration mechanism that, for some $r \in \mathsf{R}$ satisfying two (contradictory) concepts, decides which concept $r$ is assigned to. Inconsistency, on the other hand, is often based on the reduction to unsatisfiability. This arises when there is some $s \in \mathsf{S}$ with $\mathsf{C}(s) = \emptyset$. This may be due to a contradiction with another concept or a contradiction within its own concept function. It can also arise from more subtle forms of contradictions such as sufficient and necessary conditions conflicting with the first implying the latter.

**Traceability** We are concerned with traceability of the symbols and their concept functions arising during the congruence procedure. As our practical experience shows, this allows to iteratively detect and resolve reasons for misinterpretations during alignment, hence facilitating congruence. Without traceability support, keeping track of the large amount of traffic rules and their artifacts is almost impossible. Traceability thus affects the entire process of Fig. 6.

**Example 7** *A definition of* vehicle *was used for labeling training data of a machine learning component. During testing, we find that the ADS does not hold enough distance around rail vehicles. Traceability allows to pinpoint the mismatch between the machine learning engineer's and legal expert's concept. We can now align the semiotic triangles of the engineer and legal expert.*

Traceability is formally concerned with the storage and accessibility of all artifacts $\mathsf{C}$ and $\mathsf{S}$ of all semiotic triangles involved in the system life cycle.

### 4.3 Solutions to be Developed in Future Work

We close the discussion by sketching open or partially examined ends, as to identify future research directions. To that effect, we propose investigations of the following topics, some of which the related work of Sect. 2 has touched on:

- Methods for alignment and uncertainty reduction in development processes.
- Suitable formalisms to express concepts as validly as possible, e.g. ternary, fuzzy, temporal, probabilistic, defeasible, and deontic logics.
- Tools for the formal specification of traffic rules within these formalisms, including the handling of complex interrelations, such as priorities.
- Verification technology demonstrating the system's adherence to traffic rules, which is suitable for the usage in a rigorous, ISO-compliant safety case.
- Means of ontology mapping for e.g. mapping simulator on legal ontologies.
- Traceability tooling that overarches all processes, where all artifacts concerning traffic rule compliance are represented and interlinked.

## 5 Conclusion

We presented a report of experience collected during the formalization of a non-trivial subset of the StVO. Based on this experience, we extrapolated the congruence problem and connected it to a typical development process. The formal framework of semiotic triangles was utilized to delineate which challenges arise when addressing congruence. Their practical relevancy was highlighted using the exemplary distance rule taken from the StVO. We finally sketched how some of these challenges may be addressed in future research and industrial efforts.

**Acknowledgements** We thank Tom Bienmüller and Tino Teige from *BTC Embedded Systems AG* for support and discussions during the formalization. This work was partially funded by the *Federal Ministry of Education and Research* (BMBF) as part of *MANNHEIM-AutoDevSafeOps* (reference no. 01IS22087C).

## References

1. Athan, T., Boley, H., Governatori, G., Palmirani, M., Paschke, A., Wyner, A.: OASIS LegalRuleML. In: Proc. 14th Int. Conf. on Artificial Intelligence and Law (ICAIL 2013). pp. 3–12 (2013)
2. Baader, F., Calvanese, D., McGuinness, D., Patel-Schneider, P., Nardi, D., et al.: The description logic handbook: Theory, implementation and applications. Cambridge university press (2003)
3. Bhuiyan, H., Olivieri, F., Governatori, G., Islam, M.B., Bond, A., Rakotonirainy, A.: A methodology for encoding regulatory rules. In: MIREL@ JURIX (2019)
4. Buechel, M., Hinz, G., Ruehl, F., Schroth, H., Gyoeri, C., Knoll, A.: Ontology-based traffic scene modeling, traffic regulations dependent situational awareness and decision-making for automated vehicles. In: 2017 IEEE Intelligent Vehicles Symposium (IV). pp. 1471–1476. IEEE (2017)
5. Costescu, D.M.: Keeping the autonomous vehicles accountable: Legal and logic analysis on traffic code. In: Conference Vision Zero for Sustainable Road Safety in Baltic Sea Region. pp. 21–33. Springer (2018)
6. Den Haan, N., Breuker, J.: A tractable juridical KBS for applying and teaching traffic regulations. Legal knowledge-based systems. JURIX **91**, 5–16 (1991)

7. Esterle, K.: Formalizing and Modeling Traffic Rules Within Interactive Behavior Planning. Ph.D. thesis, Universität München (2021)
8. Gansch, R., Adee, A.: System theoretic view on uncertainties. In: 2020 Design, Automation & Test in Europe Conference & Exhibition (DATE). pp. 1345–1350. IEEE (2020)
9. International Organization for Standardization: ISO 26262: Road vehicles – Functional safety. Standard, Geneva, Switzerland (2018)
10. International Organization for Standardization: ISO 21448: Road vehicles – Safety of the intended functionality. Standard, Geneva, Switzerland (2022)
11. Karimi, A., Duggirala, P.S.: Formalizing traffic rules for uncontrolled intersections. In: 2020 ACM/IEEE 11th International Conference on Cyber-Physical Systems (ICCPS). pp. 41–50. IEEE (2020)
12. Maierhofer, S., Moosbrugger, P., Althoff, M.: Formalization of intersection traffic rules in temporal logic. In: 2022 IEEE Intelligent Vehicles Symposium (IV). pp. 1135–1144 (2022)
13. Maierhofer, S., Rettinger, A.K., Mayer, E.C., Althoff, M.: Formalization of interstate traffic rules in temporal logic. In: 2020 IEEE Intelligent Vehicles Symposium (IV). pp. 752–759. IEEE (2020)
14. Neurohr, C., Westhofen, L., Henning, T., de Graaff, T., Möhlmann, E., Böde, E.: Fundamental considerations around scenario-based testing for automated driving. In: 2020 IEEE Intelligent Vehicles Symposium (IV). pp. 121–127. IEEE (2020)
15. Ogden, C.K., Richards, I.A.: The meaning of meaning: A study of the influence of thought and of the science of symbolism. Harcourt, Brace & World, Inc. (1923)
16. Riemer, N.: Meaning in the empirical study of language, p. 1–44. Cambridge Introductions to Language and Linguistics, Cambridge University Press (2010)
17. Rizaldi, A., Keinholz, J., Huber, M., Feldle, J., Immler, F., Althoff, M., Hilgendorf, E., Nipkow, T.: Formalising and monitoring traffic rules for autonomous vehicles in Isabelle/HOL. In: International conference on integrated formal methods. pp. 50–66. Springer (2017)
18. Salem, N.F., Haber, V., Rauschenbach, M., Nolte, M., Reich, J., Stolte, T., Graubohm, R., Maurer, M.: Ein Beitrag zur durchgängigen, formalen Verhaltensspezifikation automatisierter Straßenfahrzeuge. preprint arXiv:2209.07204 (2022)
19. Schwammberger, M., Alves, G.V.: Extending urban multi-lane spatial logic to formalise road junction rules. In: Farrell, M., Luckcuck, M. (eds.) Proceedings Third Workshop on Formal Methods for Autonomous Systems, FMAS 2021, Virtual, October 21-22, 2021. EPTCS, vol. 348, pp. 1–19 (2021)
20. Sergot, M.J., Sadri, F., Kowalski, R.A., Kriwaczek, F., Hammond, P., Cory, H.T.: The british nationality act as a logic program. Communications of the ACM **29**(5), 370–386 (1986)
21. Teige, T., Bienmüller, T., Holberg, H.J.: Universal pattern: Formalization, testing, coverage, verification, and test case generation for safety-critical requirements. In: Workshop Methoden und Beschreibungssprachen zur Modellierung und Verifikation von Schaltungen und Systemen. Albert-Ludwigs-Universität Freiburg (2016)
22. UNECE, Geneva, Switzerland: UN Regulation No. 157: Uniform provisions concerning the approval of vehicles with regard to ALKS (2022)
23. Villata, S., et al.: Traffic rules encoding using defeasible deontic logic. In: Legal Knowledge and Information Systems: JURIX 2020: 33rd Annual Conference. vol. 334, p. 3. IOS Press (2020)

# Traffic Rule Formalization for Autonomous Vehicle

Hanif Bhuiyan [1,2] [0000-0002-6064-8525], Guido Governatori [3] [0000-0002-9878-2762],
Andry Rakotonirainy [2] [0000-0002-2144-4909], Meng Weng Wong [3] [0000−0003−0419−9443],
and Avishkar Mahajan [3] [0000−0002−9925−1533]

[1] Data61, CSIRO, Brisbane, Australia
[2] Queensland University of Technology, CARRS-Q, Queensland, Australia
[3] Centre for Computational Law, Singapore Management University, Singapore

**Abstract.** This study devised and implemented a Defeasible Deontic Logic (DDL)-based formalization approach for translating traffic rules into a machine-computable (M/C) format and thus solving rule issues: rule vagueness (open texture expressions) and exceptions in rules. The resulting M/C format of traffic rules can be utilized for automatic traffic rule reasoning to assist the Autonomous Vehicle (AV) in making legal decisions. The method incorporates the components and behaviour of regulations based on the rule's obligation, prohibition, and permission activities.

The need for the encoding methodology is motivated by the desire for automated reasoning over Autonomous Vehicle information involving traffic rules. A Queensland (QLD) overtaking traffic rule is used as a use case to illustrate this proposed encoding methodology's mechanism and usefulness.

**Keywords:** Traffic Rules, Norms, Defeasible Deontic Logic.

## 1 Introduction

Over the last few decades, intelligent systems have been a widely accepted technology with various degrees of interaction. However, despite the constructive and promising impact, this advancement of technology has some negative impacts. For example, while we know that the technological advancement of vehicles is necessary and advantageous for society, it is also known that road crashes are one of the major concerns of global public health due to the growth of road fatalities and human disabilities. Every day, more than 3,700[1] people die due to road crashes, and it was found that the driver's behaviour is solely responsible for 90% of these crashes. From January 2011 to January 2020, in Australia, 12274[2] people died in road crashes. From 2013-2017[3], in Queensland, the average number of deaths due to high speed was 58 per year. Therefore, it can be assumed that if drivers drive according to traffic rules, there might be less chance of fatalities and injuries.

---

[1] https://www.who.int/violence_injury_prevention/road_traffic/en/
[2] https://www.bitre.gov.au/statistics/safety
[3] https://streetsmarts.initiatives.qld.gov.au/speeding/factsheet

An Autonomous Vehicle (AV) can be introduced to automatically make the driving decision according to road rules [1]. As AVs are designed and programmed to follow traffic rules [2], therefore, it is suggested that AVs would be the immediate solution to traffic violations [3].

However, it is still unclear how AVs will fit into the existing regulatory framework. There is no separate and comprehensive regulatory framework for Avs [4]. Leenes and Lucivero [2] mentioned that the current traffic rule model for the AV might be incomplete for some scenarios of the road. For example, in the current traffic rules, there are some vague expressions (e.g. "can safely overtake", "over- take when there is a clear view", etc.), which are almost impossible for an AV to follow [5]. Also, it may not be possible for AVs to properly follow the rules related to exceptions [5, 6]. Therefore, it seems conceivable to formalize traffic rules into the machine-computable format by resolving the above-mentioned issues to make them followable for AVs. This formalization can bridge the gap between traffic rules and further knowledge processing for AVs.

In this paper, we intend to discuss the methodology for making the legal decision for AVs according to Queensland overtaking traffic rules. We choose overtaking traffic rules as it is one of the most challenging traffic rules, which has several complicated and varied conditions. The formalization is designed using DDL to successfully handle the exceptions and resolve the vague terms in rules. Our contributions to this work are:

- We have formalized the Queensland overtaking traffic rules using Defeasible Deontic Logic (DDL).
- We evaluate this formalization through a comprehensive experiment.

## 2 Related Work

Rule formalization (i.e., the representation of legal rules in a machine-computable format) is a crucial requirement for compliance checking, automatic reasoning, and legal validation [7]. However, it is a difficult task because of the domain-specific sentence length, clause embedding, and structure. Rules contain thousands of provisions and norms, making formalizing work even more challenging . Several studies have been conducted to address the issues of rule formalization. In addition, several languages and products have been proposed to formally represent them. For example, LegalRuleML is a novel XML standard for the representation of norms [8]. Several commercial products, such as Oracle Policy Automation , offer services to translate rules into executable language and give a user-friendly natural language online interface.

Traffic rules are primarily written in natural language. Due to the complex and varying nature of traffic rules, it is a challenging task to formalize them. Any incorrect rule formalization might have a negative impact on the reasoning process. Despite challenges, several studies focused on formalizing traffic rules for different purposes. Some significant recent studies regarding traffic rule formalization are:

Zhao, et al. [9] introduced an ontology model for making a fast decision according to traffic rules at the intersection. Ontologies in this proposed method represent knowledge of the sensory data. Traffic rules were represented through SWRL rules. However, the methodology was limited to working only on specific traffic rules. Hence, to include new regulations, further work was needed.

Zhang, et al. [10] proposed an expert system to formalize traffic rules to develop the driving knowledge base, which can play a significant role in making intelligent driving decisions for AV. A knowledge acquisition method was applied to build traffic rules knowledge base. This method defined traffic rules concepts, quantitatively explained rule characteristics, and created a logical relation between rule terms.

Rizaldi and Althoff [11] formalized traffic rules to identify which vehicle was liable for the collision. This work aimed to prove that the AV always obeys the traffic rules, and therefore, there was no chance for the AV to become responsible for the collision. The overall process was conducted mainly in three steps. First, a subset of Vienna traffic rules was formalized and concretized in the Higher-Order-Logic (HOL). Second, a black box recorded the behaviour of the AV. Later, the author extended the work and tried to solve the rule vagueness issue [5].

Censi et al. (2019) proposed a rulebook, a formalism methodology for UK & Singapore rules, to make self-driving vehicle behaviour compatible with the current traffic regulations. Rulebooks define driving behaviour by defining rules precisely and establishing a hierarchy of rules. The author experimented with this rulebook on three specific traffic scenarios: unavoidable collision, lane change near intersection and clearance and lane-keeping based on 15 rules. However, the rulebook was domain-specific, and for a different nation, the priorities of the rules needed to be changed.

McLachlan, et al. [12] proposed a method for deconstructing the traffic rule and representing it with the necessary needs and flow for the decision making of autonomous driving. Using legal vocabulary, the technique deconstructs road regulations, which were then specified in structured English logic, Boolean Logic. The key points: the chronology of operation and processes of rules were represented through the boolean logic. The evaluation of this rule representation was conducted on 23 UK road rules. However, this methodology of deconstructing traffic rules works only for simple traffic laws, such as traffic lights, seat belt wearing, speed limit etc.

None of these studies addresses the combination of vague (open texture) terms and exceptions in rules, which might create problems for AV's driving decision-making. Comparing these works in terms of handling and resolving vague terms and exceptions in rules, a Defeasible Deontic Logic (DDL) based formalization mechanism is proposed that can effectively handle and resolve these issues. DDL is a formalism that provides a conceptually approach to the formalize of the norm and, at the same time, exhibits a computationally feasible environment to reason about them. DDL has been successfully used in legal reasoning to handle norms and exceptions, and it does not undergo problems affecting other logic used for reasoning about compliance and norms [13, 14].

## 3   Traffic Rule Formalization

This paper introduces a novel formalization methodology to translate the traffic rules into the machine-computable (M/C) format. This proposed methodology works in four steps (Figure 1).

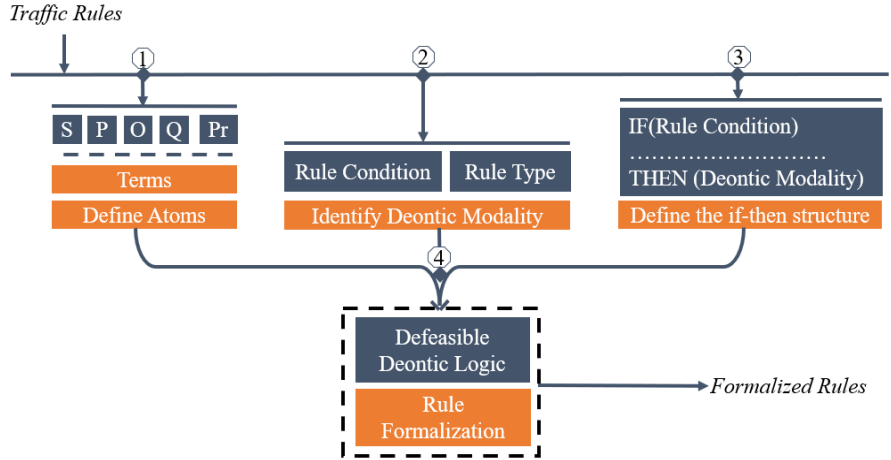

Figure 1. Traffic Rule Formalization Workflow

These steps are done manually. The methodology's input is a set of traffic rules in natural language. In the first step, atoms are defined from the rules. In the second step, norms are determined. Then, the if-then structure is identified from the rules in the third step. Finally, DDL is applied to the atoms, deontic modalities, and if-then structure to make the machine-computable (M/C) rules format. The steps are explained below.

### 3.1 Define Atom

This section briefly outlines how atoms are defined from rules. An atom is a predicate symbol including constants or variables that contain no logical connectives [15]. Here, the atoms are extracted based on the occurrences of terms/expressions in the sentences or textual provision of the relevant traffic rules. A *term* is a *variable* or an (individual) *constant* in the textual provision . This work deals with expressions (predicates, variables, and constants) that refer to subject (s), predicate (p), property (pr), object (o), and qualifier (q) in a rule sentence.

In natural language, a subject (or entity) refers to the term about which something is said in the sentence. The something which is said about something is the predicate of the sentence. The predicate of a subject-predicate sentence indicates a relation or a property. The object is what that subject does something to. In other words, the object is the result of the action. Qualifiers are terms that usually enhance or limit another word's meaning. In one sense, a qualifier can be thought of as an adverb of the sentence. Before generating terms, some article pre-processing is done on the text. For example, verbs (auxiliary, principal, modal, etc.) are not considered as terms. In logic, subjects and objects are variable or constant in the rule sentence correspondence to the entity [16]. A predicate is a constant in the rule sentence that always refers to the properties or actions of entities. Properties indicate the relationship between subject and predicate. Object refers to the properties of the entities. Qualifiers refer to the variables that enhance or limit entities.

An atom is a combination of these terms/expressions that form a (primitive) Boolean expression. For example, "*the bus breaks the traffic rule*". According to the linguistic perspective, the *term* "bus" is the subject of the sentence as this sentence is about this bus. The *term* "traffic rule" is the object of the sentence as the subject is doing something to it. The *term* "breaks" is the predicate of the sentence as it expresses the relation between the subject (bus) and object (traffic rule). In the logical approach, "bus" is that variable which is referring to the entity (subject) of the sentence. "breaks" is the predicate constant which is referring the action of the entity. "traffic rule" is a (individual) constant of the sentence which is referring to the properties of the entity. So logically, the above example can be represented as a predicate (subject, object): B(b,t) ≡ Breaks (bus, truck): Subject-Predicate-Object: *thebus_Breaks_theTrafficRule*.

Another example is Queensland Overtaking Traffic Rule 140, states "the driver can safely overtake the vehicle". In this rule, "driver" is the variable corresponding to the subject (entity) of the sentence. "overtake" is the predicate that refers to the action of the subject (entity). "can safely" is the qualifier that defines the predicate of this rule. "vehicle" is an individual variable that refers to the object of the rule. Integrating these four terms, an atom is defined as:

*Predicate (subject, qualifier, object) ≡ Overtake (driver, CanSafely, vehicle): driver_CanSafelyOvertake_vehicle.*

The current traffic rules use natural language to define the cases (events and facts) they are meant to regulate (terms, conditions, and legal provisions). Depending on the events, the description of these cases varies. There is no general structure in how traffic rules are written. Due to this heterogeneity of the rule information, the atom structure varies. Throughout the empirical study of the Queensland Overtaking Traffic Rules, atoms are defined in five patterns: Subject-Predicate-Object, Subject-Predicate-Qualifier-Object, Subject-Property, Subject-Predicate-Object-Object, and Subject-Qualifier-Predicate-Object. Based on these patterns, examples of atoms are shown in Table 1.

Table 1. Examples of defining atom from QLD traffic rules

| **Example 1:** Part 11, Division 3 rule 151: 1 (a) — driver must not overtake a vehicle |||
|---|---|---|
| Subject | Predicate | Object |
| driver | Overtake | vehicle |
| Defined Atom: NEG (driver_Overtake_vehicle). Pattern: Subject-Predicate-Object |||

| **Example 2:** Part 11, Division 3 rule 140: b — "the driver can safely overtake the vehicle". ||||
|---|---|---|---|
| Subject | Predicate | Qualifier | Object |
| the driver | overtake | safely | the vehicle |
| Defined Atom: driver_CanSafelyOvertake_vehicle. ||||

| **Example 3:** part 2, division 2, Rule 15 — "A vehicle includes a bicycle". ||
|---|---|
| Subject | Property |
| vehicle | is a bicycle |
| Defined Atom: vehicle_Isabicycle. Pattern: Subject-Property. ||

| **Example 4:** Part 11 Division 3 rule 151A-1: a — "motorbike between two adjacent lines". ||||
|---|---|---|---|
| Subject | Predicate | Object | Object |
| motorbike | between | adjacentLine1 | adjacentLine2 |
| Defined Atom: motorbike_InBetween_adjacentLine1_adjacentLine2. ||||
| **Example 5:** part 19 rule 305 — "A police vehicle is allowed not to display the light". ||||
| Subject | Qualifier | Predicate | Object |
| A police | vehicle | is allowed | not to display the light |
| Defined Atom: police_vehicle_allowed_NotToDisplayTheLight. ||||

These patterns are sufficient to cover the cases of Queensland overtaking traffic rules we considered for this reserach. While more patterns are possible, the patterns presented also offer guidance to capture more complex cases if needed.

The proposed methodology aims at improving the uniformity, consistency and repeatability of formalization efforts. Witt, et al. [17] report very high (syntactic) variability of formalization when they are done by a team of coders; moreover, they report that adopting a common naming convention and sharing a formalization methodology greatly increase the agreement among the formalizations by the different coders. Also, fitting textual provisions in the patterns allows us to identify expressions that could have different syntactic structures but the same semantic meaning; such expressions will be formalized by the same atoms.

### 3.2 Identify Deontic Modalities

Deontic modalities are expressions in the traffic rule that (legally) qualify terms and actions. They help us determine the types of norms we are going to formalize. Each norm is represented by one or more constitutive or prescriptive rules. Constitutive rules define terms specific to legal documents. Prescriptive rules prescribe the "mode" of the behaviour using deontic modalities: obligation, permission, and prohibition. Here we follow the definition given by LegalRuleML for obligation, prohibitions and permission [8]. An obligation is an action or course that the subject must perform, whereas prohibition is an action or course that the subject must not perform. Permission is the state of an action that is not subject to a prohibition or an obligation.

The prescriptive rules are determined based on conceptual semantic understanding and some special keywords, which are "must", "must not", "should", "ought", etc. Constitutive rules are identified through the semantic and syntactic mapping of descriptive notions in the sentences. Some examples of descriptive notions in Queensland overtaking traffic rules are "is", "means", "does not", etc.

This research identifies norms based on both constitutive and prescriptive forms of rules. For example (Figure 2), in the Queensland Traffic Rules, part 11, division 3, rule 141 states that "a driver must not overtake a vehicle to the left of the vehicle unless ….". Here "must not" (prohibition) is identified as a prescriptive norm within this statement. Another example in the Queensland traffic rule part 3, section 23, states, "A school zone is —". Here, this phrase is meant to define the school zone, which is a constitutive norm (Count-As) within this statement.

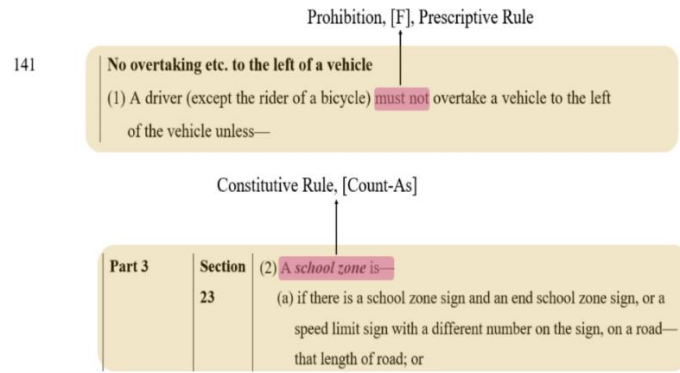

Figure 2. Examples of identifying deontic modalities (norms).

### 3.3 Define the If-then structure

Traffic rules specify the actions of the subject (or the conditions the subject must ensure hold). They consist of deontic modalities and conditions that control the subject's behaviour. A rule comprises two parts: if (antecedent or premise) and then (consequent or conclusion).

If the premise becomes true, then the consequent part of the rule triggers. A rule may have multiple antecedents joined by logical operators: OR, AND, and XOR. From a legal perspective, rules use conditions on some actions to achieve/mandate specific behaviours. Therefore, the if-then structure is identified from rules using atoms and deontic modalities (norms) for formalizing rules. For example, in the Queensland traffic rules, Part 11, Division 3, Rule 140 states:
A driver must not overtake a vehicle unless—
    (a) the driver has a clear view of any approaching traffic, and
    (b) the driver can safely overtake the vehicle".

The rule expresses an obligation norm for the driver to overtake. In this rule, there are two components. The first ($r_{140}$) is the obligation of not overtaking, and the second ($r_{140-exception}$) is the exception of the first rule, which is giving permission (negation of obligation) to overtake. By defining atoms and identifying norms, the if-then structure of this rule can be defined as:

---
  **r_{140}:**
  // the driver is not allowed to overtake
  IF
    \emptyset
  THEN[F] // *norm / deontic modality*
    %% must not overtake a vehicle %%
    NEG (driver_Overtake_vehicle)

---

---

**r_{140-exception}:**
  // the exception (unless) of the rule is that the driver is allowed to overtake if he/she meet the following conditions.//
  IF
    %% the driver has a clear view of any approaching traffic %%
    driver_HasClearViewOf_approachingTraffic // *atom*
  AND
    %% the driver can safely overtake the vehicle %%
    driver_CanSafelyOvertake_vehicle // *atom*
  THEN [P] // *norm / deontic modality*
    %% A driver must not overtake a vehicle unless %%
    driver_Overtake_vehicle // *atom*

---

  **r_{140- exception} ≫ r_{140}** // *Superiority relation*

---

### 3.4 Rule Formalization

After defining and identifying atoms, deontic modalities and if-then structures for rules, the expressions are converted into a Defeasible Deontic Logic (DDL). DDL is an extension of Defeasible Logic (DL) with Deontic Operators and compensatory obligation operators introduced [18]. DDL is a formalism that provides a conceptually approach to the formalizing of norm and, at the same time, exhibits a computationally feasible environment to reason about them. DDL has been successfully used in legal reasoning to handle norms and exceptions, and it does not undergo problems affecting other logic used for reasoning about compliance and norms [13, 14]. Below is a brief overview of Defeasible Logic and Deontic modalities and how we used them to represent traffic rules.

**Defeasible Logic:** Defeasible Logic (DL) is a non-monotonic, sceptical logic that prevents the derivation of contradictory conclusions. For example, suppose there is a piece of information that supports the conclusion A, but also there is a second piece of information that supports not A, preventing thus concluding A. DL recognized the opposite conclusions and does not derive them. However, if A's support has priority over ¬ A, then it might be possible to conclude A.

Defeasible Logic is made up of five separate knowledge foundations: strict rules, facts, defeasible rules, defeaters, and superiority relations. [19]. Knowledge is organized in a theories, where a *theory* D is a triple (F, R, ≫) where F is a set of facts, R is a set of rules, and ≫ is a superiority relation in R.

Expressions in Defeasible Logic are built from a finite set of *literals*, where a *literal* can be either an atomic statement or its negation. Given a literal A, ~A denotes its complement. That is, if A=B, then ~ A= ¬ B and if A = ¬ B, then ~ A = B. Facts (F) are unequivocal and conclusive statements. A fact represents a state of affairs (literal) or an act that has been performed and are believed to be true.

A *rule* (an element of R) specifies the relationship between premises and conclusion and can be characterized as its strength. Strict rules, defeasible rules and defeaters can be distinguished based on the relationship strength of the rules [20]. The following expressions describe these rules:

$A_1, \ldots, A_n \rightarrow B$ (Strict Rules),
$A_1, \ldots, A_n \Rightarrow Y$ (Defeasible Rules) and
$A_1, \ldots, A_n \rightsquigarrow$ (Defeaters) B,

where $A_1, \ldots, A_n$ is the antecedent or premises (clauses), and Y is the consequent or conclusion (effect) of the rule.

*Strict rules* are rules in the classical sense: if the premises are unarguable (for example, a fact), then is the conclusion. For example, "a motorbike is a vehicle," formally:

$$\text{Motorbike} \rightarrow \text{Vehicle};$$

*Defeasible rules* are rules that can be defeated by contrary evidence. For example, "a motorbike can edge filter", formally, can be written as:

$$\text{Motorbike} \Rightarrow \text{Edge\_Filtering\_Vehicle}.$$

*Defeaters* are rules that cannot be used to derive any conclusions on their own. Their purpose is to preclude some conclusions, i.e., to undermine some defeasible rules by supplying opposite evidence. For example, suppose a rule state that: "if a rider does not hold O type license, then the rider cannot edge filter".

$$\neg \text{ rider\_HoldOTypeLicence} \rightsquigarrow \neg \text{ Edge\_Filtering\_Vehicle}.$$

From this statement (and the previous one), it can be stated that a motorbike is an edge filtering vehicle, but if the rider of motorbike does not hold an O type licence, then it cannot edge filter on the road. This statement can prevent the conclusion of edge filtering. This is not also supporting the 'no edge filtering'.

The *superiority relation* ($\gg$) used the priority set among the rules, where one rule may override the other rule's conclusion. No conclusion can be made in such scenarios unless the rules are prioritized. For example, based on the following defeasible rules:

$r_1$:   Motorbike $\Rightarrow$ Edge_Filtering_Vehicle.
$r_2$:   Vehicle $\Rightarrow$ ¬ Edge_Filtering_Vehicle

No conclusive decision can be made about whether a vehicle can edge filter. However, if we establish a superiority relation $\gg$ with $r_1 \gg r_2$, then we can state that the vehicle cannot edge filter. A complete definition of defeasible logic reasoning mechanism can be found in [13].

**Deontic Operators:** In addition to defeasibility, traffic rules contain extensive occurrences of deontic concepts. This research considered Obligation [O], Prohibition [F] and Permission [P] deontic operators to formalize traffic rules. The deontic operators are modal operators. A modal operator applies to a proposition to create a new proposition where the modal operator qualifies the "truth" of the proposition to which the operator is applied. For instance, a proposition from QLD Overtaking Traffic Rule 141: *driver_OvertakeToTheLeftOf_vehicle* means that the "driver is overtaking the front vehicle from its left side". We can distinguish this proposition based on the above deontic operators:

— *OvertakeLeft:* this is a factual statement that is true if the vehicle overtakes from the vehicle's left and false otherwise (¬ OvertakeLeft is true).
— [O]*OvertakeLeft:* this is a deontic statement meaning that the vehicle must overtake the left of the vehicle. The statement is true if the obligation to overtake is in force in the particular case.
— [F]*OvertakeLeft:* this is a deontic statement meaning that the vehicle is prohibited from overtaking the vehicle on the left-hand side. The statement is true if the prohibition to overtake is in force in the particular case.
— [P]*OvertakeLeft:* this is a deontic statement meaning that the vehicle has permission to overtake the left of the vehicle. The statement can be evaluated as true if the permission to overtake is in force in a particular case.

Moreover, to formalize traffic rules using DDL, we consider the traffic system as a normative system, which has a set of clauses (norms), where the causes/norms are represented as if...then rules. Every clause/norm is represented by one (or more) rule(s) with the following form: where, $X_1, …, X_n$ are the conditions of applicability of the norm, and Y is the "effect" of the norm.

$$X_1, …, X_n \Rightarrow Y$$

The above-mentioned ([O], [P], [F]) deontic modalities modelled the normative effects. We take the standard deontic logic relationships between these deontic modalities. These are described below (taking the concept of overtake, atom: overtake).

$$[F]\ Overtake \equiv [O]\ \neg\ Overtake$$

$$[O]\ Overtake \equiv [F]\ \neg\ Overtake$$

$$[P]\ Overtake \equiv \neg\ [O]\ \neg\ Overtake$$

Now, a complete example of traffic rule (QLD rule 141 – Table 2) formalization using DDL is shown in below Figure 3.

Table 2. Queensland Traffic Rule 141

**Rule 141: No overtaking etc. to the left of a vehicle**
(1) A driver (except the rider of a bicycle) must not overtake a vehicle to the left of the vehicle unless—
(a) the driver is driving on a multi-lane road and the vehicle can be safely overtaken in a marked lane to the left of the vehicle; or
(b) the vehicle is turning right, or making a U-turn from the centre of the road, and is giving a right change of direction signal and it is safe to overtake to the left of the vehicle; or
(c) the vehicle is stationary and can be safely overtaken to the left of the vehicle; or

(d) the driver is lane filtering in compliance with section 151A or edge filtering in compliance with section 151B.

```
Atom driver_OvertakeToTheLeftOf_vehicle "Overtake Left"
Atom driver_Of_bicyle "Bicycle Rider"
Atom driver_IsDrivingOn_MultiLaneRoad "Driver driving in Multi-Lane"
Atom vehicle_CanBeSafelyOvertakenIn_markedLane "the vehicle can be safely overtaken in a marked lane"
Atom markedLane_IsToTheLeftOf_vehicle "marked lane to the left of the vehicle"
Atom vehicle_IsTurningRight "the vehicle is turning right"
Atom vehicle_IsGivingRightChangeOfDirectionSignal "the vehicle is giving a right change of direction signal"
Atom IsSafeToOvertakeToTheLeftOf_vehicle "it is safe to overtake to the left of the vehicle"
Atom vehicle_IsMakingUturn "making a U-turn"
Atom vehicle_IsOn_centreOfRoad "from the centre of the road"
Atom vehicle_IsStationary "the vehicle is stationary"
Atom driver_IsLawfullyLaneFiltering "the driver is lane filtering in compliance with section 151A"
Atom driver_IsLawfullyEdgeFiltering "the driver is edge filtering in compliance with section 151B"

r141: => [F] driver_OvertakeToTheLeftOf_vehicle
r141_bicycle: driver_Of_bicyle => [P] driver_OvertakeToTheLeftOf_vehicle
r141_a: driver_IsDrivingOn_MultiLaneRoad & vehicle_CanBeSafelyOvertakenIn_markedLane
        & markedLane_IsToTheLeftOf_vehicle => [P] driver_OvertakeToTheLeftOf_vehicle
r141_b_1: vehicle_IsTurningRight & vehicle_IsGivingRightChangeOfDirectionSignal
        & IsSafeToOvertakeToTheLeftOf_vehicle => [P] driver_OvertakeToTheLeftOf_vehicle
r141_b_2: vehicle_IsMakingUturn & vehicle_IsOn_centreOfRoad & vehicle_IsGivingRightChangeOfDirectionSignal
        & IsSafeToOvertakeToTheLeftOf_vehicle => [P] driver_OvertakeToTheLeftOf_vehicle
r141_c: vehicle_IsStationary & vehicle_CanBeSafelyOvertakenIn_markedLane
        => [P] driver_OvertakeToTheLeftOf_vehicle
r141_d_a: driver_IsLawfullyLaneFiltering => [P] driver_OvertakeToTheLeftOf_vehicle
r141_d_b: driver_IsLawfullyEdgeFiltering => [P] driver_OvertakeToTheLeftOf_vehicle

r141_bicycle >> r141
r141_a >> r141
r141_b_1 >> r141
r141_b_2 >> r141
r141_c >> r141
r141_d_a >> r141
```

Figure 3. Formalization of Queensland traffic rule 141

## 4   Experiment & Evaluation

A large-scale experiment is carried out to evaluate the proposed traffic rule formalization. Forty cases of overtaking maneuvers are evaluated based on eight realistic Queensland overtaking traffic scenarios. Every case is a specific overtaking maneuver. First, a compliance checking framework (ATRCCF) is built based using this formalization methodology. For details about the framework, please see [21]. Then this framework was applied to these overtaking maneuvers to validate AV behaviour. Then participants (general drivers and domain experts) were asked to assess these maneuvers. After that, the proposed framework's performance (effectiveness) was determined based on how many participants agreed with the framework's evaluation. This performance essentially shows the performance of the proposed formalization. From this analysis, it can be identified how well the formalization is. If the performance is promising, then it can be stated that the proposed formalization methodology effectively formalizes traffic rules for AVs. The evaluation was conducted based on two aspects:

1) legal/illegal validation of every maneuver, and
2) reason identification if the maneuver is illegal.

Five different overtaking maneuvers are designed for each traffic scenario. Two of these maneuvers are examples of explicit legal and illegal driving actions. The other three are borderline maneuvers, which may not be directly classified as traffic violations. One of the main reasons to make these three different types of maneuvers is that traffic rules contain vague terms (e.g., safe distance, approaching vehicle, clear view, etc.) requiring judgment by the drivers. Clearly, AVs need a deterministic and algorithmic approach. For example, determine whether the distance between two vehicles is safe. The parameters for the borderline situations are placed near the calculated threshold, whilst the values for the clear cases are considerably away.

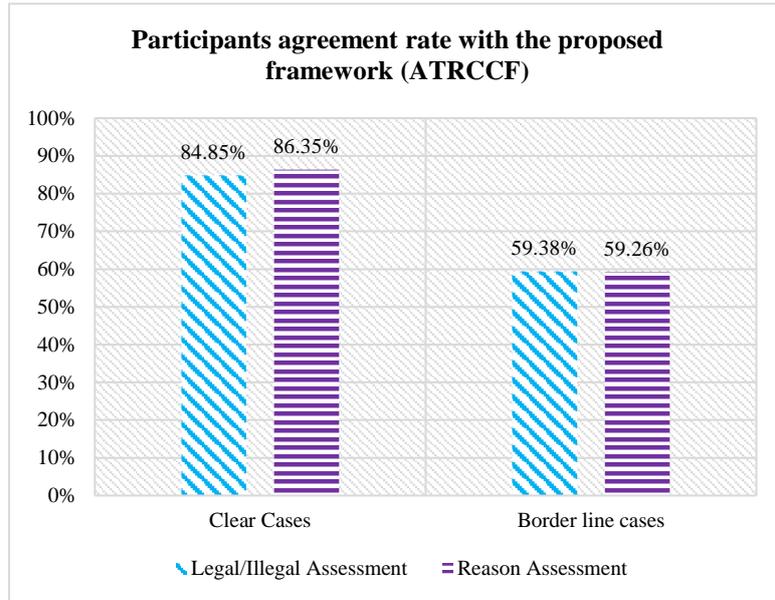

Figure 4. Performance of the compliance checking framework (ATRCCF).

Figure 4 shows the performance of the ATRCCF. In clear overtaking maneuver cases, on average, there is 84% legal/illegal and 86% reason identification agreement between participants and the framework. In borderline overtaking maneuver cases, participant average agreement rates with the framework's legal/illegal decision and reason identification are almost identical, which is 59%. The borderline cases are designed to test the human perception of the maneuvers with a very close threshold between legal and illegal in terms of a maneuver. According to the 50% outcome is truly indicative that the borderline cases are really borderline. Based on these agreement rates of clear and borderline cases, it can be stated that the compliance checking framework (ATRCCF) is promising. Essentially this indicates that the proposed formalization methodology is a promising approach to formalizing traffic rules for AVs.

## 5  Conclusion

This formalization methodology uses defeasible deontic logic to make the machine-computable (M/C) representation of traffic rules; however, there are some issues regarding the accuracy and completeness of the representation. Given the sophisticated and varied nature of traffic rules, identifying all the terms, norms, rule types, and conditions is challenging as these components are found in explicit or implicit linguistic forms. Also, traffic rule is written in natural language, which is more expressive than any other formal language. As noted, atoms are defined semantically in terms of five patterns. These patterns are sufficient to cover most cases of Queensland overtaking traffic rules. While more patterns are possible, the patterns presented also offer guidance to capture more complex cases if needed. However, there might be some subtle issues. In terms of the norm (deontic modality) determination from rules, only explicit types of norms (obligation, permission, and prohibition) are considered, although there might be different types of permission and other normative effects . A question may arise regarding the use of DDL in this work—that the formalizing methodology is not evaluated with any gold standard or other approaches. According to the Australian constitution, only the judiciary have the authority to provide the interpretation (gold standard) of the rules. Here, it is our best effort to do the formalization. When we look at the rule, and apparently, our formalization seems meaningful, and it is evaluated through the experiment. Due to the authoritative interpretative role of the courts within Australia's constitutional system, we are unable to draw a final decision about the degree to which formalized rules coincide with the legislative language . This is a limitation that applies to any effort to formalize legislation. Despite these limitations, the proposed formalization methodology's significant advantages are domain independence and scope of applicability. This methodology can be used in other domains such as anti-money laundering, university, etc. This formalization research also gives significant insights into the origins of legal formalization, its concerns, and possible remedies which could be useful for the rac (rules as code— representation of the rules so that computers can understand) movement.

## Acknowledgements

This research is supported by the National Research Foundation, Singapore under its Industry Alignment Fund – Pre-positioning (IAF-PP) Funding Initiative. Any opinions, findings and conclusions or recommendations expressed in this material are those of the author(s) and do not reflect the views of National Research Foundation, Singapore.

## References

1. P. Koopman and M. Wagner, "Autonomous vehicle safety: An interdisciplinary challenge," *IEEE Intelligent Transportation Systems Magazine,* vol. 9, pp. 90-96, 2017.
2. R. Leenes and F. Lucivero, *Laws on Robots, Laws by Robots, Laws in Robots: Regulating Robot Behaviour by Design* vol. 6, 2015.
3. G. Khorasani, A. Tatari, A. Yadollahi, and M. Rahimi, "Evaluation of Intelligent Transport System in Road Safety," *International Journal of Chemical, Environmental & Biological Sciences (IJCEBS),* vol. 1, pp. 110-118, 2013.
4. J. S. Brodsky, "Autonomous vehicle regulation: How an uncertain legal landscape may hit the brakes on self-driving cars," *Berkeley Technology Law Journal,* vol. 31, 2016.

5. A. Rizaldi, J. Keinholz, M. Huber, J. Feldle, F. Immler, M. Althoff, *et al.*, "Formalising and Monitoring Traffic Rules for Autonomous Vehicles in Isabelle/HOL," in *Integrated Formal Methods*, 2017, pp. 50-66.
6. H. Bhuiyan, G. Governatori, A. Bond, and A. Rakotonirainy, "Validation of Autonomous Vehicle Overtaking under Queensland Road Rules," presented at the 16th International Rule Challenge and 6th Doctoral Consortium@ RuleML+ RR, Germany, 2022.
7. A. Wyner and W. Peters, "On rule extraction from regulations," in *Legal Knowledge and Information Systems*, ed: IOS Press, 2011, pp. 113-122.
8. P. Monica, G. Guido, A. Tara, B. Harold, P. Adrian, and W. Adam, "LegalRuleML Core Specification Version 1.0," 2021.
9. L. Zhao, R. Ichise, Y. Sasaki, L. Zheng, and T. Yoshikawa, "Fast decision making using ontology-based knowledge base," in *Intelligent Vehicles Symposium (IV)*, 2016, pp. 173-178.
10. Z. Zhang, Q. Jiang, P. Li, L. Song, R. Wang, B. Yu, *et al.*, "The Visual Representation and Acquisition of Driving Knowledge for Autonomous Vehicle," *IOP Conference Series: Materials Science and Engineering,* vol. 235, p. 012011, 2017.
11. A. Rizaldi and M. Althoff, "Formalising Traffic Rules for Accountability of Autonomous Vehicles," in *18th International Conference on Intelligent Transportation Systems*, 2015, pp. 1658-1665.
12. S. McLachlan, M. Neil, K. Dube, R. Bogani, N. Fenton, and B. Schaffer, "Smart Automotive Technology Adherence to the Law:(De) Constructing Road Rules for Autonomous System Development, Verification and Safety," *International Journal of Law and Information Technology,* vol. 29, pp. 255--295, 2021.
13. G. Antoniou, D. Billington, G. Governatori, and M. J. Maher, "Representation results for defeasible logic," *ACM Trans. Comput. Logic,* vol. 2, pp. 255–287, 2001.
14. G. Governatori, "The Regorous Approach to Process Compliance," in *2015 IEEE 19th International Enterprise Distributed Object Computing Workshop*, 2015, pp. 33-40.
15. O. Wolfson and A. Silberschatz, "Distributed processing of logic programs," *SIGMOD Rec.,* vol. 17, pp. 329–336, 1988.
16. L. Gamut, L. van Benthem, and L. Gamut, *Logic, language, and meaning, volume 1: Introduction to logic* vol. 1: University of Chicago Press, 1991.
17. A. Witt, A. Huggins, G. Governatori, and J. Buckley, "Converting copyright legislation into machine-executable code: interpretation, coding validation and legal alignment," presented at the 18th ICAIL, São Paulo, Brazil, 2021.
18. G. Governatori and A. Rotolo, "Logic of violations: a gentzen system for reasoningwith contrary-to-duty obligations," *The Australasian Journal of Logic,* vol. 4, 2006.
19. M. J. Maher, "Propositional defeasible logic has linear complexity," *Theory and Practice of Logic Programming,* vol. 1, pp. 691-711, 2001.
20. G. Governatori, "Practical Normative Reasoning with Defeasible Deontic Logic," in *Reasoning Web.* , C. d'Amato and M. Theobald, Eds., ed Cham: Springer International Publishing, 2018, pp. 1-25.
21. H. Bhuiyan, G. Governatori, A. Bond, and A. Rakotonirainy, "Traffic Rules Compliance Checking of Automated Vehicle Maneuvers," *Artificial Intelligence and Law,* vol. In press, 2022.

# A validation process for a legal formalization method

Abdelhamid Abidi[1][0000−0002−5362−0348] and Tomer Libal[2][0000−0003−3261−0180]

[1] Sciences Po, France
[2] University of Luxembourg, Luxembourg

**Abstract.** The formalization of legal texts is an important step in order to be able to improve law's consistency, or to automate several legal tasks such as reasoning or answering legal questions. In this paper, we present a new approach for legal texts' formalization, based on a two-step method, and explain why it meets several requirements that existing approaches cannot satisfy, such as bi-directionality or isomorphism. We use an excerpt of the GDPR as an example, translate it into a logical formula, and then translate it back in order to evaluate, successfully, the accuracy of the process.

**Keywords:** Legal formalization · Knowledge representation · Domain specific languages

## 1 Introduction

A main aspect of the law is to govern the normative behavior of the population. As such, it is of high interest that the understanding of the law is not only available to a selected number of people, such as lawyers, but is widely accessible. At the same time, we would like its application to be fair and objective.

An interesting approach for attaining both these goals is by making the law understandable to computers. The ability of a computer to reason over the law will allow it both to explain it to people, via for example expert systems, and to reason over it in an objective way.

In this paper, we are considering one of the biggest problems of legal formalization, which is the ability of legal experts to participate in the process and their ability to confirm the quality and faithfulness of the formalization.

The need for legal experts when formalizing legal text stems from the need to have a translation which preserves the legal meaning of the text. Since the process requires two different formats - the legal text on the one side and the formal representation on the other - it traditionally required two domain experts, one for each format. Among the various approaches to this problem, one should mention Bertolini et al approach [3], where an agile methodology is introduced which involves both domain experts via short sprints. The goal of this approach is to allow iterations between the two experts and in this way, to increase the confidence in the translation.

The approach taken in this paper proposes to have a language close to the original text, therefore allowing an easier way to validate. At the same time, the chosen language enjoys properties which make it easily computable (second contribution). This is achieved by defining two target languages, one close to the original text and one denoted in first-order logic, which is a highly computable format. We then define a translation between the two levels which allows the automatic translation from the high level to the technical one. Moreover, we require that this translation is isomorphic, which allows the translation back from the formal to the legal levels (third contribution), those allowing the legal expert to validate the result. Our claim is that this approach enjoys the good properties of both categories, without any of the undesired ones.

The paper is organized as follows. In Section 2, we present our main contribution, the LLTs. After a short introduction, we survey the current state-of-the-art and continue to the presentation of the various elements of LLTs. In Section 3, which demonstrates one of the main advantages of the approach, which is the ability to get back the legal text from the translations in order to allow the legal expert to easily validate their translations. We conclude in Section 4.

## 2 LLTs

LLT stands for "Legal Logical Template". It is a formal grammar, that helps describing the structure of a legal text. Informally, LLTs are formal representations of legal concepts and capture the way a legal expert interprets the law.

Let us first note that the LLTs we present later were created for a specific legal context, and are not necessarily universal, it is the method that matters more. And we do not claim that we objectify legal interpretation, since the interpretation of a legal text cannot be absolute, but we do give lawyers all the necessary tools to formalize the law themselves, the way they think is the best.

LLTs allow us to dodge the necessity to translate directly legal texts into pure logical formulae, and gives us an intermediate step in the process, which is fortunate for several reasons :

Firstly, we want law formalization to become an accessible tool for lawyers, and we cannot expect from them to be experts in pure logic, so we must not rely on direct translation from legal text to logical formula as a methodology.

Secondly, it allows us to be more accurate through this intermediate process. Since there is a big gap between a legal text and its computable formalization, it would be too ambitious to do it all in one go, like many other attempts. The smaller the steps are, the easier it is to check and evaluate them, so with the two-step translation, we can check the correctness of each one and then compose them at the end.

Another key-point will be to create a step-by-step isomorphism between the legal text and the code, the relevancy of isomorphisms for knowledge based systems has already been highlighted in the past, for example by T.J.M.Bench-Capon & F.P.Coenen in 1992, in their paper [5]. Their relevancy still holds today as it has been reconfirmed almost 20 years later by Trevor J. M. Bench-Capon and Thomas F. Gordon. in 2009 [6].

In this paper, we are trying to go the closest possible to an ideal isomorphism by separating the process into two steps : creating LLTs from the legal texts, then creating logical formulae from the LLTs. Indeed, with an intermediate step, it will be easier to make the correspondence between the items.

The need for an intermediate step for a legal formalization has already been considered, for example by the authors of "An Interdisciplinary Methodology to Validate Formal Representations of Legal Text Applied to the GDPR"[4], saying that "A key element in the methodology is an intermediate representation".

Many attempts at formalizing law already exist, we could cite as examples "Logical english"[10], Akoma-Ntoso [18], DAPRECO [19], PENG$^{ASP}$ [21], Catala [17] etc.

These existing tools fall within two categories. The first category, syntax-based languages, contains those approaches which aim to be as close as possible to the legal text, the language used there is normally very general and less structured, which results in translations which are not easily computable. Since the main goal of legal formalization is to allow computers to reason over the text, this is undesirable. There is the advantage though that the correctness and faithfulness of the translation are easier to validate. The second category, semantics-based languages, contains those approaches which aim to be computable, as close as possible to the machine. These approaches require a translation which is much farther than the original text and therefore, its correctness is harder to validate.

We want the formalization to be both high-level and formal, in two steps, compatible and faithful, understandable and computable.

Indeed, we need it high-level enough for lawyers to use it without any scientific background, we need it also bidirectionally isomorphic so we can both check it and update it. And the existing formalizations we cited do not seem to meet this duality we seek.

The scope of our paper here is not to do a full and detailed comparison of the existing languages, but let us just express some requirements we think are essential, and most of our criteria are not new.

In "The Grammar of PENG$^{ASP}$ Explained"[21], the author comes up with requirements for a controlled language : "a controlled language specification should be translatable into an executable answer set program", "the grammar for this language should be highly configurable for different application scenarios", "the grammar should be bi-directional".

We could also cite the overlapping criteria of Routen and Bench-Capon in "Hierarchical formalizations"[20] : "a formalization should not only (1) be a faithful representation of what is expressed by the legislation; (2) be compu-tationally adequate, i.e. should permit us to make all relevant derivations by machine; but also (3) be easy to validate and maintain".

Our approach satisfies the previous demands as we will see through this paper, since it is usable by the lawyers without a scientific background, it is based on an isomorphic (and bi-directional) methodology, it is computable, and allows a faithful translation of the legal texts' structure and content, finally the isomorphic characteristic allows us to validate and maintain it.

In order to formally denote the LLTs, we will use a Context Free Grammar [2].

Our goal here will therefore be to create two isomorphic functions, one between the grammar and the law, and another between the grammar and the semantics.

To reach it, we need the grammar's syntax to be both abstract enough to include as much legal situations as possible, and precise enough to avoid any confusion regarding its relevancy when used. An example of a user-friendly interface for such a language is described in [12].

**Definition 1.** *LLTs The following is the CFG of the LLTs. The terminals set $\Sigma$ contains ATOM and all lower case words. An ATOM is a group of words. The starting symbol S is LLT.*

```
LLT :=              EXCEPTION | REF | CONDITION | LIST | PROHIBITION
EXCEPTION :=        {LLT | REF} + CONDITION
CONDITION :=        LIST | ATOM
LIST :=             ATOM | LIST + ATOM
REF :=              LABEL + LLT
LABEL :=            ATOM
PROHIBITION := LLT
```

In order to help our purpose, we can add as an even higher level a lawyer-friendly interface supporting both the accessibility, the annotations and the validation of our method : the LegAi tool presented in [15]. You can see an annotation example of articles 44 and 45 of the GDPR in the top image below.

We want at the end to establish an isomorphic correspondence between legal texts and logical formulae, in two steps. Now that we have the syntax of the grammar, let's build some corresponding formal semantics. These will not be used on their own, but will be automatically generated from the LLTs syntax established at the previous level.

We base our semantics on the First Order Logic, to which we add the classic deontic Obligation operator, and a second negation "$\sim$"the negation as failure (the first order negation is noted "-").

The adding of the obligation operator is necessary and sufficient to capture every deontic prescription. As for the negation as failure [7], it is very needed for the exceptions or the conditioned modalities, that are frequent in law. Indeed, the negation as failure (NAF) consists in deriving not A if you cannot derive A, and it is important to consider that by default, exceptions do not apply in law, unless provable. The most general case in law is generally the default one.

Our semantics are computable, thanks to existing methods that already handle First Order Logic, First Order Modal Logics and the addition of Negation-as-failure ; various theorem provers like SPINdle[11] can handle our semantics, finalizing the step-by-step path from law to code. We can already automate trees creation as you see in the last image. Rolf Schwitter, in [21], recommends the use of syntax trees as both a strategy, and a way of checking if the requirements are met.

**Definition 2.** *Semantics We define* $\phi : \{LLTs\} \rightarrow \{FOL \vee \{\sim\} \vee \{Ob\}\}$ , *associating to each LLT its semantics :*

$$\begin{aligned}
\phi(EXCEPTION) &:= \sim \phi(CONDITION) \Rightarrow \text{-}\phi(LLT/REF) \\
\phi(REF) &:= \phi(LABEL), \phi(LLT) \\
\phi(CONDITION) &:= \phi(LIST \mid ATOM) \\
\phi(LABEL) &:= \phi(ATOM) \\
\phi(LIST) &:= \phi(ATOM \mid LIST) , \phi(ATOM) \\
\phi(PROHIBITION) &:= Ob(\text{-}\phi(LLT)) \\
\phi(ATOM) &:= ATOM
\end{aligned}$$

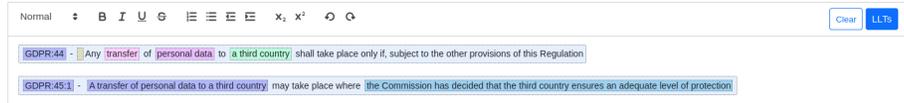

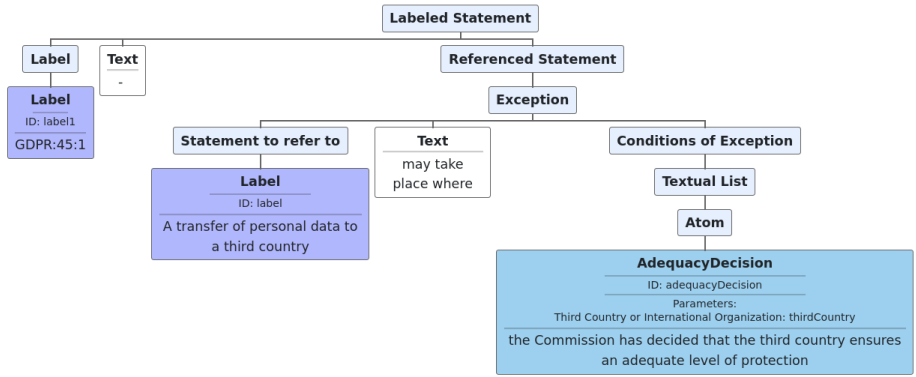
45

## 3 Functional evaluation

A an important property of legal formalization is the ability of the legal expert to validate its quality [14] ; that is thanks to our intermediate step. The evaluation we use is a functional one, consisting of a reverse translation of LLTs. If we automate this reverse translation, we just have to check then if the new legal text has the same meaning and the same legal substance (modulo synonyms). And the closer the new text is, the finer is the LLTs system.

We could cite T.J.M.Bench-Capon & F.P.Coenen's paper [5], in which they mention several advantages of isomorphism-based knowledge representations, among which is validation. With the reverse translations, we are going even closer to the creation of and "ideal"isomorphism, by making it bijective.

Indeed, in his paper [21], Schwitter supports this idea, using the bidirectionality of the grammar as an argument of its executability, unlike the other attempts of controlled languages back then.

**Definition 3.** *Reverse translation In order to translate the LLTs back, we must think of a standard translation regardless of the LLT's terminals, which is recursive and begins at the treetop. Here is an attempt of doing so :*
```
EXCEPTION  :=   LLT/REF is allowed if CONDITION
PROHIBITION := LLT is prohibited
```

Now we do have a method for validation via reverse translation, we can observe an example of it (modulo synonyms) in the middle image above.

## 4 Conclusion

In this paper, we have introduced two languages. One of them is novel and can be used by legal experts. The second can be used for computation. We have introduced an algorithm to automatically translate between the two languages. In addition, we have given an algorithm for the automatic translation of one language back into legal text. The resulted approach helped by the LegAi tool allows a legal expert to use the methodology and validate the quality of the translation without the need for a second domain expert (a programmer or a logician). The formalization can be used for computation, and we argue that our multi-layered approach solves some of the issues with the state-of-the-art.

## References

1. Publications Office of the European Union (2021). AKN4EU Guidelines and modeling. https://op.europa.eu/en/web/eu-vocabularies/akn4eu.
2. Aho, A. V., & Ullman, J. D. (1971). Translations on a context free grammar. Information and Control, 19(5), 439-475.
3. Bartolini, C., Lenzini, G., Santos, C.: An Agile Approach to Validate a Formal Representation of the GDPR. In: Kojima, K., Sakamoto, M., Mineshima, K., and Satoh, K. (eds.) New Frontiers in Artificial Intelligence. pp. 160—176. Springer International Publishing, Cham (2019).

4. Bartolini, C., Lenzini, G., & Santos, C. (2018). An interdisciplinary methodology to validate formal representations of legal text applied to the GDPR.
5. Bench-Capon, T. J., & Coenen, F. P. (1992). Isomorphism and legal knowledge based systems. Artificial Intelligence and Law, 1(1), 65-86.
6. Bench-Capon, T., & Gordon, T. F. (2009, June). Isomorphism and argumentation. In Proceedings of the 12th international conference on artificial intelligence and law (pp. 11-20).
7. Clark, K. L. (1978). Negation as failure. In Logic and data bases (pp. 293-322). Springer, Boston, MA.
8. European Parliament regulation 2016/679 (2016). https://eur-lex.europa.eu/eli/reg/2016/679/oj
9. Gen, K., Akira, N., Makoto, N., Yasuhiro, O., Tomohiro, O., & Katsuhiko, T. (2016). Applying the Akoma Ntoso XML schema to Japanese legislation. Journal of Law, Information and Science, 24(2), 49-70.
10. Kowalski, R. (2020). Logical english. Proceedings of Logic and Practice of Programming (LPOP).
11. Lam, H. and Governatori, G. The Making of SPINdle. Proceedings of International Symposium on Rule Interchange and Applications (RuleML 2009), Springer-Verlag. 2009.
12. Libal, T., & Steen, A. (2019). The NAI Suite-Drafting and Reasoning over Legal Texts. In JURIX (pp. 243-246).
13. Libal, T. (2020). A meta-level annotation language for legal texts. In International Conference on Logic and Argumentation (pp. 131-150). Springer, Cham.
14. Libal, T., & Novotná, T. (2021). Towards Transparent Legal Formalization. In International Workshop on Explainable, Transparent Autonomous Agents and Multi-Agent Systems (pp. 296-313). Springer, Cham.
15. Libal, T. (2022) The LegAi Editor: An Annotation Tool for the Construction of Legal Knowledge Bases. JURIX.
16. Lokhorst, G. J. C. (1996). Reasoning about actions and obligations in first-order logic. Studia Logica, 57(1), 221-237.
17. Merigoux, D., Chataing, N., & Protzenko, J. (2021). Catala: a programming language for the law. Proceedings of the ACM on Programming Languages, 5(ICFP), 1-29.
18. Palmirani, M., & Vitali, F. (2011). Akoma-Ntoso for legal documents. In Legislative XML for the semantic Web (pp. 75-100). Springer, Dordrecht.
19. Robaldo, L., Bartolini, C., Palmirani, M., Rossi, A., Martoni, M., & Lenzini, G. (2020). Formalizing GDPR provisions in reified I/O logic: the DAPRECO knowledge base. Journal of Logic, Language and Information, 29(4), 401-449.
20. Routen, T., & Bench-Capon, T. (1991). Hierarchical formalizations. International Journal of Man-Machine Studies, 35(1), 69-93.
21. Schwitter, R. (2021, September). The Grammar of PENG ASP Explained. In Proceedings of the Seventh International Workshop on Controlled Natural Language (CNL 2020/21).

# A Survey on Logic-Based Approaches to the Formalisation of Legal Norms

## With Particular Focus on the Formalisation of Traffic Rules

Diogo Sasdelli

Institute of Legal Informatics, Saarland University, 66123 Saarbrücken, Germany
`diogo.campos_sasdelli@uni-saarland.de`

**Abstract.** The survey at hand offers a brief overview of the interdisciplinary (Philosophy, Law, Computer Science) discussion concerning logic-based approaches to the formalisation of legal norms, while particularly discussing the advantages and challenges offered and posed by them with respect to the formalisation of traffic rules. Four main archetypes of logic of norms are identified, depending on whether the formalism is based on (1) propositional, (2) modal, (3) predicate or (4) many-valued-logic.

**Keywords:** Logic of Norms, Deontic Logic, Legal Logic, Formalisation of Traffic Rules.

## 1 Introduction

> I see no reason, why that Law and Logike should not bee
> The nearest and the dearest freends, and therfore best agree.
> (Abraham Fraunce, 1588 [25])

Literature offers countless examples for attempts at using logic-based methods for the formalisation of legal or moral norms. Historically, these attempts can be traced back to medieval logic, and arguably even to Greek antiquity, particularly to Aristotelian or to stoic logic.[1] Following the remarkable developments in the fields of mathematical logic and computability theory in the first half of the 20th century, as well as the more recent achievements in Artificial Intelligence, especially in the area of Natural Language Processing (NLP), logic-based approaches to the formalisation of normativity have been enjoying new bursts of popularity, leading to the development and the proposal of several logic systems and models for the formalisation of (legal) norms. The survey at hand briefly discusses the main archetypes for the development of a logic of norms, while particularly focussing on the advantages and challenges offered and posed by each of them with respect to the task of formalising traffic rules. It attempts to provide a reasonable overview of the discussion within Philosophy, Law and Computer Science.

---

[1] For historical aspects cf. particularly [1-5]; for practical syllogisms in Aristotelian logic cf. [6]; for the practical dimension of stoic logic, cf. [7], p. 6 and particularly [8], pp. 144 and 150-151. Unfortunately, the history of deontic logic and related approaches is often ignored by current research. The authors of [9] claim, e.g., that so far there have not been many attempts of translating law into a formal, logic-based language. Similarly [10] claims that [11] "could be considered as the first to propose a serious effort in formalising law".

Based on the fundamental structure of the formal language respectively employed, it is possible to group most approaches to the development of a logic of norms in four main archetypes: (1) models based on propositional logic (PL); (2) models based on modal logic (ML) – including so-called *deontic logic* proper –; (3) models based on first or higher order predicate logic (FOL/HOL); (4) models based on many-valued-logic (MVL). These approaches are discussed in more detail in the following sections.

## 2   PL-based Logic of Norms

### 2.1   Outline

The use of PL and closely related systems to formalise (legal) norms has been proposed by many scholars (for direct examples cf., e.g., [12-16]; other systems, like the ones proposed in [17] and [18] have been shown to be reducible to classic PL [19]; another equivalent approach is the so-called "satisfaction-logic" [18, 20-22]). Semantically, PL-systems introduce a simple Boolean truth-assignment function $\beta_{PL}$ which leads to the assignment of one of the values "true" (T) or "false" (F) to each PL-formula. In a PL-based logic of norms, norms are usually represented as being (reducible or equivalent to) atomic apophantic propositions. Sometimes, the classic truth-values T and F are replaced by values such as "valid" and "not-valid" or "satisfied" and "not-satisfied". This replacement, however, although often motivated by fundamental logical-philosophical problems, e.g., the so called *Jørgensen's Dilemma* [22-23], generally has no practical consequences for the logical structure of the respective system [95].

In order to distinguish between normative and descriptive propositions, one often introduces, as it is done below in Table 1, a unary normative operator, usually represented by the exclamation sign ! (cf., e.g., [17-18, 24-26]). However, due to the semantic limitations of PL, this normative operator is equivalent to the so-called *empty modality*, i.e., the normative proposition !Φ is always equivalent to the descriptive proposition Φ. In other words, !Φ~Φ is a tautology.

Table 1 shows how the rule 1. a), accompanying the traffic *sign 295* of the second annex to the German Road Traffic Regulation (*StVO*) could be formalised in PL:

**Table 1.** PL-formalisation of *sign 295*.

| Legal text (raw) | Norm (in conditional form) | Possible formalisation in PL |
| --- | --- | --- |
| 1.a) A person operating a vehicle must not cross or straddle the continuous line [103] | If a person is operating a vehicle in a street with a continuous line, this person must not cross or straddle this line. | Φ→!Ψ<br>Φ: a Person is operating a vehicle in a street with a continuous line.<br>!Ψ: this person must not cross or straddle this line |

## 2.2 Advantages and Challenges

PL has the advantages of being decidable and relatively intuitive: It offers a basic formalisation framework that is both human- and machine-friendly. PL also provides a simple structure for the detachment of a norm's condition or context of application and the actual concrete command. In the formalisation provided above in Table 1, one can easily infer the concrete command !Ψ from the general conditional norm Φ→!Ψ, and from the normative condition Φ.

However, PL-based formalisations offer no suitable means for inferring concrete commands for situations which, although related, are still slightly different from the one described in the application context (condition) of a norm. Consider, e.g., the process of overtaking. It follows from the norm accompanying *sign 295*, that one may only overtake another vehicle if it is possible to do so without crossing or straddling the continuous line. Intuitively, this inference relies on the fact that overtaking requires space; in case there is not enough space to the right of the continuous line, overtaking will result in the line being crossed or straddled, which violates the above-mentioned norm. Since PL's atomic propositions are truth-functionally independent from one another, if the act of overtaking is simply formalised as Ω, it will not be possible to infer any conclusions concerning Ω from Φ→!Ψ. What the formalisation is lacking is a means of grasping the logical relationship between the concepts of "overtaking" and "crossing" or "straddling" the line with respect to the available space on the street. While PL-based approximations are possible – e.g., one could introduce a sentence Γ meaning "there is enough space on the street" and set Ω∧¬Γ→¬Ψ, so that Φ∧(Ω∧¬Γ), i.e., overtaking in a street with a continuous line when there is not enough space, would violate the norm Φ→!Ψ –, a proper formalisation of these relations requires, especially in more complex contexts, a framework based on predicate logic, e.g., first-order predicate logic (FOL).

Since !Φ~Φ is a tautology, formalisation methods based on single PL-systems offer no reasonable way of checking whether a norm was satisfied or not. Particularly, neither the conflict between two norms nor the violation of a norm can be properly formalised within such frameworks. Put briefly, this is due to the fact that the formalisation of these situations would culminate in a logical contradiction, which in turn, due to the monotonic, two-valued, conflict-intolerant nature of classic PL, would enable the inference of any arbitrary proposition, rendering the whole formalisation useless. The literature offers a wide scope of proposed solutions to this issue, most of which involve replacing the basic PL-framework with more expressive paradigms, such as the ones offered by non-monotonic [51], paraconsistent [97], or many-valued logics [84-87]. A somewhat simpler solution to the problem of checking violations in a PL-based formalisation consists in using two distinct PL-systems for respectively representing norms and facts. One can then verify whether a set of PL-formulae corresponding to a description of a factual situation is satisfiable with respect to the set of PL-formulae corresponding to norms. In its essence, this task is nothing but a *satisfiability modulo theories* (SMT) problem, which can be solved, e.g., by employing an adapted *SAT-solver* or an *SMT-solver* [27]. Although verifying Boolean satisfiability is generally a computationally expensive problem, new SAT-algorithms such as *conflict-driven clause* *learning* (CDCL) and its further developments, which make clever use of human understanding of the properties of PL, have enabled an effective employment of SAT-solvers to several real-world applications. Since (at least) two distinct PL-valuations are employed, this approach is, from a semantical point of view, closely related to approaches based on modal logic (ML), which shall be discussed in more detail in Sec. 3.

Finally, PL can only represent the truth-functional dimension of semantics; it cannot differ between sentences that, albeit of different meaning in natural language, are true (or false) under the same conditions. The classic example is the distinction between adversative and conjunctive copulas. Consider the sentences A: "It is sunny and cold", B: "It is sunny, but cold". These sentences are both true if and only if it is in fact sunny, and, at the same time, also cold. In natural language, however, sentence B also implies a kind of opposition between "being sunny" and "being cold", which goes beyond the mere extensional, truth-functional dimension of semantics.

## 3    ML-Based Logic of Norms

### 3.1    Outline

Historically, the development of modern modal logic (ML), was closely connected to the research in the field of logic of norms [28]. Proponents of an ML-approach to logic of norms can be found as early as 1939 [29], and similar ideas were already proposed by Leibniz in his *logic of love* [30-31]. After the Second World War, and following the publication of works from G. Kalinowski (1953) [32], O. Becker (1952) [33] and especially of G. H. v. Wright's seminal paper *Deontic Logic* (1951) [34], which introduced the now widely used term *deontic logic*, the discussion surrounding the development of a logic of norms gained new momentum. Since then, ML-based approaches have become mainstream in the field.

Syntactically, ML-systems usually expand PL's alphabet by introducing the signs $\Box$ and $\Diamond$. In alethic modal logic, $\Box\Phi$ is intended to correspond to "$\Phi$ is necessary", $\Diamond\Phi$ to "$\Phi$ is possible". In an analogous way, one can define a normative interpretation of these signs in which $\Box\Phi$ is taken to mean "$\Phi$ is obligatory", and $\Diamond\Phi$ is read as "$\Phi$ is permitted". Semantically, the most usual ML systems can be described as consisting in the structure $<\beta_{ML}, R, U>$. Following this so-called *possible world semantics* or *Kripke-semantics*, ML expands PL by introducing a set of sets of formulae (or universe) U and a dyadic, so-called *accessibility relation* R, defined among the elements of U. Intuitively, each set of formulae $W_x$ corresponds to a description of a *possible world*; hence the introduced set of sets of formulae U corresponds to the set of possible worlds. Through the accessibility relation R, each possible world $W_x$ is assigned to a set $U_x \subseteq U$ containing those possible worlds $W_y$ for which $R(W_x, W_y)$ holds. These worlds $W_y$ represent the worlds which are accessible or conceivable for $W_x$. Finally, like in PL, $\beta_{ML}$ is a function that assigns to each formula $\Phi$ in a world $W_x$ a value T or F according to the usual Boolean definitions and with the added truth-condition concerning the operator $\Box$: $\beta_{ML}(\Box\Phi, W_x)=T$ if and only if, for all $W_y$ for which $R(W_x, W_y)$ holds, $\beta_{ML}(\Phi, W_y)=T$. In other words: $\Box\Phi$ is true in a world $W_x$ if and only if $\Phi$ is true in all worlds $W_y$ accessible to $W_x$.

The intuitive motivation for this semantical framework is the idea that one can conceive many different possible worlds. The notion of alethic necessity is then represented by a formula that is true in all conceivable worlds; the notion of possibility, in its turn, is represented by a formula that is true in at least one conceivable world. For the representation of normative notions such as obligation, prohibition and permission as *deontic modalities*, one must further introduce the idea of *deontically perfect worlds*, i.e., roughly speaking, of worlds in which no norms are violated. Hence, something is obligatory if it is true in all deontically perfect worlds; it is permitted if it is true in at least one of these worlds.[2]

To guarantee that deontic modalities only depend on the truth-values of formulae in deontically perfect worlds, one has to introduce further restrictions on the properties of the accessibility relation R. The goal is to define R so that it connects a world $W_x$ to only (and all) its deontically perfect alternatives $W_y$. By altering the properties of R, different logical models can be generated. For instance, if R is reflexive, i.e., if $R(W_x, W_x)$ holds for all worlds $W_x$, then the formula $\Box\Phi\rightarrow\Phi$, which is often called (axiom) T, is logically valid. Under an alethic interpretation, this formula corresponds to the intuitively valid notion that "if $\Phi$ is necessary, then $\Phi$ is true". If, however, the formula is interpreted normatively, it corresponds to "if $\Phi$ is obligatory, then $\Phi$ is true". This would imply that all commands are logically always fulfilled, which is forbiddingly counterintuitive. Moreover, if one assumes that R only connects a world $W_x$ to its deontically perfect alternatives $W_y$, the reflexivity of R would imply that every world is a deontically perfect alternative to itself, which is also counterintuitive. Thus, the accessibility relation R of an ML-based system of logic of norms cannot be reflexive.[3] On the other hand, the formula $\Box\Phi\rightarrow\Diamond\Phi$, often called (axiom) D, is usually considered to be a valid principle of logic of norms. Under normative interpretation, it corresponds to "if $\Phi$ is obligatory, then $\Phi$ is permitted". The property of the accessibility relation R associated with the validity of D is called *seriality*: R is serial if for all worlds $W_x$ there is a $W_y$ such that $R(W_x, W_y)$. Since reflexivity implies seriality, D is implied by T, but not the other way around. The term *deontic logic* is usually employed to denote ML-based systems of logic of norms that contain D, but not T, i.e., systems with serial, but irreflexive R. In particular, so-called *standard deontic logic* (*SDL*) is the smallest ML-system containing, besides the axioms of classic PL, the axiom D, the distribution axiom K: $\Box(\Phi\rightarrow\Phi)\rightarrow(\Box\Phi\rightarrow\Box\Phi)$, as well as the inference rules *modus ponens* and the so-called *necessitation rule*. Extensions of SDL that include the axiom SH: $\Box(\Box\Phi\rightarrow\Phi)$ are sometimes called *SDL$^+$*. SH is a somewhat weaker alternative to T. Normatively, it is usually interpreted as stating that it is obligatory that obligations are fulfilled. The property of R related to SH is the so-called *quasi-reflexiveness*: $R(W_x, W_x)$ holds if, for all $W_y$, $R(W_y, W_x)$ also holds, i.e., worlds that are deontically perfect alternatives to all other worlds are deontically perfect alternatives to themselves (for a detailed account of such systems, cf. [38-42]).

---

[2] The main structure of the truth-definition in possible worlds semantics is usually associated with the philosophy of G. W. Leibniz (cf. [33], p. 18). In fact, Leibniz uses the idea of conceivable possible worlds to explain the nature of logical truth as being a kind of truth valid with respect to possible worlds [35]. Interestingly, in his famous *Theodizee*, Leibniz claims that our world was created by god as the best of all possible worlds. Hence, the idea of a deontically perfect alternative to our world seems to contradict Leibniz' own ideas (cf. [36]).

[3] This was already noticed by Kripke in [37], p. 95. In his normative-legal interpretation of ML, O. Becker tried to interpret T as meaning that if $\Phi$ is obligatory, then it factually happens in a legal way ([33], p. 44). Somewhat similarly, [10] proposes to address the problem of clarifying liabilities in a collision involving autonomous vehicles by "ensuring that autonomous vehicles always comply with the traffic rules so that they cannot be held liable for a collision". Both attempts ignore the fact that it is generally always possible to violate norms.

T is not the only problematic ML-formula under normative interpretation. Formulae such as $\Box\Phi\rightarrow\Box(\Phi\vee\Psi)$ or $\Box\Phi\rightarrow((\Phi\rightarrow\Psi)\rightarrow\Box(\Psi))$, which are valid in common ML-systems, are also counterintuitive when interpreted normatively. If, e.g., in the first formula, $\Box\Phi$ is read as "it is obligatory to deliver the letter" and $\Box\Psi$ as "it is obligatory to burn the letter", then the complete formula states "if it is obligatory to deliver the letter, then it is also obligatory to either deliver it or burn it". Hence, if one does not fulfil the obligation $\Box\Phi$, the only way to fulfil the obligation $\Box(\Phi\vee\Psi)$, which is logically implied by $\Box\Phi$, is by doing $\Psi$, i.e., by burning the letter. This is the so-called *Ross' Paradox* [43]. This and other problematic formulae are known in the literature as *the paradoxes of deontic logic* [44].[4] From a model-theoretical perspective, one strategy to overcome such paradoxes is to modify the accessibility relation R so as to avoid the validity of the respective formulae. The search for a paradox-free system has motivated the development of countless different systems of deontic logic.

Several paradoxes (particularly the so-called *Chisholm's paradox* [48]) seem to be avoidable by dropping monotonicity with respect to deontic modalities. Intuitively, this allows for the assumption of new premises to alter the derivability of a conclusion, i.e., for the defeasibility of normative inferences through counterarguments. Defeasibility seems particularly important for the formalisation of so-called *contrary-to-duty* obligations and of normative conflicts. These situations cannot be properly represented with the framework described above, because they presuppose or imply the violation of a norm, which is not possible in a deontically perfect world.

Defeasibility can be achieved, e.g., by replacing the accessibility relation R by a partial ordering of the universe set U of all possible worlds. Thus, a model for defeasible deontic logic can be described as consisting in the structure $<\beta, \geq, U>$, where $\beta$ and U are defined as above, and $\geq$ is a transitive relation over the elements of U. Intuitively, $\geq(W_x, W_y)$ represents the idea that the world $W_x$ is better than the world $W_y$. Hence, instead of conditioning the truth-value of an obligation $\Box\Phi$ in a world $W_x$ on the truth-value of $\Phi$ in its deontically perfect alternatives, the value of $\Box\Phi$ in $W_x$ is defined as depending on the value of $\Phi$ in the worlds which are better (or at least not worse) than $W_x$ according to $\geq$. For systems based on this approach, see [49-51]

Another approach is based on so-called *neighbourhood frames* [52]. While in classic *Kripke frames* each world $W_x$ in U is assigned to a set of alternatives worlds $W_y$ for which R($W_x$, $W_y$) holds, neighbourhood frames introduce a relation $R_n$ to assign to each world $W_x$ a set of sets $N_i$ (which are subsets of U) of alternatives worlds $W_y$, for which $R_n(W_x, N_i)$ holds. Intuitively, $N_i$ is the *neighbourhood* of $W_x$. Through this approach, the truth-value of a norm $\Box\Phi$ in a world $W_x$ depends on the truth-value of $\Phi$ in the worlds $W_y \in N_i$. Neighbourhood frames are particularly suitable for the representation of (relevant) similarities, analogy, and exceptions, because a neighbourhood $N_i$ can be seen as a set of situations sufficiently similar (according to some predetermined criteria) to the situation contained in $W_x$. By introducing a transitive relation $\geq$ to the model, one can also determine that the validity of $\Box\Phi$ in $W_x$ is only influenced by the 'better' worlds in $W_x$'s neighbourhood. For systems based on this and closely related approaches, see [53-56], for an automation approach with a HOL-based theorem prover, see [57].

---

[4] A similar problem is raised in the AI-discussion by [46], p. 24: "Take the two sentences: 'Helmets must be worn' and 'Dogs must be carried'. They are identical in form, yet they mean totally different things. The would-be traveller on the London Underground needs to know several unspoken things about the way the world works to be confident that he does not have to acquire a dog before boarding the escalator." Related problems also seem to appear in connection with Reinforcement Learning [47].

Sometimes, instead of using a monadic (unary) operator such as $\Box$, systems of so-called *dyadic deontic logic*, introduce a *dyadic* normative operator, which in classic systems of modal logic is usually represented by the so-called *fish tail* operator $\prec$. An expression such as $\Phi \prec \Psi$ is to be interpreted as "$\Psi$ is obligatory under the condition of $\Phi$". Usually, $\Box\Phi$ is defined as being equivalent to $\top \prec \Phi$, with $\top$ being the logical *verum*, i.e., basically any logical tautology.

Table 2 provides some examples of possible ML-formalisations of sign 295.

**Table 2.** ML-Formalisation of *sign 295*

| Norm (in conditional form) | Possible formalisations in ML |
|---|---|
| If a person is operating a vehicle in a street with a continuous line, this person must not cross or straddle this line. | $\Phi \to \Box\Psi$ $\quad$ $\Box(\Phi \to \Psi)$ $\quad$ $\Phi \prec \Psi$ <br> $\Phi$: a Person is operating a vehicle in a street with a continuous line. <br> $\Psi$: this person neither crosses nor straddles this line |

### 3.2 Advantages and Challenges

Due to its ability to express intensionality, ML-based systems have been employed, with varying degree of success, for the formalisation of various linguistic structures. Several implementation frameworks have been developed for ML-based deontic logic: *LegalRuleML* [64-65] is a markup language based on *RuleML* designed to be able to represent many particularities of normativity and deontic logic. There are also several reasoning engines and theorem provers available, e.g., Turnip [66], NAI [67], Leo-III [68] and MleanCoP [69], HOL-based theorem-solvers, e.g., [79], can also be employed.

Classic ML-based systems of logic of norms are simple and intuitive, but also prone to paradoxes. Furthermore, they fail to provide a solid formalisation framework for contrary-to-duty situations. On the other hand, more complex systems, e.g., those based on preference logic or on neighbourhood frames, usually lack a clear or intuitive justification for their semantical structure. Finally, while ML is more expressive than PL, it is still not as semantically rich as predicate logic. While technically feasible, introducing quantification to ML, might lead to semantical problems [98].

## 4 FOL/HOL-Based Logic of Norms

### 4.1 Outline

The use of predicate logic to formalise norms is commonly found in the tradition of so-called *legal logic* (cf., e.g., [12-16, 23, 45, 58-62]). Legal logic emerged independently from its philosophical cousin *deontic logic* in the early post-war period. Instead of trying to develop a new kind of logic to specifically represent norms, adepts of legal logic propose the use of established logical systems, especially FOL, to formalise norms. The modern field of *legal informatics* originated from legal logic.

Predicate logic expands PL by adding quantifiers such as the universal quantifier $\forall$ and the existential quantifier $\exists$. Semantically, predicate logic formulae have to be interpreted on the basis of a domain D, i.e., a (usually non-empty, countably infinite) set of objects. FOL can be further extended by adding specific predicates (e.g., equality) and, e.g., temporal operators. Second and higher-order predicate logic (HOL) expands FOL by allowing for the predication and quantification of predicates.

More recently proposed formalisations of norms based on predicate logic usually expand classic FOL to introduce so-called *temporal operators*. The result is a system of *temporal logic* (TL), such as, e.g., *linear temporal logic* (LTL) [71-72], *metric temporal logic* (MTL) [73-74] and *signal temporal logic* (STL) [75-76]; see also [77]. This expansion is trivial in the sense that these operators are usually defined as being equivalent to classic FOL-formulae, so that they can be dropped from any TL-formula to obtain an equivalent FOL-formula.

Formalisations based on predicate logic can be implemented by using programming languages such as PROLOG or PROLEG [78] or HOL-based theorem-solvers such as *Isabelle* [79]. In models with restricted semantics, adapted SAT-solvers can also be employed to verify fulfilment or violation of norms [27, 80]. Table 3 provides an example for a FOL-formalisation of sign 295.

**Table 3.** FOL-formalisation of *sign 295*.

| Norm | Possible formalisation in simple FOL | |
|---|---|---|
| If a person is operating a vehicle in a street | $\forall x \forall y \forall z((P(x) \land V(y) \land S(z) \land C(z) \land (Cr(x, y, z) \lor St(x, y, z))) \to Vio(x))$ | |
| with a continuous line, | P(x): x is a person. | Cr(x, y, z): x crosses z while operating y |
| this person must not | V(x): x is a vehicle | St(x, y, z): x straddles z while operating y |
| cross or straddle this | C(x): x has a cont. line | Vio(x): x violated the law |
| line. | S(x): x is a street | |

### 4.2 Advantages and Challenges

Thanks to their higher expressivity, systems based on FOL/HOL can deliver more accurate formal theories and so-called *ontologies* [99], which also consider logical properties concerning relations among individual objects, as well as their properties. HOL-formalisations can also represent second and higher-order properties, i.e., properties of properties. From a practical perspective, FOL is a well-studied formal language – it has been used as the logical-philosophical *lingua-franca* for about a century. Furthermore, FOL-deductions also bare structural similarities to legal subsumption (cf. [81], pp. 71-72 and 176-183). Hence, FOL-based approaches tend to be more accessible to both legal practitioners and technicians.

On the other hand, their higher expressivity leads to undesired metalogical properties. As mentioned above, FOL does not allow for decision procedures concerning the semantic status of its formulae. HOL, in its turn, is not even complete: no algorithm can be employed to generate all HOL-truths. From a practical point of view, these problems can be dealt with in two ways: One either (1) introduces (severe) semantic limitations to the models in order to achieve decidability or completeness, e.g., by employing so-called *Henkin-semantics*; or one (2) develops heuristically driven programs (possibly while also employing machine learning) that seem to work 'well enough' in a 'sufficient' number of cases. How to define when a program works 'well enough' for a 'sufficient' number of cases is a particular challenge, for there will necessarily always be infinitely many cases for which the program will not deliver the correct result.

## 5  MVL-Based Logic of Norms

### 5.1  Outline

A somewhat rarer approach to formalising legal norms consists in employing systems of so-called *many-valued logic* (MVL). Systems of classic two-valued logic, like the systems discussed above, assign to each formula a value from a set containing two members, e.g., {T, F} or {0, 1}. MVL breaks with this paradigm, replacing this set with a set containing three or more elements. Particularly, systems that adopt a set containing infinitely many truth-values are often called systems of *fuzzy logic*. Historically, modern MVL can be traced back to the works of J. Łukasiewicz [83-84]. More recently, due to the many applications of MVL, particularly of *fuzzy logic*, in AI and machine learning, this type of logic has been enjoying increasing popularity.

Overall, any of the above discussed logic archetypes can be transformed into an MVL-system by altering the range of the respective truth-assignment function $β_x$, i.e., by replacing this range with a set containing three or more elements. Since this modification is purely semantic, MVL-systems can be built on the same syntax as classic two-valued ones; hence, there is no need to introduce, in this section, a specific table containing examples of possible MVL-formalisations of *sign 295*. Semantically, the introduction of new values leads to an exponential increase in the number of possible operations. For instance, while classic two-valued systems of extensional logic only allow for $2^2=4$ distinct unary operations, a three-valued system contains $3^3=27$ different unary operations.

The use of MVL to formalise normativity can be traced back to the very origins of modern logic of norms in the late 1920s and early 1930s. Karl Menger, who also worked on the related field of *fuzzy set theory*, was likely the first to propose a three-valued system of logic to serve as basis for the representation of norms [84-85]. In fact, Menger argued that the problems with E. Mally's *Deontik* originated from it being a system of two-valued logic (cf. [84], p. 59). For Menger, the object of a norm can be neither necessary nor impossible, but only contingent, or, as he puts it, *doubtful*. That is why he tries to develop a *logic of the doubtful*, which would serve as a proper formalism to represent normativity. Another attempt at using MVL to develop a logic of norms was undertaken in 1946 by T. Storer [86]. Storer develops a system with a total of five values: besides the usual apophantic values *true* and *false*, he introduces the values *moral*, *amoral* and *immoral* (for which he uses the values 0, 1 and 2). Analogously to how MVL can be used to build alethic logic, Kalinowski proposed, in 1952, the use of a three-valued system as a basis for the development of his system *K1* [32]. M. Fisher's system from 1961 can be considered an expansion of Kalinowski's original ideas [87-88]. More recent attempts at employing MVL for the development of a logic of norms usually build on previously established systems based on ML or on FOL/TL [89-91].

### 5.2 Advantages and Challenges

Due to their higher expressivity, MVL-systems have also been proposed as a basis for the development of alethic logic (cf. [92], p 254; [94], pp. 469-472; [83]). Furthermore, the addition of extra truth values allows for a better representation of probabilities, degrees of satisfaction, thresholds etc. While semantically richer than two-valued systems, it is trivially possible to develop decision procedures for MVL systems with finitely many values, e.g., by building the respective truth-tables. More complex systems of MVL only allow for decision procedures under specific conditions [94].

The main weakness of MVL-formalisations is the fact that MVL has not been nearly as well studied as classic two-valued logic. Besides, MVL breaks with – or at least strongly relativises – fundamental principles of classic logic, e.g., the law of the excluded middle (*tertium non datur*) or even the principle of non-contradiction.

## 6 Conclusions

The various approaches to the formalisation of norms are connected to different advantages and challenges. In general, the less expressive systems enable a more intuitive justification and are easier to understand and to implement. However, they are prone to paradoxes, and lack the ability to formalise more complex normative (e.g., contrary-to-duty) situations. To properly formalise such situations, more complex systems are needed, which in their turn are less intuitive and more difficult to implement. Hence, determining the best logical formalism to employ in a concrete case will generally depend on the specific requirements involved. After choosing a basic formalism to represent norms, one can further fine-tune the model by employing a kind of *meta-formalism* for representing the relations between the various norms in a system, such as *rulebooks* [102], *I/O-logic* [95, 96], *normative systems* [100] and *imperative semantics* [101].

Almost a century has passed since E. Mally's *Grundgesetze des Sollens* [17], arguably the first modern attempt at developing a system of symbolic logic of norms. While substantial progress has been made, many challenges – perhaps the most fundamental ones – are yet to be solved.

## 7 Acknowledgment

The research leading to these results is funded by the German Federal Ministry for Economic Affairs and Energy within the project "KI Wissen – Automotive AI powered by Knowledge." The author would like to thank the consortium for the successful cooperation.

## References

1. Knuutilla, S.: The Emergence of Deontic Logic in the Fourteenth Century. In: Hilpinen, R. (ed.), New Studies in Deontic Logic. Norms, Actions, and the Foundations of Ethics, pp. 225-248. D. Reidel, Dordrecht (1981).
2. Knuuttila, S., Hallamaa, O.: Roger Roseth and Medieval Deontic Logic. Logique & Analyse, 149, 75-87 (1995).
3. Hilpinen, R., McNamara, P.: Deontic Logic: A Historical Survey and Introduction. In: Gabbay, D., Horty, J., Parent, et al. (eds.) Handbook of Deontic Logic and Normative Systems, Vol 1, pp. 3-136. College Publications, London (2013).
4. Kalinowski, G.: La logique des norms. Presses Universitaires de France, Paris (1972).
5. Horovitz, J.: Law and Logic. A critical Account of Legal Argument. Springer, Wien (1972).
6. Price, A.: The Practical Syllogism in Aristotle: A New Interpretation. Logical Analysis and History of Philosophy, 11, 151-162 (2008).
7. Rescher, N.: The Logic of Commands. Routledge, London (1966).
8. Vaz, H.: Escritos de Filosofia, IV. Introdução à Ética Filosófica 1. Loyola, São Paulo (2002).
9. Nikol, D., Althoff, M.: Die Formalisierung von Rechtsnormen am Beispiel des Überholvorgangs. InTeR 1/19, 12-16 (2019).
10. Rizaldi, A., Althoff, M.: Formalising Traffic Rules for Accountability of Autonomous Vehicles. In Proceedings of the 18[th] IEEE International Conference on Intelligent Transportation Systems, IEEE, Gran Canaria, Spain (2015).
11. Buchanan, B., Headrick, T.: Some speculation about artificial intelligence and legal reasoning. Stanford Law Review, 23, 1, 40-62 (1970).
12. Klug, U.: Logische Analyse rechtstheoretischer Begriffe und Behauptungen. In Käsbauer, M., Kutschera, F, (eds.), Logik und Logikkalkül, pp. 115-125. Karl Alber, München (1962).
13. Schreiber, R.: Logik des Rechts. Springer, Berlin (1962).
14. Tammelo, I., Schreiner, H.: Grundzüge und Grundverfahren der Rechtslogik. Bd. 1. Dokumentation, Pullach bei München (1974).
15. Tammelo, I., Schreiner, H.: Grundzüge und Grundverfahren der Rechtslogik. Bd. 2. Dokumentation, Pullach bei München (1977).
16. Tammelo, I.: Rechtslogik und materiale Gerechtigkeit. Athenäum, Frankfurt a.M (1971).
17. Mally, E.: Grundgesetze des Sollens. Leuschner & Lubensky, Graz (1926).
18. Hofstadter, A., McKinsey, J.: On the Logic of Imperatives. Philosophy of Science 6, 4, 446-457 (1939).
19. Centrone, S.: Notes on Mally's deontic logic and the collapse of Seinsollen and Sein. Synthese, 190, 4095-4116 (2013).
20. Dubislav, W.: Zur Unbegründbarkeit der Forderungssätze. Theoria, 3, 330-342 (1937).

21. Grue-Sørensen, K.: Imperativsätze und Logik. Begegnung einer Kritik. Theoria, 5, 195-202 (1939).
22. Jørgensen, J.: Imperatives and Logic. Erkenntnis, 7, 288-296 (1938).
23. Weinberger, O.: Rechtslogik. Springer, New York (1970).
24. Ledent, A.: Le statut logique des propositions imperatives. Theoria, 8, 262-271 (1942).
25. Fraunce, A.: The lawyer's logic. Scolar Pr., Menston (1969).
26. Bergström, L.: Imperatives and Ethics. A Study of the logic of imperatives and of the relation between imperatives and moral judgements. Stockholm (1962).
27. Lin, Y., Althoff, M.: Rule-Compliant Trajectory Repairing using Satisfiability Modulo Theories. In: Proceedings of the 2022 IEEE Intelligent Vehicles Symposium (IV), pp. 449-456. Curran, Red Hook (2022).
28. Woleński, J.: Deontic Logic and Possible Worlds Semantics. A Historical Sketch. Studia Logica, 49, 273-282 (1990).
29. Grelling, K.: Zur Logik der Sollsaetze. Synthese, 4, 44-47 (1939).
30. Leibniz, G.: Frühe Schriften zum Naturrecht. Meiner, Hamburg (2003).
31. Busche, H. Die innere Logik der Liebe in Leibnizens Elementa Juris Naturalis. Studia Leibnitiana, 23, 2, 170-184 (1991).
32. Kalinowski, G.: Théorie des Propositions Normatives. Studia Logica, 1, 147-182 (1953).
33. Becker, O.: Untersuchungen über den Modalkalkül. A. Hain, Meisenheim am Glan (1952).
34. Wright, G. H. v.: Deontic Logic. Mind, 60, 237,1-15 (1951).
35. Leibniz, G.: Vom höchsten Gute. In: Leibniz, G.: Deutsche Schriften. 2. Bd. Ed. by G. E. Guhrauer, pp. 35-47. Olms, Hildesheim (1966).
36. Leibniz, G.: Die Theodizee. Meiner, Hamburg (1968).
37. Kripke, S.: Semantical analysis of modal Logic I, normal modal propositional calculi. Zeitschrift für mathematische Logik und Grundlagen der Mathematik, 9, 67-96 (1963).
38. Åqvist, L.: Introduction to Deontic Logic and The Theory of Normative Systems. Bibliopolis, Napoli (1987).
39. Smiley, T.: Relative Necessity. The Journal of Symbolic Logic, 28, 2, 113-134 (1963).
40. Hansson, B.: An Analysis of Some Deontic Logics. Nous 3, 373-398 (1969).
41. Hintikka, J.: Some Main Problems of Deontic Logic. In: Hilpinen, R. (ed.): Deontic Logic: Introductory and Systematic Readings, pp. 59-104. D. Reidel, London (1981).
42. Hintikka, J.: Models for Modalities. Selected Essays. D. Reidel, Dordrecht (1969).
43. Ross, A.: Imperatives and Logic. Theoria, 7, 53-71 (1941) (= Philosophy of Science, 11, 1, 30-46 (1944).
44. Stranzinger, R.: Die Paradoxa der deontischen Logik. In: Tammelo, I., Schreiner, H. (eds.): Grundzüge und Grundverfahren der Rechtslogik. Bd. 2, pp. 142-159. Dokumentation, Pullach bei München (1977).
45. Máynez, E. Introducción a la lógica jurídica. Fondo de Cultura Economica, Mexico (1951).
46. Michie, D., Johnston, R.: The Creative Computer. Machine Intelligence and Human Knowledge. Harmondsworth, Penguin (1985).
47. Sasdelli, D.: KI-Sicherheit, Reward Hacking und die Paradoxa der Normenlogik. In: Schweighofer, E. et al.: Recht DIGITAL - 25 Jahre IRIS, pp. 409-416 (2022). Weblaw, Bern.
48. Chisholm, R.: Contrary to Duty Imperatives and Deontic Logic. Analysis 24, 33-36 (1963).
49. Cornides, T.: Ordinale Deontik. Springer, Wien (1974).
50. Åqvist, L.: Some Results on Dyadic Deontic Logic and The Logic of Preference. Synthese, 66, 95-110 (1986).
51. Nute, D. (ed.): Defeasible Deontic Logic. Kluwer, Dordrecht (1997).
52. Hughes, G., Cresswell, M.: A New Introduction to Modal Logic. Routledge, New York (1996).

53. Nortmann, U.: Deontische Logik: Die Variante der lokalen Äquivalenz. Erkenntnis, 25, 275-318 (1986).
54. Nortmann, U.: 1989, Deontische Logik ohne Paradoxien. Semantik und Logik des Normativen. München: Philosophia.
55. Carmo, J., Jones, A.: Deontic Logic and Contrary-to-Duties. In: Gabbay, D., Guenthner, F. (eds.): Handbook of Philosophical Logic. 2nd Edition. Volume 8, pp. 265-343. Springer, Dordrecht (2002).
56. Carmo, J., Jones, A.: Completeness and Decidability Results for a Logic of Contrary-to-Duty Conditionals. Journal of Logic and Computation, 23, 585-626 (2013).
57. Benzmüller, C., Farjami, A., Parent, X.: A dyadic Deontic Logic in HOL. In: Broersen, J. et al. (eds.): Deontic Logic and Normative Systems. 14th Deon, pp. 33-50. (2018).
58. Klug, U.: Juristische Logik. Springer, Berlin (1951).
59. Vilanova, L.: Lógica Jurídica. Bushatsky, São Paulo (1976).
60. Tammelo, I.: Sketch for a Symbolic Juristic Logic. Journal of Legal Education, 8, 3, pp. 277-306 (1955).
61. Weinberger, O.: Die Sollsatzproblematik in der modernen Logik. Nakladatelství Československé akademie věd, Prag (1958).
62. Reisinger, L.: Rechtsinformatik. De Gruyter, Berlin (1977).
63. Palmirani M. et al.: LegalRuleML: XML-Based Rules and Norms. In: Olken F. et al (eds.): Proceedings of the 5[th] International Symposium on Rule-Based Modelling and Computing on the Semantic Web (RuleML2011), pp. 298-312 (2011).
64. Athan T. et al.: LegalRuleML: Design Principles and Foundations. In: Reasoning Web. Proceedings on the 11[th] International Summer School on Web Logic Rules, p. 151-188. Springer, Berlin (2015).
65. Athan, T. et al.: LegalRuleML: From Metamodel to Use Cases. In: Morgenstern, L. et al. (eds.): Theory, Practice and Applications of Rules on the Web. Springer, Berlin (2013).
66. TurnipBox Homepage: https://turnipbox.netlify.app/, last accessed 2022/10/21.
67. Libal T., Steen, A.: NAI: the normative reasoner. In: Proc. of ICAIL, pp. 262–263. (2019).
68. Steen, A., Benzmüller, C.: The higher-order prover Leo-III. In: Galmiche, D. et al. (eds.) Proceedings of IJCAR 2018, pp. 108–116 (2018).
69. Otten, J.: MleanCoP: a connection prover for first-order modal logic. In: Demri, S. et al. (eds.): Proceedings of IJCAR 2014, pp. 269–276 (2014).
70. Esterle, K. et al.: Formalizing Traffic Rules for Machine Interpretability. 2020 IEEE 3[rd] Connected and Automated Vehicles Symposium (CAVS). IEE (2020).
71. Rizaldi, A. et al.: Formalising and Monitoring Traffic Rules for Autonomous Vehicles in Isabelle/HOL. In: Polikarpova N., Scheinder S., (eds.): Integrated Formal Methods. Proceedings of the 13[th] International Conference, IFM 2017. Turin, Italy (2017).
72. Krasowski, H., Althoff, M.: Temporal Logic Formalization of Marine Traffic Rules. In: Proceedings of the 2021 IEEE Intelligent Vehicles Symposium (IV). IEEE (2021).
73. Maierhofer, S. et al.: Formalization of Interstate Traffic Rules in Temporal Logic. In: Proceedings of the 2020 IEEE Intelligent Vehicles Symposium (IV). IEEE (2020).
74. Gressenbuch, L., Althoff, M.: Predictive Monitoring of Traffic Rules. In: Proc. of the 2021 IEEE International Intelligent Transportation Systems Conference (ITSC). IEEE (2021).
75. Hekmatnejad, M., et al.: Encoding and Monitoring Responsibility Sensitive Safety Rules for Automated Vehicles in Signal Temporal Logic. In: Proceedings of the 17th ACM-IEEE International Conference on Formal Methods and Models System Design, pp. 1-11 (2019).
76. Xiao, W., et al.: Rule-based Optimal Control for Autonomous Driving. arXiv:2101.05709v1 [cs.RO] (2021).

77. Karimi, A., Duggirala, P.: Formalizing traffic rules for uncontrolled intersections. In: Proceedings of the ACM/IEEE 11[th] Inter. Conf. on Cyber-Physical Systems (ICCPS) (2020).
78. Satoh, K., et al.: PROLEG: An Implementation of the Presupposed Ultimate Fact Theory of Japanese Civil Code by PROLOG Technology. In: Onoda, T. et al (eds.): New Frontiers in Artificial Intelligence, pp. 153-164. Springer, Berlin (2011).
79. Nipkow, T., et al: Isabelle/Hol. A Proof Assistant for Higher-Order Logic. (2002).
80. Zhang, Q., et al.: A Systematic Framework to Identify Violations of Scenario-dependent Driving Rules in Autonomous Vehicle Software. In: Proc. of POMACS, pp. 1-25 (2021).
81. Raabe, O., et al.: Recht ex Machina. Formalisierung des Rechts im Internet der Dienste. Springer, Berlin, (2012).
82. Łukasiewicz, J.: On Three-Valued Logic. In: Łukasiewicz, J.:, Selected Works. Ed. by L. Borkowski. North-Holland, Amsterdam, pp. 87-88 (1970). (= O logice trójwartościowej. Ruch Filozoficzny, 5, 170-171.
83. Łukasiewicz, J.: Philosophische Bemerkungen zu mehrwertigen Systemen des Aussagenkalküls. Comptes Rendus Séances Société des Sciences et des lettres Varsovie, cl. III, 23. 51-77 (1930).
84. Menger, K.: A Logic of the Doubtful. On Optative and Imperative Logic. Reports of a Mathematical Colloquium. University of Notre Dame, 2, 1, 53-64 (1939).
85. Menger, K.: Moral, Wille und Weltgestaltung. Grundlegung zur Logik der Sitten. Suhrkamp, Frankfurt a.M. (1997).
86. Storer, T.: The Logic of Value Imperatives. Philosophy of Science, 13, 1, 25-40 (1946).
87. Fisher, M.: A three-valued calculus for deontic logic. Theoria, 27, 107-118 (1961).
88. Åqvist, L.: Postulate sets and decision procedures for some systems of deontic logic. Theoria, 29, 2, 154-175 (1963).
89. Jobczyk, K.: Multi-Valued Deontic Halpern-Shoham Logic for Fuzzy Deontic-Temporal Expressions. Journal of Intelligent and Fuzzy Systems, 36, 5, pp. 1-12 (2019).
90. Dellunde, P., Good, L.: Introducing Grades in Deontic Logics. In: Deontic Logic in Computer Science. Proc of the 9[th] DEON 2008, pp. 248-262 (2008).
91. Sadegh Zadeh, K.: Fuzzy Deontics. In: Seising, R., Sanz González, V. (eds.): Soft Computing in Humanities and Social Sciences, pp. 141-156. Springer, Berlin (2012).
92. Gottwald, S.: Mehrwertige Logik. Akademie-Verlag, Berlin (1989).
93. Bocheński, J.: Formale Logik. Karl Alber, München (2015).
94. Stachniak, Z.: Many-Valued Computational Logics. In: Proceedings of the nineteenth International Symposium on Multiple-Valued Logic, pp. 391-397. IEEE (1989).
95. Makinson, D.; Torre, L. v. d.: What is Input/Output Logic? In: Löwe, B. et al. (eds.): Foundations of the Formal Sciences II, pp. 163-174. Kluwer, Philadelphia (2003).
96. Parent, X., Torre, L. v. d.: Introduction to Deontic Logic and Normative Systems (2018).
97. Costa, N./Carnielli, W.: On Paraconsistent Deontic Logic. Philosophia, 16, 293-305 (1986).
98. Garson, J.: Modal Logic. In: Zalta, E. (ed.): The Stanford Encyclopedia of Philosophy (2021) URL = https://plato.stanford.edu/archives/sum2021/entries/logic-modal/.
99. Engers, T. v., et al.: Ontologies in the Legal Domain. In: Chen, H., et al. (eds.): Digital Government, pp. 233-261. Springer, Berlin (2008).
100. Alchourrón, C., Bulygin, E.: Normative Systems. Springer, New York (1971).
101. Hansen, J.: Imperatives and Deontic Logic. Leipzig (2008).
102. Censi, A. et al.: Liability, Ethics, and Culture-Aware Behavior Specification using Rulebooks. In: Proceedings of the 2019 ICRA, pp. 8536-8542. IEEE (2019).
103. German Law Archive, StvO-Translation: https://germanlawarchive.iuscomp.org/?p=1290, last accessed 2022/10/21.

# Logical English for Law

Robert Kowalski[1], Jacinto Dávila[2,3], Galileo Sartor[4], and Miguel Calejo[5]

[1] Department of Computing, Imperial College, London, UK
[2] Contratos Lógicos. C.A.
[3] Universidad de Los Andes Mérida, Venezuela
[4] Department of Computing, University of Turin, Turin
[5] Logical Contracts, Lisbon, Portugal

**Abstract.** In this paper we summarise the key features of Logical English (LE) as syntactic sugar for logic programming languages such as pure Prolog, ASP and s(CASP); and we illustrate LE with examples from the Italian citizenship legislation and the US Tax Code.

**Keywords:** Logic Programming, Prolog, Controlled Natural Language, Legal Rule Modelling, Logical English

## 1 Introduction

Logical English (LE) exploits the unique feature of Prolog-like logic programming (LP), that LP is the only programming paradigm based on the use of logic for human thinking and communication. By exploiting this feature, LE becomes a general-purpose programming language, which can be understood with only a reading knowledge of English and without any technical training in computing, mathematics or logic.

LE is not only a Turing-complete computer programming language. It has the potential to represent and reason with a broad range of human knowledge, as shown by its ability to codify the language of law.

Consider the following example, written in LE, and its translation into Prolog.
Ordinary English:

> *All meetings with unvaccinated people are prohibited unless they are excused.*

Logical English:[6]

> *A meeting is prohibited*
> *if a person attends the meeting*
> *and the person is unvaccinated*
> *and it is not the case that the meeting is excused.*

Prolog:

```
is_prohibited(A) :-
attends(B, A), is_unvaccinated(B), not is_excused(A).
```

The implementation of LE in SWISH[8] translates LE programs and queries into Prolog, uses Prolog to answer queries, and translates answers and explanations into LE English syntax.

The example illustrates some of the following characteristics of LE:

– LE avoids pronouns, which are a major source of ambiguity, as in the case of "they", which in this example could refer either to meetings or to people.

– LE represents variables by common nouns prefixed by a determiner such as "a", "an" or "the". The indefinite determiner, "a" or "an", introduces the first occurrence of a variable in a rule. The definite determiner, "the" is used for all later occurrences of the same variable in the same rule. As in Prolog (with some exceptions), all variables are implicitly universally quantified with scope being the rule in which they occur. This means that variables in different rules have no relationship with one another.

– Sentences in LE are either facts, or rules, as in Prolog. Rules have the Prolog-like conditional form *conclusion if conditions*, where the *conclusion* is an atomic sentence and the *conditions* are a combination of atomic sentences, typically connected by *and*. But *conditions* can also be connected by *or* and negation, written in the form *it is not the case that.* The relative precedence of the logical connectives is indicated by indentation (not illustrated in this example).

– As a matter of style and in the interests of greater precision, common nouns are preferably expressed in the singular, and verbs are expressed in the present tense. The temporal relationship between events and time-varying facts can be expressed, if necessary, by referring to time explicitly.

LE inherits the feature of Prolog that propositions can occur as arguments of higher-order or meta-level predicates. LE uses this to represent deontic modalities (obligation, prohibition, permission) and other propositional attitudes (notification, belief, desire, dislike). For example, here the keyword *that* introduces the proposition *a meeting is prohibited at a time T1* as an argument of the meta-predicate *the person is notified*:

> *a person violates the rules at a time T2*
> *if the person is notified that a meeting is prohibited at a time T1*
> *and the person attends the meeting at T2*
> *and T1 is before or at the same time as T2.*

As this rule also shows, a variable can be given a symbolic name.

Atomic sentences, which are facts, the conclusions of rules, or constituents of the conditions of rules, are instances of predicates declared by means of templates, such as:

> *a person* violates the rules at *a time*,
> *a person* is notified that * message*,
> *an eventuality* is prohibited at *a time*.

where the asterisks identify the arguments of the predicates.

We have used the implementation of LE in SWISH to represent a wide range of legal texts, helping to identify ambiguities, to explore modifications and alternative representations of the same text, and to compare the logical consequences of the alternatives.

## 2 The Italian Citizenship Example

We are also developing analogues of LE for other natural languages, such as Spanish and Italian. Figure 1 shows both an LE representation and a corresponding LI representation of Article 1 of Act No. 91 of 5 February 1992:

> 1. E' cittadino per nascita: a) il figlio di padre o di madre cittadini; b) chi e' nato nel territorio della Repubblica se entrambi i genitori sono ignoti o apolidi, ovvero se il figlio non segue la cittadinanza dei genitori secondo la legge dello Stato al quale questi appartengono.

Google translate gives the following translation into English:

> Citizen by birth: a) the child of a citizen father or mother; b) who was born in the territory of the Republic if both parents are unknown or stateless, or if the child does not follow the citizenship of the parents according to the law of the state to which these belong.

Here both the English condition "the child does not follow the citizenship of the parents according to the law of the state to which these belong" and its Italian counterpart, taken literally, seem to cover only the case where both parents have the same citizenship. Moreover, both the Italian "ovvero se" and the corresponding English "or if" seem to relate to a separate alternative from the alternatives that precede it. These readings of the natural language texts leave uncovered such deserving cases as the child having one parent who is stateless or unknown, and another parent who cannot pass on its citizenship(s) to its child. It seems doubtful that this would have been the intention of the law.

The LE and LI representations in figure 1 1.1 [6] incorporate one intended interpretation of Article 1.1. Of course, other interpretations are possible, and they could also be represented in LE.

Figure 1 also illustrates two further features of LE: the use of indentation to represent the relative strength of binding of the logical connectives, and the LE construction "for all cases in which ... it is the case that ... ", which translates into "forall" in Prolog.

---

[6] https://logicalenglish.logicalcontracts.com/p/italian_citizen_new.pl
https://logicalenglish.logicalcontracts.com/p/cittadinanza_italiana.pl

```
16    the knowledge base italian_citizen_new includes:     16    la base di conoscenza cittadinanza_italiana include:
17                                                         17
18    a person A is an italian citizen                     18    una persona A ha la cittadinanza italiana
19    if the person A is an italian citizen by birth.      19    se A ha la cittadinanza italiana per nascita.
20                                                         20
21    a person A is an italian citizen by birth            21    una persona A ha la cittadinanza italiana per nascita
22    if a person B is the parent of A                     22    se una persona B è genitore di A
23    and B is an italian citizen.                         23    e B ha la cittadinanza italiana.
24                                                         24
25    a person A is the parent of a person B               25    una persona A è genitore di una persona B
26    if A is the father of B                              26    se A è madre di B
27        or A is the mother of B.                         27        o A è padre di B.
28                                                         28
29    a person A is an italian citizen by birth            29    una persona A ha la cittadinanza italiana
30    if A is born in italy                                30    se A è nato in italia
31    and for all cases in which                           31    e per tutti i casi in cui
32        a person B is the parent of A                    32        una persona B è genitore di A
33        it is the case that                              33        è provato che
34        B is stateless                                   34        B è sconosciuto/a
35            or B is unknown                              35            o B è apolide
36            or A does not follow the citizenship of B.   36            o A non segue la cittadinanza di B.
```

**Fig. 1.** The Italian Citizenship Example

In this representation of Italian citizenship, the possibility that a parent is unknown is left explicit, to more closely reflect the wording of the original legal source. This possibility could also be expressed implicitly, using negation as failure, to conclude that a parent of a person is unknown if information about that parent is missing from the knowledge base or from the scenario. It is in fact possible, with the current representation, to say that a person is born in italy, and not to give any information about the parents at all. The "forall" condition would be satisfied (vacuously), and the person would be granted Italian citizenship.

The presence or absence of a fact that a person A is the parent of a person B may depend on such circumstances of the birth as whether the parent A (mother) decides to be recognized as a parent if B, or instead decides to abandon the child B. Both possibilities, that a parent is unknown (implicitly through negation as failure or explicitly by means of an fact that the parent is unknown), are compatible with the legal source. Moreover, the ability to represent both possibilities in LE may help to remove the ambiguity of the rule and it assist in its automation.

## 3  A Tax Law Example

LE and its counterparts for other natural languages can be used to codify legal rules to support their complete or partial automation. But they can also be used to assist with the drafting of legal rules, to help ensure that the rules actually express their intended interpretation. Used in this way, for both purposes, LE can provide powerful support for both drafting and applying the law, as envisaged in the campaign to represent Rules as Code (RAC) [11].

Within this RAC context, we have investigated §121 of the US Internal Revenue Code, following the lead of the Catala project [9]. Figure 2 shows an LE representation of a portion of §121, which deals with the exclusion of gain from the sale of a principal residence:

> (Subsection a) Exclusion
> Gross income shall not include gain from the sale or exchange of property if, during the 5-year period ending on the date of the sale or exchange, such property has been owned and used by the taxpayer as the taxpayer's principal residence for periods aggregating 2 years or more.

Notice that the condition of the sentence is ambiguous.

- It could mean that the periods during which the taypayer both owns and uses the property within the 5-year period aggregate to 2 years.
- Or it could mean that the periods during which the taypayer owns the property within the 5-year period aggregate to 2 years, and the periods during which the taypayer uses the property within the 5-year period aggregate to 2 years.

The authors of [9] do not mention that the sentence is ambiguous. However, the LE representation in figure 2 follows the Catala implementation of §121, which assumes that the drafters of the Tax Code intended the second interpretation[7]. Of course, the alternative interpretation could also be represented in both LE and Catala.

Subsection (b)1 of §121 defines a cap of $ 250,000 on the amount of gain that can be excluded from a sale or exchange of property. But the Code itself does not express the common sense understanding of the cap as limiting the amount that can be excluded. The Caltala implementation builds this understanding into the representation of subsection (a) itself. However, the LE implementation expresses this understanding separately in lines 126-132 of figure 2.

As is well-documented in the field of AI and law, a typical legal document consists of rules and exceptions, as well as exceptions to exceptions, etc. In this regard, the US Tax Code is not exceptional. In particular, subsection (a) is subject to the exception:

> (3) Application to only 1 sale or exchange every 2 years
> (A) In general
> Subsection (a) shall not apply to any sale or exchange by the taxpayer if, during the 2- year period ending on the date of such sale or exchange, there was any other sale or exchange by the taxpayer to which subsection (a) applied.

To represent this exception, we add an extra, explicit condition (in lines 45-46 of figure 2) to the rule for subsection (a), expressing that it is not the case that the exception holds. We also interpose an intermediate conclusion (on line 42) and an intermediate condition (on line 40) in the representation of subsection (a), expressing explicitly that subsection (a) applies. The exception itself is on lines 118-124.

---

[7] https://catala-lang.org/en/examples/us-tax-code

In general, rules and exceptions are represented in Prolog and most other logic programming languages by adding such otherwise implicit conditions (that the contrary of the conclusion does not hold) explicitly. However, Satoh [10] has argued that rules written in this Prolog form are hard for lawyers to understand. It is easier for lawyers to understand rules written more simply with unwritten implicit conditions, as in the Prolog-based Legal reasoning support system, Proleg [10]. We agree with this approach, and plan to incorporate such implicit conditions in a future version of LE.

Figure 3 shows an example scenario and query, both written in LE syntax. The SWISH implementation of LE allows several such scenarios and several such queries in the same document. Figure 4 shows the answer and proof tree obtained by combining scenario one with query one. The proof tree is an explanation of the answer to the query, given the scenario and the more general rules in the knowledge base. The last part of the proof, highlighted in red, displays the conditions that could not be proved, justifying the conclusion that the exception does not hold in the given scenario.

## 4 Conclusions

Our experience with using LE for many practical, proof-of-concept applications suggests that LE has many valuable applications, which are not restricted to the automation of legal rules. These applications include the disambiguation of legal rules written in natural language, as well as the exploration of the logical consequences of the rules, in the context of different scenarios.

These applications are facilitated by the fact that users can read, understand and use LE without any technical training in logic, computing or mathematics. But, although LE may be easy to read, at this stage in its development, it is not easy to write. The fact that the drafters of the Italian citizenship law and of the US Tax Code did not identify the ambiguities in their legal texts proves how difficult it can be to express information clearly and unambiguously in natural language.

Even if LE were never used to help automate the application of legal rules, it would serve a useful purpose as a discipline in training writers to express themselves in terms that readers can more readily understand.

```
39   gross income of a taxpayer excludes gain from a sale or exchange of a property at a date if
40       subsection (a) applies to the sale or exchange of the property by the taxpayer at the date.
41
42   subsection (a) applies to a sale or exchange of a property by a taxpayer at a date if
43       the taxpayer meets the ownership requirements of subsection (a) with respect to the sale or exchange of the property at the date
44       and the taxpayer meets the use requirements of subsection (a) with respect to the sale or exchange of the property at the date
45       and it is not the case that
46           subsection (a) shall not apply to the sale or exchange of the property by the taxpayer.
47
48   a taxpayer meets the ownership requirements of subsection (a) with respect to a sale or exchange of a property at a date
49       if the sale or exchange of the property occurs at the date
50       and a period of 5 years ends at the date
51       and the property has been owned by the taxpayer for periods aggregating 2 years or more during the period.
52
53   a taxpayer meets the use requirements of subsection (a) with respect to a sale or exchange of a property at a date
54       if the sale or exchange of the property occurs at the date
55       and a period of 5 years ends at the date
56       and the property has been used by the taxpayer as principal residence for periods aggregating 2 years or more during the period.
118  subsection (a) shall not apply to a sale or exchange of a property by a taxpayer
119      if the sale or exchange of the property occurs at a date
120      and an other sale or exchange of a second or the same property occurs at a second date
121      and the other sale or exchange is different from the sale or exchange
122      and a period of 2 years ends at the date
123      and the second date is included in the period
124      and subsection (a) applies to the other sale or exchange of the second or the same property by the taxpayer at the second date.
125
126  the amount of gain to be excluded for a taxpayer from a sale or exchange under subsection (a) is an amount G
127      if the received income for the sale or exchange is an amount I
128      and the amount of gain excluded for the taxpayer from the sale or exchange under subsection (a) shall not exceed a cap
129      and      the cap >= I
130          |        and G is I
131          or   the cap < I
132          |        and G is the cap.
```

**Fig. 2.** §121 of the US Internal Revenue Code subsection (a)

```
134    scenario one is:
135        the sale of the house occurs at 2022-06-20.
136        the received income for the sale is 300000.
137        the given period of 5 years ends at 2022-06-20.
138        first set of periods of the taxpayer owing the house aggregates to 2 of years.
139        first set of periods is contained in the given period.
140        second set of periods of the taxpayer using the house as principal residence aggregates to 3 of years.
141        second set of periods is contained in the given period.
142
143  > scenario two is: ⋯
156  > scenario three is: ⋯
180    query one is:
181        gross income of which taxpayer excludes gain from which exchange of which property at which date.
```

**Fig. 3.** A scenario and query for subsection (a)

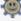

**Fig. 4.** Explanation of the answer of a query in LE

## References

1. Kowalski R (1990) English as a logic programming language. New Generation Comput 8(2):91–9.
2. Kowalski R (1992) Legislation as Logic Programs In: Logic Programming in Action (eds. G. Comyn , N. E. Fuchs, M. J. Ratcliffe), Springer Verlag, 203-230.
3. Kowalski R (2020) Logical English, In Proceedings of Logic and Practice of Programming (LPOP).
4. Kowalski, R. and Datoo, A., 2022. Logical English meets legal English for swaps and derivatives. Artificial Intelligence and Law, 30(2), pp.163-197.
5. Kowalski, B., Dávila, J. and Calejo, M., Logical English for legal applications. XAIF, Virtual Workshop on Explainable AI in Finance (2021).
6. LE unvaccinated: https://le.logicalcontracts.com/p/unvaccinatedupdate.pl, last accessed 2022/09/06.
7. Sartor, G., Dávila, J., Billi, M., Contissa, G., Pisano, G. and Kowalski, R., 2022. Integration of Logical English and s (CASP), 2nd Workshop on Goal-directed Execution of Answer Set Programs (GDE'22), August 1, 2022.
8. Jan Wielemaker, Torbjörn Lager , and Fabrizio Riguzzi, 2015. SWISH: SWI-Prolog for sharing. arXiv preprint arXiv:1511.00915.
9. Huttner, L., Merigoux, D. & Chataing, N. Catala: Moving Towards the Future of Legal Expert Systems. (2022), https://hal.inria.fr/hal-02936606
10. Satoh, K., Asai, K., Kogawa, T., Kubota, M., Nakamura, M., Nishigai, Y., Shirakawa, K. and Takano, C., 2010, November. PROLEG: an implementation of the presupposed ultimate fact theory of Japanese civil code by PROLOG technology. In JSAI international symposium on artificial intelligence (pp. 153-164). Springer, Berlin, Heidelberg.
11. Mohun, James, and Alex Roberts. "Cracking the code: Rulemaking for humans and machines." (2020).

# Making the Implicit Explicit:
## The Potential of Case Law Analysis for the Formalization of Legal Norms

Georg Borges[a], Clara Wüst[a], Diogo Sasdelli[a], Salome Margvelashvili[a]
and Selina Klier-Ringle[a]

[a] Institute of Legal Informatics, Saarland University, 66123 Saarbrücken, Germany
georg.borges@uni-saarland.de

**Abstract.** Representing legal norms through a logical formalism is not only one of the most traditional subjects of legal informatics, but also an essential prerequisite for the development of law-compliant autonomous systems, e.g., autonomous vehicles (AVs). Formalizing legal norms adequately requires more than merely choosing or defining a logical system by the means of which the norms are to be represented. Instead, as a first step, it is essential to assess the entire set of legal norms relevant for the legal question at hand. In addition to norms expressly laid down by statutory law, implicit ('unwritten') norms, which are revealed by courts and other authorities, have to be taken into account. In the paper at hand, we present a method for systematically identifying and compiling applicable implicit legal norms, thus significantly adding to the set of norms explicitly put into statute and applicable to a concrete case. We illustrate the method, which was designed as a first step towards formalizing legal norms applicable to a concrete case, by applying it to the written legal rules defining Zeichen 295 [traffic sign 295] of the Deutsche Straßenverkehrsordnung (StVO) [German Road Traffic Regulation]. It is shown that this method can significantly increase the corpus of applicable legal norms. Particularly, it leads to closing eventual legal gaps and solving legal antinomies, thus providing clarity to many difficult cases.

**Keywords:** Systematic Compilation of Implicit Norms, Autonomous Driving, Formalization of Legal Norms, Legal Logic, Traffic Law.

## 1 Introduction – Reviving an Old Dream of Legal Informatics

Using logical methods to represent legal norms in a mathematically precise way is an old dream of legal informatics, or rather of *legal cybernetics*, as the field was originally known. As a research discipline, legal logic emerged independently from its philosophical counterpart, i.e., so-called *deontic logic*, in the early post-war period. It is closely related to German legal theory; among its founders are names such as Julius Stone [1], Ulrich Klug [2], Eduardo García Máynez [3] and Ilmar Tammelo [4]. Over the following decades, however, as the true extent of the arduous task of formalizing legal systems became clearer, and following the so-called AI-Winters, the interest in this original goal of legal informatics and, *a fortiori*, in legal (and deontic) logic steadily dwindled.

Today, following groundbreaking results in the field of AI-research, especially with respect to neural networks and machine learning, approaches to formally represent norms and legal knowledge are slowly regaining the attention of legal scholars and computer scientists.

Recently, research on autonomous driving seems to revitalize the idea of legal logic, and with it of legal informatics in its original form. The success of autonomous driving as a mobility concept requires developing algorithms that not only steer a car on real-world streets in a safe and efficient manner, but are also able to comply with traffic rules. This becomes particularly obvious where autonomous vehicles (AVs) and human drivers, as well as pedestrians share the same streets and also have to comply with the same traffic rules. Incidentally, in Germany, which, in 2019, pioneered amending its *Straßenverkehrsgesetz* (*StVG*) [Road Traffic Act] to include regulations on autonomous driving, explicitly requires AVs to be able to autonomously comply with traffic rules (§ 1e para. 2 s. 2 German Road Traffic Act).

For a computer to be able to comply with traffic rules, it is necessary to reduce all the tasks involved herein – e.g., recognizing the rules applicable to a given case, identifying violations and evaluating exceptions – to a purely syntactic method. This syntactic reduction, which can also be called a *formalization for computational purposes*, consists of three main phases: (1) determining the *object of the formalization*, i.e., in the case at hand, traffic regulations; (2) determining the *formalization framework*, e.g., a logic system to serve as base for the formalization of traffic rules; (3) determining the *implementation framework*, i.e., the concrete form by which a computer is supposed to represent and process the respective formalization of traffic rules (e.g., a machine-readable formal language).

So far, research on formalizing traffic rules for autonomous driving has focused almost exclusively on the last two phases, leaving phase (1), i.e., determining the very object of the formalization, largely untouched.

With respect to phase (2), current research on formalizing traffic rules for autonomous driving is focusing either on frameworks based on modal logic (ML), particularly so-called deontic logic [5], or on frameworks based on first-order predicate logic (FOL), including different types of temporal logic (TL) [6, 7, 8].

Concerning phase (3), depending on which of the formalization approaches discussed above is being used, several different implementation frameworks can be applied. ML-based formalizations can be represented, e.g., by using *LegalRuleML* [9], [10], a markup language based on *RuleML* designed to be able to grasp the particularities of legal rules. Furthermore, arguments and inferences built on ML-formalizations can be represented and verified by using normative reasoning engines or frameworks capable of encoding systems of deontic logic, such as Turnip[1] and NAI [11, 12]. FOL-based formalizations, in turn, can be easily implemented in programming languages such as PROLOG [13] or PROLEG [14]. One can also use theorem provers based on higher-order-logic (HOL), such as Isabelle [15], to verify arguments based on FOL-formalizations [16, 17]. To verify whether a rule has been fulfilled or violated, adapted SAT-solvers can be employed to both ML- and FOL-based systems [18, 19, 20].

---

[1] https://turnipbox.netlify.app/ (accessed 10.11.2022).

Evidently, the decision of which implementation framework to employ in phase (3) depends on the formalization framework used in phase (2). Similarly, determining which formalization framework should be employed depends on the nature of the object of the formalization, which is determined in phase (1). In other words: before a proper formalization of traffic regulations can be implemented, it is necessary to clarify what exactly these regulations are and what semantic particularities traffic rules might have.

Hence, the disregard for phase (1) can be considered as a fatal flaw of current research on the formalization of traffic rules for autonomous driving. To fill this gap, we propose a method for systematically identifying and compiling implicit legal norms, thus significantly expanding the corpus of explicitly stipulated norms applicable to a concrete case. We illustrate this method by applying it to the written legal rules defining traffic sign 295 of the German Road Traffic Regulation.

## 2     Systematic Compilation of Implicit Norms – Overview

With respect to formalizing legal norms, the main issue concerning the determination of the object of the formalization (phase (1) in the model above) arises from the fact that law can be regarded as a corpus of valid legal norms which is considerably larger than the set of written norms explicitly stipulated in statutes and other legal texts promulgated by the legislator, such as regulations etc.

Hence, before one can properly formalize the set of norms applicable to a case, one must first determine the corpus of valid legal norms. The corpus of valid legal norms consists out of the set of explicit norms derived from legislation and the set of implicit norms derived from case law and legal dogmatics. By compiling the implicit norms as well, it is possible to considerably add to the set of explicitly formulated norms. Moreover, implicit norms are particularly valuable, for they often contain clarifications for solving borderline cases and solutions for legal gaps and normative conflicts. As a result, by adding to the set of explicitly formulated norms, difficult cases can be addressed and solved more efficiently.

While the method proposed here largely builds on classic legal hermeneutics and legal methodology, this systematic compilation of implicit norms is intended to be a preliminary stage for logical formalization. Hence, it has a particular focus on the way in which its results are compiled. Particularly, it aims at revealing compact, unambiguous, and consistent sets of implicit norms applicable to a case.

The method proposed here is based on the concept that legal norms can be extracted from explicitly formulated statutes as well as from court decisions and even from legal literature. Although different approaches to and concepts of sources of law have been discussed for centuries, there is a consensus that legal norms are not only put down in black and white in statutory law by the legislator, but also in decisions by courts, in so-called case law. Even the opinions of legal scholars interpreting statutory or case law can be considered as part of the lawmaking process as such opinions may enjoy the rank of an authority and be followed by applying the law. For this reason, we consider written (statutory) and 'unwritten' norms formulated and applied by courts and by experts as potential sources of traffic rules.

However, for a (explicit or implicit) legal norm to be binding, it is necessary for it to be *recognized*. In this sense, legal experts recognize statutory law once it has become effective and as long as it has not been repealed by statute or, in some jurisdictions, such as Germany, by a competent court (e.g., in Germany, the Federal Constitutional Court (*Bundesverfassungsgericht*)). In the case of case law, legal experts consider a norm as recognized if it has been put down by the highest court or a majority of courts. In the same way, norms put down by legal scholars can be considered as recognized if a strong majority of experts agrees on the existence of such a norm.

Generally, within the method proposed here, legal experts identify 'unwritten', implicit norms connected with 'written', explicit norms in three steps:

(1) Finding relevant court decisions and/or expert literature
(2) Extracting the relevant norm therein
(3) Verifying whether that norm is recognized. A norm is recognized in this sense if a court of highest authority, or a majority of courts has formulated that norm down with the same normative content. In the absence of an opposing view, a norm can be considered as recognized if one source has suggested this norm and another has confirmed it.

## 3 Illustrating the Method – The Solid Line

In this section, we illustrate the method proposed here by applying it to the solid line, a well-known example from the field of autonomous driving.

### 3.1 The Solid Line

From a purely syntactical perspective, road signs and markings constitute a system of symbols and can therefore be considered as a type of (formal) language [21]. Nevertheless, programming an AV to comply with the law requires more than just encoding these signs as mere symbols. Rather, the information which has to be formalized and encoded (i.e., the *object of formalization* (phase (1))) are the traffic rules contained in the meanings of these signs and markings. To properly understand these meanings, it is necessary to also consider their historical development.

The need to create rules and signs governing motorized traffic went hand in hand with the flourishing development of the automotive industry at the beginning of the 20th century. The induced change in mobility resulted in a need for new safety requirements. To meet these, in 1909, the *Convention with Respect to the International Circulation of Motor Vehicles* attempted, for the first time, to establish and standardize traffic regulations at an international level. Further international agreements followed: the 1926 *International Convention Relative to Motor Traffic* already included six road signs. The *Geneva Convention on Road Traffic*, prepared by the *United Nations Conference on Road and Motor Transport* in 1949, included a *Protocol on Road Signs and Signals* with an extensively developed system of road signs. Finally, the *Vienna Convention on the Signs and Signals* and the *Vienna Convention on Road Traffic*, signed in October 1968, replaced the *Geneva Convention*. Notwithstanding, some countries, including the United States, China and Japan, are yet to sign or ratify the aforementioned conventions.

At the national level, however, such rules had existed even before the first international convention on traffic regulations came into force. In Germany, e.g., the first regulation concerning motorized traffic dates back to 1906.

Traffic signs and road markings had already been used in railway and shipping traffic in the 19th century. The marking of the solid line can be traced back to Edward N. Hines (1870-1938) and was the very first road marking. It spread quickly since it was useful for providing a clear marking of the roadway. In Germany, the solid line was established as a standard marking in the 1930s, during the expansion of the motorway (*Reichsautobahn*) [21].

Being a convention under international law, the Vienna Convention has to be transformed by the contracting parties into their national law. In Germany, these rules were transformed in the German Road Traffic Act [*Straßenverkehrsgesetz* (*StVG*)]. The *StVG* is a federal law granting the Federal Government the competence to issue regulations. On this legal basis, the Federal Government enacted the Road Traffic Regulation [*Straßenverkehrsordnung* (StVO)] containing a set of precise rules on traffic regulations and road signs.

In an annex to § 41 *StVO*, this regulation introduces the sign 295, which contains a picture of the solid line and the explicit legal text stating the legal norms indicated by the sign.

### 3.2    Identification of the Implicit, 'Unwritten' Norms

**Search for the relevant court decisions.** German courts are not bound by any strict rule on whether they have to publish their decisions. The Federal Constitutional Court [*Bundesverfassungsgericht (BVerfG)*] and the Federal Supreme Court [*Bundesgerichtshof BGH)*] publish most of their decisions, yet judges at other courts publish their decisions only on a case-to-case basis, according to their discretion. In many cases, the parties themselves publish decisions on data bases or in law journals.

For our research, qualified legal experts[2] searched through the relevant case law in the leading databases on German law, *juris* and *beck-online*, and in an extensive collection of traffic law, the so-called Traffic Law Collection [*Verkehrsrechtsammlung (VRS)*], comprising 141 volumes and dating back to the year 1949. The legal experts searched by using the keywords "Zeichen 295" [sign 295] and "durchgezogene Linie" [solid line]. The researchers supplemented their search with relevant decisions gained by following cross-references from expert literature. Altogether, 173 decisions on sign 295 could be collected and examined. This first research step could likely be greatly optimized by the employment of methods based on Natural Language Processing (NPL).

---

[2] All legal experts working on this project have at least a master's degree in law (LL.M.) or the German First State Examination, the latter of which equals the first.

The legal experts screened the court decisions manually and extracted the relevant sections. When doing so, they identified those parts of the decisions formulating a legal norm (i.e., those parts with normative content) flowing from sign 295.

Out of the 173 decisions examined, 35 relevant decisions with normative content have been found from the on sign 295.

**Extraction of Legal Norms.** The normative content, or the legal rule formulated in a court decision can come in various ways, and the wording used by the courts may often differ even if the normative content is the same. Therefore, the most demanding task in the method discussed here is to decide, by way of interpretation of the court decision, what the exact normative content of that decision is and whether different wordings stipulate different rules. This task is nothing less than one of the most traditional tasks of legal experts who are trained to interpret court decisions and other legal texts in order to reveal the legal norm stipulated therein. In order to create a unified wording and to prepare the logic representation of the norms, the identified norms in the court decisions have been rephrased as a conditional sentence using an "if..., then" structure.

Each of the relevant sections in the court decisions has been examined by three to five legal experts. In each case, one of the experts suggested a rephrased wording of the norm found in a section of a relevant court decision, and the other experts, who had interpreted the same section independently from each other, either agreed or suggested a different wording. In case of divergence, all of the experts discussed the interpretation of the section and the suggested wordings together. Then, they voted for one of the wordings. In almost every case, after having discussed the wordings, the experts could unanimously agree on one version. By this method, 21 rules could be identified on the basis of 35 decisions with normative content.

**Verification Whether the Norm is Being Recognized.** At last, the recognition of the identified legal norms had to be verified. To be recognized, at least two court decisions or one court of highest rank had to put that norm down in black and white. A further condition in both cases was that there were no opposite/differing decisions.

The legal experts came to the conclusion that all of the identified legal norms had to be classified as being recognized, mainly due to the fact, that almost all of the published decisions were rulings of a Court of Appeal or of the Federal Supreme Court (*BGH*).

### 3.3 Legal Dogmatics

The main particularities of the method discussed here with respect to the analysis of secondary materials containing legal dogmatics are presented in this section. As in the case of case law, any secondary materials are to be assessed with regard to their normative content; the first step is to collect the relevant sources. The analysis of sign 295 is limited to classic legal literature; with the main source being commentaries on the *StVO*. Legal commentaries are a specific type of literature in the DACH region. Amongst others, they contain explanations on how to interpret and apply the law.

Only a few commentaries explicitly address sign 295. This might be due to the fact that signs and road markings are not part of the *StVO* itself, but are only contained in the annex to § 41 *StVO*, along with with a written explanation by the legislator.

The team of legal experts evaluated a total of eleven commentaries. As a result, eight norms could be extracted from literature. Whereas four norms were identical with the ones derived from court decisions, legal scholars put down four norms which are different from those formulated by courts.

## 4   Evaluation of the Results on Sign 295

The use of the method presented here on the solid line (sign 295) led to considerable results: From two explicitly stipulated norms in the statutory provision, the analysis of court decisions led to the deduction of a total of 16 legal norms.

The most important 'unwritten' norms certainly concern exceptions to the general prohibition not to cross the solid line. Such an exception applies if there is an obstacle on the roadway which makes it impossible to go on without crossing the line. In this case, courts exceptionally allowed to cross the solid line, if further specifications applied. One of these is that crossing the line in such a situation is only permissible if this does not create any additional risk to road traffic.

Also, the courts recognized some additional norms deduced from sign 295. For example, the German Federal Supreme Court (*BGH*) deduced the norm that a driver travelling on a roadway with a solid line may rely on not being overtaken, if the carriageway is so narrow that the driver behind him can overtake only by crossing the solid line and driving on the opposite roadway.

To illustrate the results of the method, Table 1 below shows two legal norms explicitly laid down in the annex to § 41 *StVO* (1.), an exception to the rule (2.) and an additional rule (3.) recognized by the courts. With regard to the latter, first, the table shows the norm as our researcher team formulated it to prepare its representation in logic (a). Second, it shows the original section of a court decision which recognized the norm (b). Finally, the table shows the court decision from which the section named at (c) was quoted.

## 5   Result – Potential of Systematic Norm Collection

In conclusion, as it was shown through the example of sign 295, it becomes apparent that the method proposed here considerably expands the corpus of applicable legal norms by extracting implicit, 'unwritten' norms from case law as well as from legal dogmatics.

This expansion is remarkable both in quantity and quality: in addition to one norm laid down in the statutory text, the researcher team identified 21 norms recognized in court decisions and literature.

**Table 1.** Basic rule, exceptions of and addition to sign 295.

| | |
|---|---|
| **1. Legal Norm ext. of sign 295** | a) a person operating a vehicle must not cross or straddle the solid line. |
| | b) if the solid line separates the portion of the roadway for traffic travelling in the opposite direction, traffic must keep to the right of it. |
| **2. a) Norm recognized by courts** | If a vehicle driver is driving on a roadway with a solid line, then he may cross the line in order to drive past a parked vehicle or a vehicle that is not only stopping for a short time, provided that there is no danger to other traffic. |
| **b) Normative text passage from the given case** | "In principle, the prohibition to cross the solid line established by the law must be strictly followed. It cannot be left to the discretion of the individual road user whether or not a traffic law requirement or prohibition is to be followed in an individual case. Notwithstanding this principle, crossing the solid line may be permissible in exceptional cases if there is a compelling reason for doing so and a hazard to other traffic is ruled out. Thus, case law has permitted crossing the solid line in exceptional cases for the purpose of passing a stopped or parked vehicle if this does not cause any danger (OLG Hamm in JMBl NRW 1957, 209 and Verk-Mitt 1960, 62 No. 93). The AG is to be agreed that such an exceptional case justifying non-observance of the prohibition is not already given if the obstacle arises only for a short moment." |
| **c) Source (court decision) of the quoted section** | OLG Hamm, Urt. v. 14.10.60 – Az. 1 Ss 1207/60 |
| **3. a) Norm recognized by courts** | If a driver is driving on a roadway with a solid line and this roadway is so narrow that the driver behind him can overtake only by driving on the opposite roadway, then he may rely on not being overtaken. |
| **b) Normative text passage from the given case** | "On the contrary, such a marking, where it has the effect of a de facto prohibition of overtaking due to the narrowness of the roadway (cf. Senate judgment of 26.11.1974 - VI ZR 10/74 - VersR 75, 331, 332), also protects the trust of the person driving in front that he does not have to expect to be overtaken at this point. Similar to a natural narrowing of the road, the driver may rely on a following driver to act in accordance with traffic regulations, i.e. not to overtake, if this is only possible by driving over the solid line or the restricted area. This could be the decisive cause of the accident, namely that the plaintiff did not expect to be overtaken by the bus when she turned into the – as she puts it – permitted space between the barrier and therefore did not prepare herself for this either." |
| **c) The source (court decision) of the quoted section** | BGH, Urt. v. 28.04.1987 – Az.: VI ZR 66/86, Rnmn. 249 |

Above all, courts recognized essential implicit exceptions to the basic rule as well as the implicit complementary rules. The enlarged corpus of traffic rules can serve as a basis for logic representation and programming of the rules.

We assume that the use of this or similar methods to identify and compile implicit traffic rules will be an essential element of any approach to formalizing traffic rules for automated driving.

# 6      Acknowledgment

The research leading to these results is funded by the German Federal Ministry for Economic Affairs and Energy within the project "KI Wissen – Automotive AI powered by Knowledge." The authors would like to thank the consortium for the successful co-operation.

## References

1. Stone, J.: The Province and Function of Law: Law as Logic, Justice, and Social Control, a Study in Jurisprudence. Harvard University Press, Cambridge (1946).
2. Klug, U.: Juristische Logik. Springer, Berlin (1951).
3. Máynez, E. G.: Introducción a la lógica jurídica. Fondo de Cultura Economica, Mexico (1951).
4. Tammelo, I.: Sketch for a Symbolic Juristic Logic. Journal of Legal Education, **8**(3), 277-306 (1955).
5. Bhuiyan, H., Governatori, G., Bond, A., Demmel, S., Islam, M. B., Rakotonirainy, A.: Traffic Rules Encoding Using Defeasible Deontic Logic. In: Villata S, Harašta J, Křemen P, (eds.). Proceedings of the 33th Annual Conference on Legal Knowledge and Information Systems (JURIX); 2020 Dec 9-11; Brno, Czech Republic, IOS Press, Amsterdam, p. 3-12 (2020), doi: 10.3233/FAIA210330.
6. Esterle, K., Gressenbuch, L., Knoll, A.: Formalizing Traffic Rules for Machine Interpretability. In: Proceedings of the 3rd IEEE Connected and Automated Vehicles Symposium (CAVS); 2020 Oct 4-5; Victoria, Canada. Curran Associates, Red Hook (2020), doi: 10.1109/CAVS51000.2020.9334599.
7. Maierhofer, S., Rettinger, A. M., Mayer, E. C., Althoff, M.: Formalization of Interstate Traffic Rules in Tem-poral Logic. In: Proceedings of the 31st IEEE Intelligent Vehicles Symposium (IV); 2020 Oct 19-Nov 13; Virtual. Curran Assicuates, Red Hook, p.752-759 (2020), doi: 10.1109/IV51971.2022.9827153.
8. Gressenbuch, L., Althoff, M.: Predictive Monitoring of Traffic Rules. In: Proceedings of the 24th IEEE International Intelligent Transportation Systems Conference (ITSC); 2021 Sep 19-22, Indianapolis, US. Curran Associates, Red Hook, p. 915-922 (2021), doi: 10.1109/ITSC48978.2021.9564432.
9. Palmirani, M., Governatori, G., Rotolo, A., Tabet, S., Boley, H., Paschke, A.: LegalRuleML: XML-Based Rules and Norms. In: Olken F, Palmirani M, Sottara D, editors. Proceedings of the 5th International Symposium on Rule-Based Modeling and Computing on the Semantic Web (RuleML2011); 2011 Nov 3-5, Ft. Lauter-dale, US. Springer, Berlin. p. 298-312 (2011), doi: 10.1007/978-3-642-24908-2_30.
10. Athan, T., Governatori, G., Palmirani, M., Paschke, A., Wyner, A.: LegalRuleML: Design Principles and Foundations. In: Proceedings on the 11th International Summer School on

Web Logic Rules; 2015 Jul 31-Aug 4, Berlin, Germany. Springer, Berlin, p. 151-188 (2015), doi: 10.1007/978-3-319-21768-0_6.
11. Libal, T., Steen, A.: NAI: the normative reasoner. In: Proceedings of the Seventeenth International Conference on Artificial Intelligence and Law (ICAIL); 2019 Jun 17-21, Montréal, Canada. The Association for Computing Machinery. New York, p. 262–263 (2019), doi: 10.1145/3322640.3326721.
12. Libal, T., Steen, A.: Towards an Executable Methodology for the Formalization of Legal Texts. In: Proceedings of the 3rd International Conference, CLAR 2020; 2020 Apr 6-9; Hangzhou, China. Springer, Berlin. p. 151-165 (2020), doi: 10.1007/978-3-030-44638-3_10.
13. Sergot, M. J., Sadri, F., Kowalski, R. A., Kriwaczek, F., Hammond, P., Cory, H. T.: The British Nationality Act as a logic program. Communications of the ACM, **29**(5), 370–386 (1986).
14. Satoh, K., Asai, K., Kogawa, T., et al.: PROLEG: An Implementation of the Presupposed Ultimate Fact Theory of Japanese Civil Code by PROLOG Technology. In: Onoda, T., Bekki, D., McCready, E. (eds.): New Frontiers in Artificial Intelligence. Springer, Berlin, p. 153-164 (2011).
15. Nipkow, T., Paulson, L. C., Wenzel, M.: Isabelle/Hol. A Proof Assistant for Higher-Order Logic. Springer, Berlin (2002), doi: 10.1007/3-540-45949-9.
16. Rizaldi, A., Althoff, M.: Formalising Traffic Rules for Accountability of Autonomous Vehicles. In: Proceedings of the 18th IEEE International Conference on Intelligente Transportation Systems, 2015 Sep 15-18, Gran Canaria, Spain. Curran Associates, Red Hook, p. 1658-1665 (2015), doi: 10.1109/ITSC.2015.269.
17. Rizaldi, A., Keinholz, J., Huber, M., Keinholz, J., Feldle, J.: Formalising and Monitoring Traffic Rules for Autonomous Vehicles in Isabelle/HOL. In: Proceedings of the 13th International Conference on Integrated Formal Methods (IFM), Turing, Italy, 2017 Sep 20-22. Springer, Berlin, p. 50-66 (2017), doi: 10.1007/978-3-319-66845-1_4.
18. Giunchiglia, E., Tacchella, A., Giunchiglia, F.: SAT-Based Decision Procedures for Classical Modal Logics. Journal of Automated Reasoning, 28, 143-171 (2022).
19. Lin, Y., Althoff, M.: Rule-Compliant Trajectory Repairing using Satisfiability Modulo Theories. In: Proceedings of the 33rd IEEE Intelligent Vehicles Symposium (IV); 2022 Jun 5-9 Aachen, Germany. Curran Associates, Red Hook, p. 449-456 (2022), doi: 10.1109/IV51971.2022.9827357.
20. Zhang, Q., Hong, D. K., Zhang, Z., et al.: A Systematic Framework to Identify Violations of Scenario-dependent Driving Rules in Autonomous Vehicle Software. Proceedings of the ACM on Measurement and Analysis of Computing Systems; **5**(2), art. No. 15, p. 1-25 (2021), doi: 10.1145/3460082.
21. Stephan, P.: Die Straßenverkehrszeichen in Deutschland und seinen Nachbarstaaten: Vergleichende Untersuchung der Geschichte und Struktur eines kulturellen Zeichensystems. Berlin (2008).
22. Lafontaine, C. In: Freymann H, Wellner W. jurisPK-Straßenverkehrsrecht, 2. Aufl., § 41 StVO (01.12.2021).

# Handling Inconsistent and Uncertain Legal Reasoning for AI Vehicles Design

Yiwei Lu[1], Zhe Yu[2*], Yuhui Lin[3], Burkhard Schafer[1], Andrew Ireland[3], and Lachlan Urquhart[1]

[1] Edinburgh Law School, University of Edinburgh, Edinburgh, UK
[2] Institute of Logic and Cognition, Department of Philosophy, Sun Yat-sen University
[3] School of Mathematical and Computer Sciences, Heriot-Watt University
Y.Lu-104@sms.ed.ac.uk, zheyusep@foxmail.com,
{B.Schafer, lachlan.urquhart}@ed.ac.uk, {y.lin,ceeai}@hw.ac.uk

**Abstract.** As AI products continue to evolve, increasingly legal problems are emerging for the engineers that design them. For example, if the aim is to build an autonomous vehicle (AV) that adheres to current laws, should we give it the ability to ignore a red traffic light in an emergency, or is this merely an excuse we permit humans to male? The paper argues that some of the changes brought by AVs are best understood as necessitating a revision of law's ontology. Current laws are often ambiguous, inconsistent or undefined when it comes to technologies that make use of AI. Engineers would benefit from decision support tools that provide engineer's with legal advice and guidance on their design decisions. This research aims at exploring a new representation of legal ontology by importing argumentation theory and constructing a trustworthy legal decision system. While the ideas are generally applicable to AI products, our initial focus has been on Autonomous Vehicles (AVs).

**Keywords:** Legal ontology · Autonomous vehicle · Legal detection · Argumentation theory · Explainable AI.

## 1 Introduction

Advances in artificial intelligence (AI) and the automated systems that they enable pose challenges to the legal system that in some jurisdiction have led to far reaching reform proposals. Law reform, especially when it affects the more abstract and foundational legal concepts such as "person" or indeed "law", can create significant uncertainty. Sometimes, developers and manufacturers have to make decisions "in the shadow of"a law-reform debate, anticipating different trajectories that the law may take. Even when a new law has finally been enacted, it often takes time for the courts to clarify contested concets or concrectise general rules. In this paper, we will use autonomous vehicles (AVs) as a case study to show how the challenges to the law can also be translated into a challenge (and opportunity) for legal AI.[4]

---

* Corresponding Author.

[4] The current paper is an extended version of [27]. Work on this paper was supported by Trustworthy Autonomous Systems EP/V026607/1 and AISEC EP/T026952/1As per University of

Concerns about the safety of AVs, and a recognition that widespread lack of trust in them will impede their uptake, have resulted in a plethora of legislative and regulatory activity, including a recent ambitious law reform proposal by the Law Societies of England and Scotland [1]. These "third generation" proposals for technology regulation regularly share two strategies that foreground the concept of design, and leverage its regulatory potential: on the one hand, they ask for engineering solutions that can provably (in the mathematical sense) demonstrate that the product will be law compliant. This extends the concept of "privacy by design" from data protection law to other fields of engineering, and also revives discussions around the legal status of formal software and hardware verification that first rose to prominence in the 1980s, for instance the sage of the Viper microprocessors [12]. Secondly, they create new obligations for developers and manufacturers to document and justify the design choices that they made to achieve this goal. Taken together, this creates significant compliance burdens that require that engineers, not trained in the law, to consider, legal issues and to document and communicate the resulting design choices at every step of the development and design stage.

Our contention is that legal AI, and more specifically a combination of legal ontologies and argumentation systems, can help support with these compliance obligations, by providing intelligent design environments that help the engineer to reason through the legal implications of their design choices, and in a second step helps with creating the type of documentation that the law requires.

Our approach shares the underlying philosophy with the Smarter Privacy project developed at the KIT, which aims at assisting developers of smart grids to comply with data protection law [30]. Their approach models its subject domain using the Sumo ontology, enriched with concepts from data protection law, and combined it with a rule-based reasoner about the relevant legal domain. While sympathetic to this approach, our proposal differs in a crucial jurisprudential assumption: for them the law and its categories are taken as a given and stable, and the reasoner merely subsumes new facts under the old categories. The result is a "Dworkinian one-right-answer"[34], the position that the system takes that of a judge. By contrast, we argue that the legal analysis of new technologies takes place under uncertainty not just of the facts but also the law, whose categories can become unstable in response to external change, contested and open to revision.A more appropriate approach in these cases is in our view therefore to take the position of an inhouse lawyer whose task it is to advice on all possible outcomes (or future interpretations of the law) and to generate arguments that, if challenged, could be used to argue in court for the lawfulness of the design choice that was taken.

Our hypothesis, in a nutshell, is this: the introduction of AVs and other autonomous systems creates fundamental challenges to the legal system that can't any longer be resolved by mere analogous application of existing categories to these new objects. Rather, they potentially "break" the underlying ontology and conceptual divisions of the law, creating systematic inconsistencies and gaps, which are then in need of "ontology repair". Because law, like language, is self-reflective, this process of ontology repair in

---

turn uses legal arguments – in one and the same decision, the judge may e.g. propose an interpretation that subsumes the facts under an existing legal category, while also making an argument that some higher-order legal principle requires to amend, delete or add to the existing categories. This ability of lawyers to reason *about* legal categories in addition to *using* them is particularly visible when more fundamental changes in the external world create problems when old categories are applied to new realities. These exercises in ontology repair and ontology evolution inevitably create legal uncertainty. As we will see in the examples below, this can create an unmet need for engineers (or other members of society) to make legally informed decisions under uncertainty. We hope to show how building on existing approaches to legal ontologies and legal reasoning that make ontology repair explicit can help to address this.

A simple example may help to explain this more abstract notion. The United Kingdom Department of Transport 2015 report *The Pathway to Driverless* Cars stated that - testing of automated technologies is legally possible, provided that *the vehicle can be used compatibly with road traffic law*. In other words, the AV must observe the same rules originally addressed to human drivers. How can a developer of an AV make sense of this requirement? A starting point would be to consult the relevant road traffic rules, and treats the AV as the new norm addressee that "inherits" the legal obligations of the human driver. For some of these, this change is unproblematic and merely reinterprets the old category of "driver" as including "autonomous vehicles". An example would be "the driver has to stop the car at a red traffic light". For other rules, however, this strategy is less convincing. A candidate could be the rule that "the driver must not be drunk", or "if the driver is drunk, they must not operate the vehicle". Here the engineer has a number of possible interpretations available. They can continue to treat the AV as "the driver", and as cars are never drunk, the conditional norm: "If drunk, don't drive" is trivially true all the time, and the car is trivially compliant with this provision. Alternatively, "the driver" in this context may refer to some human inside the car who may have been assigned specific legal duties, for instance the duty to take over from the AV under certain conditions. This means the concept of "driver" has now been subdivided to "heal" the counterintuitive outcome. If this interpretation is taken, a number of follow-up questions need to be answered. In one interpretation, this human "non-driver" is responsible for being sober and faces sanctions when drunk – but this is not a concern of the car developer. However, another interpretation is also possible: here, the duty to ensure that the vehicle is operated lawfully transfers more fully to the AV, which now has to monitor if there is at least one sober passenger available, e.g. [42].

The underlying problem that leads to these three different interpretations is that AVs share some properties with the category "driver" and some properties with the disjoint category "car", creating systematic ambiguities when interpreting laws whose semantics reflect the old ontology. Even more fundamentally, the reason AI regulation is difficult is that they seem to violate some of the most basic ontological distinctions that structure the law, in particular the distinction between persons and objects.

This was a central point made by the joint report of the Law Commissions of England and Scotland. They note that "Existing law reflects a division between rules governing vehicle design on the one hand and the behaviour of drivers on the other." What the Commissions ask for in response is a new conceptual scheme that bridges these two regulatory spheres: the automated driving system is at the same time equipment fitted in a vehicle (and object) but it also determines the behaviour of the vehicle (an agent). If we think of this suggestion as a form of "repair work" to deal with emerging inconsistencies, we can repurpose research in ontology repair to model not only the reasoning of the Commission, but also how that document can in turn be made into a legal argument that informs design decisions. Ideally, at every decision that they have to document, the engineers need a system that (1) Gives feedback about whether a de- sign draft is in compliance with current or possible future laws, depending on which of several competing interpretations is chosen, (2) Answers what happens to the legal analysis if a single functionality is added, deleted or modified in the draft design; (3) Sup-ports reasoning based on how conflicting preferences and values have been resolved; (4) Gives an understandable explanation of the legal results for auditing purposes. Here, we present a legal support system for autonomous vehicles (*LeSAC*), as shown in Figure 1. It is built on top of legal ontology and a legal argumentation framework adapted from *ASPIC*$^+$ [32] based on legal reasoning, which we name *L-ASPIC* [41]. In *LeSAC* we extract necessary elements for legal content and add relevant designs such as legal principle based preference order.

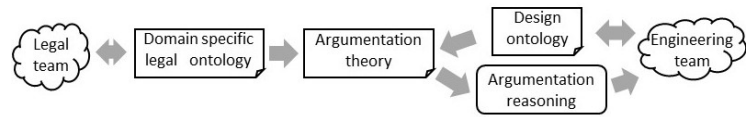

**Fig. 1.** Overview of the process of legal support system

The rest of this paper is organized as follows. Related work is discussed in §2. §3 introduce a running example and its encoding in legal ontology, as well as the background for description logic (DL) and the argumentation framework we built. §4 presents *LeSAC* and its functionality to assist AV design from the legal aspect. §5 concludes this paper.

## 2 Related Work

The most direct interaction between AV cars and the law are road traffic rules,. Because they are more detailed and precise than "top level" legal regimes such as the general law of delict, they are also a good candidate for representation in a logic framework [27]. To meet the requirements like "understandable explanation "discussed above, we choose a classical logic-based format:, legal ontologies, as the foundation, as they have proved themselves as powerful tools for legal KR in applications such as search in legal databases or documents management:

In the literature, many legal ontologies for different functions have been proposed such as the Legal Knowledge Interchange Format (LKIF) Core ontology builds on the Web Ontology Language (OWL) and LKIF rules [25,3]; the Core Legal Ontology (CLO) based on the extension of the DOLCE (DOLCE+) foundational ontology [19]; the LRI-Core ontology aimed at the legal domain grounded in common-sense [8], UOL[24] and the Functional Ontology for Law (FOLaw) [38,39]. These legal ontogies mainly focus on capturing core concepts and framework of the abstract legal theory. And there are also many legal ontology models constructed for specific legal domains like Ukraine legal ontology[22], MCO[31] and JUR-IWN[10]. In a word, legal ontology has proved itself a very strong tool for law as legal expert systems, legal database, documents manager and so on.

However, legal ontology alone is insufficient for legal reasoning. as pointed out by van Engers et al. [40]. Law is based on a dialectical process, This is the consequence of the inevitable introduction of ambiguity )Hart's "open texture" and inconsistencies that are typically resoved through an adversarial debate. However, the main description language of the current legal ontology is the Web Ontology Language (OWL) or OWL2, whose semantics are based on DL [4] and they cannot support inconsistent reasoning as a subset of first-order logic. This is also an important reason that most of the existing legal ontology focuses more on capturing abstract legal concepts, playing the role of document management or legal dictionary. To address this problem, there are works focused on detecting and repairing inconsistent parts [35,18] or extending classical logic by adding true values [43]. However, these works weaken the reasoning strength of DL [43] and require guidance outside the ontology [35,18]. Also, they lack explainability which is a desirable feature when inconsistency happens.

As discussed above, legal ontology needs a method to analyze and reason in an environment full of uncertainties, something not available inside description logic semantics. So a new formal tool supporting handling conflicts and uncertainty while expressing incomplete information in legal ontology is needed, which points to formal argumentation theory.

In [15], formal argumentation has been noticed as an approach to dealing with reasoning under inconsistent and uncertain contexts [16,20,23,29]. Formal argumentation has the merits of computational efficiency and explainability. There have been works combining argumentation theories and ontology for argument mining [37] and quantified reasoning [9] but they didn't solve inconsistent reasoning. For handling reasoning with inconsistent ontologies, several previous studies have considered applying structured argumentation systems in this field, e.g. [23,28], DL ontology is expressed as Defeasible Logic Programs (*DeLP*) [20], while paper [7] present argumentation frameworks based on the Deductive Argumentation [5] framework for dealing with inconsistent DL ontologies. These works support inconsistent reasoning but they cannot describe more complicated interactions or agents' different attitudes. They also do not connect the explanation of reasoning results with the underlying design choices of the legal semantics. However, in legal applications, explainability is even more important than in other AI tasks. The reasoning and process of how a legal result was obtained is often as important as getting the result right. Formal argumentation is capable of handling this problem. Paper [13] performed a comprehensive literature survey among explainable AI from an argumentation perspective. Research [6] constructed a flexible framework to provide explanations about why a claim is finally accepted or rejected under various extension-based semantics in argumentation frameworks. Therefore, this paper will explore how to give a formal explanation based on argumentation theory within this legal support system.

## 3  Legal ontology and argumentation & a case study

We start by introducing a scenario as a case study. Encoding of this case study is available in [2].

Consider the following scenario that engineers may face if their task is to design AV that behaves in a law-compliant way:

*Example 1.* Currently, the law stipulates a number of behaviours that a human driver has to observe after an accident has happened. This includes a duty to stay at the scene of an accident and to provide first aid if necessary and feasible. The design question now is, does a driverless car "inherit" this obligation in the same way it "inherits" from the human driver the duty to stop at a traffic light? A recent proposal by the Scottish and English Law Commissions differentiates between functions that are core to the safe operation of an AV and those that are auxiliary. As a complication, let's assume there is one passenger in the car, but he is (illegally) too drunk to do anything. In such a "contrary to duty" scenario, how should the AV car react now when somebody is hit? Should it just report to the police and keep doing its original job: Sending the passenger to destination as soon as possible? If the injured party is likely to die if not receiving medical aid in time, is it a legal requirement that the AV stops the only passenger from leaving and asks him to take the responsibility as a driver to give some help? Especially considering the passenger is drunk, what if he does second harm to the injury or put himself in danger? Here road traffic law interacts with other, more general legal provisions about duty of care.

To handle this possible situation, we refer to current and relevant legal rules. We extract and select some most relevant information from traffic law and criminal law:

(1) It is illegal to drive a motor vehicle while intoxicated. People who drive while intoxicated shall lose their driving license and may be prosecuted in criminal law.

(2) A person who commits a hit-and-run accident will be criminal responsibility, especially when the escape causes the death or the driver is intoxicated.

(3) When an accident happens, the driver should take the responsibility to transfer the injured party to a safe place and provide aid if the situation is urgent.

Obviously, current rules will cause uncertainty and inconsistency if directly applied in this situation if the vehicle is an AV, such as how to define the concept 'driver' and how will the obligation be allocated. If engineers base their designs on a given possible development, potential conflicts will also arise. For example, if we assume any future law still counts the only human in the car as driver who should take responsibility in an accident, the AI car should ask the drunk passenger to get out of the car and try to help the injury. However, what will the law say about the passenger's intoxication? On one hand, a second harm is possible to occur from a drunk person, which will lead to new legal responsibility. On the other hand, what about the passenger's safety?

If we assume it should be the AI car's job to make sure no more risk will occur to the drunk passenger, which is highly possible. It means encouraging the passenger to offer help could be against the law. To reflect how will *LeSAC* handle this problem, we import one more legal rule into this example:

(4) It's illegal to let a drunk passenger leave the car alone during the trip.

As mentioned above, DL is the basic semantic of OWL or OWL2, which are the main logic languages of current legal ontologies. DLs are a family of knowledge representation formalisms. The basic notions of DL systems are *concepts* and *roles*. A DL system contains two disjoint parts: the TBox and the ABox. TBox introduces the terminology, while ABox contains facts about individuals in the application domain. There are many DLs and the legal ontology in this paper is built upon the *ALC* expression [36,4].

In a legal ontology, legally reasoning rules such as traffic rules will be allocated into Tbox, while explicit legal designs, e.g. AVs will stop or not, will be in ABox. As for the situation in **Example** 1, a legal ontology's TBox will be:

*Example 2 (DL encoding of Example 1, TBox).*

$$T = \begin{cases} Driver \sqsubseteq Sober;\ Sober \sqcap Intoxicated \sqsubseteq \emptyset;\ Intoxicated \sqcap LeaveCar \sqsubseteq \emptyset; \\ Driver \sqcap Intoxicated \sqsubseteq BeRevokedDrivingLicense \sqcap TakeCriminalResposibility; \\ \exists hitAndRun.Injury \sqsubseteq TakeCriminalResposibility; \\ \exists hitAndRun.causeDeath \sqsubseteq AggravatedPunishment; \\ \exists hitAndRun.Injury \sqcap Driver \sqcap Intoxicated \sqsubseteq AggravatingPunishment; \\ CauseAccident \sqcap Injury \sqsubseteq \exists transferToSafePlace.Injury; \\ CauseAccident \sqcap NeedEmergencyAid.Injury \sqsubseteq doNecessaryAid; \\ (transferToSafePlace \sqcup doNecessaryAid) \sqcap \neg LeaveCar \sqsubseteq \emptyset \end{cases}$$

Assuming an AV design named "AC1", we consider the driver should take the responsibility as the current legal concept "driver". When an AV hits somebody named "Injury1" on the road, it will ask the only passenger named "PS1" to leave the car and help the injured party, no matter if he is drunk or not. The corresponding ABox is:

**Example** (Example 2 cont. DL encoding of ABox).

$$A = \begin{cases} Driver(PS1);\ Intoxicated(PS1);\ hitAndRun(PS1, Injury1);\ Injury(Injury1); \\ causeDeath(PS1, Injury1);\ CauseAccident(PS1);\ NeedEmergencyAid(Injury1) \end{cases}$$

### 3.1 Argumentation Framework for Legal Reasoning: *L-ASPIC*

*L-ASPIC* is an argumentation system for legal reasoning based on *ASPIC*[+] framework [32]. It associates with a set of legal principles for argument preference extraction. Due to space limitations, this section focuses on the settings specially adapted to the application of legal ontology. A detailed account of *L-ASPIC* can be found in [41].

**Definition 1 (Argumentation system for legal reasoning).** *An L-ASPIC argumentation system (L-AS) is a tuple $(\mathcal{L}, \mathcal{R}, n, \mathcal{P}, prin)$, where*

- $\mathcal{L}$ *is a set of formal language closed under negation (¬), where $\psi = -\varphi$ means $\psi = \neg \varphi$ or $\varphi = \neg \psi$;*
- $\mathcal{R} = \mathcal{R}_s \cup \mathcal{N}$ *is a set of strict inference rules ($\mathcal{R}_s$) of the form $\varphi_1, \ldots, \varphi_n \to \varphi$, and legal norms ($\mathcal{N}$) based on defeasible inference rules, of the form $\varphi_1, \ldots, \varphi_n \Rightarrow \varphi$ ($\varphi_i, \varphi \in \mathcal{L}$); $\mathcal{R}_s \cap \mathcal{N} = \emptyset$; n is a naming function such that $n : \mathcal{N} \to \mathcal{L}$.*

- $\mathscr{P}$ *is a set of legal principles, and prin is a total function from elements of* $\mathscr{N} \to \mathscr{P}$.

$ASPIC^+$ framework [32] defines the set of rules ($\mathscr{R}$) as consisting of two disjoint sets, i.e. a set of strict rules and a set defeasible rules. In the current paper we intend to use the argumentation theory as a tool only for dealing with normative reasoning in legal contexts, so defeasible rules for epistemic reasoning are not considered.

The argumentation system of *L-ASPIC* defines defeasible inferences as legal norms ($\mathscr{N}$), and we assume that each legal norm is associated with a legal principle in $\mathscr{P}$ (whereas one legal principle may be associated with multiple norms), which is the primary legal principle on which the norm is based. In other words, for any legal norm, $N \in \mathscr{N}$, $prin(N) \in \mathscr{P}$ is the basic legal principle upon which $N$ is constructed, whereby conflicts between arguments may be resolved based on priorities on $\mathscr{P}$. The reason we import the set $\mathscr{P}$ is twofold: 1) giving more legal semantics for reasoning results, making them more explainable; 2) supplying methods to solve complicated situations when different conflicting legal suggestions occur. Since all norms in this paper are defeasible, a very strong (but not as strict as some self-evident facts) norm can be given higher priority based on the legal principle associated with it.

Given an *L-AS*, we can construct arguments by rules starting from a set of premises (knowledge base), denoted as $\mathscr{K}$. Let $\Delta = (T,A)$ be a legal ontology for AV based on description logic, $(L\text{-}AS, \mathscr{K}^A)$ is an argumentation theory about $\Delta$ (denoted as *L-AT*$^\Delta$), where *L-AS* $= (\mathscr{L}, \mathscr{R}^T, n, \mathscr{P}, prin)$, such that $\mathscr{R}^T$ is the set of rules corresponding to $T$ (a mapping table can be found in [26,41]), and $\mathscr{K}^A$ is the set of premises based on $A$.

We denote all the formulas in $\mathscr{K}$ that are used to build an argument as Prem, all its sub-arguments as Sub, all the applied rules as Rules, and the consequent of the last rule as Conc. Formally, arguments constructed based on *L-AT*$^\Delta$ are defined as follows.

**Definition 2 (Argument).** *An argument* $\alpha$ *constructed based on L-AT*$^\Delta$ *has one of the following forms:*

1. $\varphi$, if $\varphi \in \mathscr{K}^A$, s.t. $\text{Prem}(\alpha) = \{\varphi\}$, $\text{Conc}(\alpha) = \varphi$, $\text{Sub}(\alpha) = \{\varphi\}$, and $\text{Rules}(\alpha) = \emptyset$;
2. $\alpha_1, \ldots, \alpha_n \to / \Rightarrow \psi$ if $\alpha_1, \ldots, \alpha_n$ are arguments, s.t. there exists a rule $\text{Conc}(\alpha_1), \ldots, \text{Conc}(\alpha_n) \to / \Rightarrow \psi$ in $\mathscr{R}^T$, and $\text{Prem}(\alpha) = \text{Prem}(\alpha_1) \cup \ldots \cup \text{Prem}(\alpha_n)$, $\text{Conc}(\alpha) = \psi$, $\text{Sub}(\alpha) = \text{Sub}(\alpha_1) \cup \ldots \cup \text{Sub}(\alpha_n) \cup \{\alpha\}$, $\text{Rules}(\alpha) = \text{Rules}(\alpha_1) \cup \ldots \cup \text{Rules}(\alpha_n) \cup \{\text{Conc}(\alpha_1), \ldots, \text{Conc}(\alpha_n) \to / \Rightarrow \psi\}$.

In addition, for any argument $\alpha$ constructed based on an *L-AT*$^\Delta$, let $\text{LastNorms}(\alpha) = \emptyset$ if $\text{Rules}(\alpha) \cap \mathscr{N} = \emptyset$, or $\text{LastNorms}(\alpha) = \{\text{Conc}(\alpha_1), \ldots, \text{Conc}(\alpha_n) \Rightarrow \psi\}$ if $\alpha = \alpha_1, \ldots, \alpha_n \Rightarrow \psi$, otherwise $\text{LastNorms}(\alpha) = \text{LastNorms}(\alpha_1) \cup \ldots \cup \text{LastNorms}(\alpha_n)$. And we denote the set $\{prin(N) | N \in \text{LastNorms}(\alpha)\}$ as $\text{LastPrin}(\alpha)$.

Given an *L-AT*$^\Delta$, the inconsistency of information can be reflected by conflicts (attacks) among arguments. An argument can be attacked on its uncertain parts or the consequences of these parts. The current paper assumes that all the elements in $\mathscr{K}$ are uncertain, and all the norms must be applicable, then the attack relation is defined as follows.

**Definition 3 (Attacks).** *Let $\alpha$, $\beta$, $\beta'$ be arguments constructed based on an $L\text{-}AT^\Delta = (L\text{-}AS, \mathcal{K}^A)$, $\alpha$ attacks $\beta$ on $\beta'$, iff: 1) $\beta' \in \text{Sub}(\beta)$ of the form $\beta_1'', \ldots, \beta_n'' \Rightarrow \varphi$ and $\text{Conc}(\alpha) = -\varphi$; or 2) $\beta' = \varphi$ and $\varphi \in \text{Prem}(\beta) \cap \mathcal{K}$, s.t. $\text{Conc}(\alpha) = -\varphi$.*

Since *L-ASPIC* is specifically designed to detect and handle conflicts arising from legal norms, and the context has been limited to the design of AVs, elements in $\mathcal{K}^A$ can be considered as assumptions (e.g. "$Driver(PS1)$", "$Intoxicated(PS1)$"), which should have similar priorities, therefore the premises of arguments may be conflicting/attackable.

When two arguments conflict as shown in Def. 2, whether one can defeat another is determined by some pre-defined preferences (these could e.g. be highler-order legal values such as "repect for human life").

Let $\mathscr{A}$ denote all the arguments constructed based on an $L\text{-}AT^\Delta$, $\leqslant$ denote a pre-ordering on $\mathscr{P}$ and $\trianglelefteq_{Dem}$ denote a set comparison based on the *Democratic* approach [11]. The preference ordering $\preceq$ on $\mathscr{A}$ is defined as follows.

**Definition 4 (Argument ordering).** *Let $(L\text{-}AS, \mathcal{K}^A)$ be an $L\text{-}AT^\Delta$, for all $\alpha$, $\beta$ constructed based on it, $\beta \preceq \alpha$ iff $\text{LastPrin}(\beta) \trianglelefteq_{Dem} \text{LastPrin}(\alpha)$, i.e.: 1) $\text{LastPrin}(\alpha) = \emptyset$ and $\text{LastPrin}(\beta) \neq \emptyset$; or 2) $\forall P \in \text{LastPrin}(\beta), \exists P' \in \text{LastPrin}(\alpha)$ s.t. $P \leqslant P'$.*

We write $\beta \prec \alpha$ iff $\beta \preceq \alpha$ and $\alpha \not\preceq \beta$. Based on **Def.** 4, arguments that do not apply norms always stronger than arguments that do apply norms. The reason is that in the setting of *L-ASPIC*, if an argument contains only elements from $\mathcal{K}$ and $\mathcal{R}_s$, it represents the currently acceptable epistemic knowledge (beliefs). Since beliefs are generally considered to take precedence over normative knowledge, it should be plausible that it cannot be defeated by an argument that contains norms.

For the choice of comparative principles, preferences on the set of arguments are extracted according to the *Democratic* approach for set comparison [11] and the *last-link principle* [32] for the elements selection. The legal basis of the Democratic approach is that law in real life will protect preferred rights and benefits when it has to choose. For example, when two groups of legal rules are incompatible, the one containing the principle to protect human lives defeats the one aiming at protecting money (in many, but not all contexts). Meanwhile, the last-link principle is used primarily for legal applications and is more suitable for normative reasoning [33].

In order to get an output of acceptable conclusions, first, we need to identify the justified arguments, which can be achieved by an argument evaluation process based on abstract argumentation frameworks (*AF*) and argumentation semantics [15]. Given an $L\text{-}AT^\Delta$, an $AF = (\mathscr{A}, \mathscr{D})$ can be established based on the set of all the arguments ($\mathscr{A}$) and the defeat relation ($\mathscr{D}$) on the basis of the attack relation between arguments and the ordering $\preceq$ on $\mathscr{A}$. Let *S* denote one of the basic argumentation semantics introduced in [15], $\mathscr{E}_S$ denote the set of all extensions obtained under *S*, and $E_S \in \mathscr{E}_S$ denote one of the extensions. An argument $\alpha \in \mathscr{A}$ is said to be *accepted* w.r.t. $E_S$ if $\alpha \in E_S$. In the following we say $\alpha$ is *sceptically justified* under *S* if $\forall E_S \in \mathscr{E}_S$, $\alpha \in E_S$, and $\alpha$ is *credulously justified* under *S* if $\exists E_S \in \mathscr{E}_S$ such that $\alpha \in E_S$. Then according to the accepted/justified arguments, we can identify the accepted conclusions.

## 4 Legal support system for autonomous vehicles

In §3.1 we explained a structured argumentation framework *L-ASPIC* for reasoning based on an inconsistent legal ontology. Then given a legal ontology, particularly for AVs design, we can construct a *LeSAC* system based on *L-ASPIC* :

*Example 3 (A LeSAC ).* Given a legal ontology for AV $\Delta = (T,A)$, as shown in Example 2. $LeSAC = (\mathcal{L}, \mathcal{K}^A, \mathcal{R}^T, n, \mathcal{P}, prin)$ is an argumentation theory instantiated by $\Delta$:

$$\mathcal{N} = \begin{cases} r_1 : Driver(x) \Rightarrow Sober(x); \\ r_2 : Intoxicated(x) \Rightarrow \neg LeaveCar(x); \\ r_3 : Driver(x), Intoxicated(x) \Rightarrow BeRevokedDrivingLicense(x); \\ r_4 : Driver(x), Intoxicated(x) \Rightarrow TakeCriminalResposibility(x); \\ r_5 : hitAndRun(x,y) \Rightarrow TakeCriminalResposibility(x); \\ r_6 : hitAndRun(x,y), causeDeath(x,y) \Rightarrow AggravatedPunishment(x); \\ r_7 : hitAndRun(x,y), Driver(x), Intoxicated(x) \Rightarrow AggravatedPunishment(x); \\ r_8 : CauseAccident(x), Injury(y) \Rightarrow transferToSafePlace(x,y); \\ r_9 : CauseAccident(x), Injury(y), NeedEmergencyAid(y) \Rightarrow doNecessaryAid(x,y) \end{cases}$$

$$\mathcal{R}_s = \begin{cases} r_{10} : Sober(x) \to \neg Intoxicated(x); \\ r'_{10} : Intoxicated(x) \to \neg Sober(x); \\ r_{11} : transferToSafePlace(x,y) \to LeaveCar(x); \\ r'_{11} : \neg LeaveCar(x) \to \neg transferToSafePlace(x,y); \\ r_{12} : doNecessaryAid(x,y) \to LeaveCar(x); \\ r'_{12} : \neg LeaveCar(x) \to \neg doNecessaryAid(x,y) \end{cases}$$

$$\mathcal{K}^A = \begin{cases} Driver(PS1); Intoxicated(PS1); \\ hitAndRun(PS1, Injury1); \\ Injury(Injury1); \\ causeDeath(PS1, Injury1); \\ CauseAccident(PS1); \\ NeedEmergencyAid(Injury1) \end{cases}$$

$$\mathcal{P} = \begin{cases} p_1 : Human\ lives\ should\ be\ protected\ as\ a\ priority; \\ p_2 : AI\ products\ should\ avoid\ extra\ risk\ about\ safety\ for\ their\ users; \\ p_3 : People\ should\ avoid\ putting\ others\ into\ dangerous\ by\ his\ own\ behaviours, \\ \quad\ and\ should\ bear\ corresponding\ responsibility. \end{cases}$$

$prin(r_1) = p_3$; $prin(r_2) = p_2$; $prin(r_3) = p_3$; $prin(r_4) = p_3$; $prin(r_5) = p_3$;
$prin(r_6) = p_3$; $prin(r_7) = p_3$; $prin(r_8) = p_1$; $prin(r_9) = p_1$

Properties for a well-defined argumentation theory based on $ASPIC^+$ are specified in [32]. Likewise, a well defined *LeSAC* should also meet some requirements, such as $\mathcal{R}_s$ should be closed under transposition or contraposition. closure under transposition (or contraposition). ie.: if $\varphi_1,\ldots,\varphi_n \to \psi \in \mathcal{R}_s$, then for each $i = 1\ldots n$, there is $\varphi_1,\ldots,\varphi_{i-1}, -\psi, \varphi_{i+1},\ldots \varphi_n \to -\varphi_i \in \mathcal{R}_s$. In **Example** 3, rules $r'_{10}$, $r'_{11}$ and $r'_{12}$ are the transposed rules of rule $r_{10}$, $r_{11}$ and $r_{12}$, respectively.

We now present *LeSAC* its reasoning functions using the case study. To clarify these support functions more visually, we use Example 3 to show how AV designers could solve their different problems in this situation through *LeSAC* .

**Legal compliance detection** When engineers complete a whole design draft, they could use the consistency checking function to check whether this design is fully compliant with given laws and where conflicts are.

**Definition 5 (Consistency Checking).**
*The ABox of $\Delta$ is consistent w.r.t. the TBox of $\Delta$ iff $\mathscr{A}$ is conflict-free based on attack relations, i.e., $\nexists \alpha, \beta \in \mathscr{A}$ such that $\alpha$ attacks $\beta$.*

If a design is completely consistent after reasoning, it means it is fully compliant with given laws. Otherwise, it is not. And by tracing where arguments conflict, we could know which part of the design needs modification. Based on the *LeSAC* in Example 3, we can at least construct the following two arguments.

**Example** (Example 3 cont.).
$\alpha = (CauseAccident(PS1), Injury(Injury1) \Rightarrow transferToSafePlace(PS1, Injury1)) \rightarrow LeaveCar(PS1)$
*and $\beta = Intoxicated(PS1) \Rightarrow \neg LeaveCar(PS1)$. According to Definition 3, $\alpha$ and $\beta$ attack each other, therefore the legal ontology on which this LeSAC is based is inconsistent.*

**Feedback for single change** If AV engineers want to keep the main design of an AV and only do some minimal changes, *LeSAC* can provide possible further legal consequences with these new details by instance checking. According to *LeSAC*, assertions are the conclusions of arguments. So based on the extension of arguments, we can decide whether an assertion is accepted. The definition of acceptance of assertions is:

**Definition 6 (Assertion Acceptance).** *An assertion X is sceptically/credulously accepted under certain argumentation semantics S, iff $\exists A \in \mathscr{A}$, s.t. A is sceptically/credulously justified w.r.t. $\mathscr{E}_S$ and $\text{Conc}(A) = X$.*

To determine whether a certain modification is consistent with the current design and given laws, we translate this problem into whether a legal assertion about this AV can be accepted as a conclusion of an accepted/justified argument. Consider arguments $\alpha$ and $\beta$ in Example 4, we have $\text{LastNorms}(\alpha) = \{r_8\}$, $\text{LastNorms}(\beta) = \{r_2\}$, and $\text{LastPrin}(\alpha) = p_1$, $\text{LastPrin}(\beta) = p_2$ respectively. Assume that based on $\leqslant$ on $\mathscr{P}$, $p_2 < p_1$, then according to Definition 4, $\beta \prec \alpha$. Therefore, $\alpha$ can defeat $\beta$, but not vice versa. Based on the *LeSAC* in Example 3, there are no other arguments to attack or defeat $\alpha$. As a consequence, $\alpha$ is sceptically justified w.r.t. any $\mathscr{E}_S$, and the assertion "*LeaveCar(PS1)*" is sceptically accepted. The following definition defines instance checking based on a *LeSAC* for all the possible forms of classes.

**Definition 7 (Instances Checking).** *Let $\varphi$ be an individual, sceptically or credulously, it holds that $\varphi$ is an instance of a class:*

- *C (/¬C), iff $\exists \alpha \in \mathscr{A}$, s.t. $\alpha$ is sceptically justified w.r.t. $\mathscr{E}_S$ and $\text{Conc}(A) = C(\varphi)(/\neg C(\varphi))$;*
- *$C \sqcap D$, iff $\exists \alpha, \beta \in \mathscr{A}$ such that $\alpha$ and $\beta$ are both sceptically/credulously justified w.r.t. $\mathscr{E}_S$ and $\text{Conc}(A) = C(\varphi)$, $\text{Conc}(B) = D(\varphi)$;*
- *$C \sqcup D$, iff $\exists \alpha, \beta \in \mathscr{A}$ s.t. at least one of $\alpha$ and $\beta$ are sceptically/credulously justified w.r.t. $\mathscr{E}_S$ and $\text{Conc}(\alpha) = C(\varphi)$, $\text{Conc}(\beta) = D(\varphi)$;*
- *$\exists P.D$, iff $\exists \alpha, \beta \in \mathscr{A}$ such that $\alpha$ and $\beta$ are both sceptically/credulously justified w.r.t. $\mathscr{E}_S$ and $\text{Conc}(\alpha) = P(\varphi, x)$, $\text{Conc}(\beta) = D(x)$ (x is an individual);*
- *$\forall P.D$, iff $\exists \alpha \in \mathscr{A}$ s.t. $\text{Conc}(\alpha) = P(\varphi, x)$; and $\forall \alpha \in \{\alpha | \text{Conc}(\alpha) = P(\varphi, x)\}$, $\exists \beta \in \alpha$, s.t. $\text{Conc}(\beta) = D(x)$.*

**Giving legal explanations** How to best use argumentation theory to generate understandable explanations has become an increasingly important topic in AI regulation and AI design. There have been works [14,17] discussing which standards should an argumentation-based explanation meet and some formal explanations [6,17,21] for different argumentation frameworks are given. For legal systems, an agent needs explanation for a certain reasoning result as an assertion rather than an acceptable set with understandable legal information. Considering AV engineers' needs, the explanation of reasoning results from *LeSAC* should show how a legal conclusion is obtained and which content in this situation makes it accepted or not. For any agent $y$, let $\leqslant_y$ denote $y$'s priority orderings over set $\mathscr{P}$, we propose a formal definition of explanation:

**Definition 8 (Explanation).** *Let X be an assertion in a LeSAC that is sceptically accepted under certain argumentation semantics S by a rational agent y, then $\exists \alpha \in \mathscr{A}$ s.t. $\text{Conc}(\alpha) = X$. The explanation for y to accept X is $Exp_y = \mathscr{C}(\alpha) \cup \mathscr{C}(\beta) \cup \{\leqslant_y\}$, where:*

- $\mathscr{C}(\alpha) = \text{Prem}(\alpha) \cup \text{Rules}(\alpha)$, *which explains how X is reached;*
- $\mathscr{C}(\beta) = \text{Prem}(\beta) \cup \text{Rules}(\beta)$ *such that $\beta$ defends $\alpha$ according to $\leqslant_y$ and the defeat relation $\mathscr{D}$, which explains why X is justified.*

**Def.** 8 provides a formal explanation of why a legal conclusion $X$ is accepted for certain design requirement. It consists of two parts. The first part explains how $X$ is reached by presenting all the premises contained in $\mathscr{K}^A$ and all the legal rules contained in $\mathscr{R}^T$ that are applied to construct argument $\alpha$. The second part explains why this legal conclusion is accepted by presenting all the legal information and relevant legal principles applied to construct the arguments that defend $\alpha$. Consider our running examples, for the acceptance of the assertion "*LeaveCar(PS1)*",

$Exp_y = (\{Injury(Injury1), CauseAccident(PS1), NeedEmergencyAid(Injury1)\} \cup \{r_8, r_9\}) \cup \{p_2 < p_1\}$

and for the acceptance assertion "¬*LeaveCar(PS1)*", it is:

$Exp'_y = (\{Intoxicated(PS1)\} \cup \{r_2\}) \cup (\{Intoxicated(PS1)\} \cup \{r'_{10}\}) \cup \{p_1 < p_2\}$

## 5 Conclusion and future work

This paper constructed a legal support system able to help engineers of AVs improve their designs' legal compliance by importing argumentation theory into legal ontology. We discussed the emerging legal problems brought by AI products and the new responsibility engineers are facing from the perspective of both legal theory and social reality. The conclusion was that a new kind of legal technology is useful to help fix the gap between AI products and the legal world. We extended ASPIC+ to a new argumentation framework particularly for legal reasoning called *L-ASPIC*. Based on *L-ASPIC* we constructed the legal support system *LeSAC*. We showed how this system performs different relevant reasoning functions with both formalized definitions and an example. What's more, we also encoded the test case to show the feasibility of this system. In future work, we will explore deeper the legal discussion and try to include some important ethical issues that delineate the duties of the designer and that of the legislator.

We will also improve the legal representation and import machine learning methods such as representation learning to match the empirical information in real cases. The database for further research and experiments is also under construction. Another plan is to integrate it into a conventional engineering workflow.

## References

1. https://www.lawcom.gov.uk/project/automated-vehicles/
2. LN2FR paper resource webpage. https://colab.research.google.com/drive/1BibJ5mVdoL7AvBRqmBQzL2XISaj6WICa?usp=sharing, accessed: 2022-09-30
3. Alexander, B.: LKIF core: Principled ontology development for the legal domain. Law, ontologies and the semantic web: channelling the legal information flood **188**, 21 (2009)
4. Baader, F., Calvanese, D., McGuinness, D., Patel-Schneider, P., Nardi, D., et al.: The description logic handbook: Theory, implementation and applications (2003)
5. Besnard, P., Hunter, A.: Constructing argument graphs with deductive arguments: a tutorial. Argument & Computation **5**(1), 5–30 (2014)
6. Borg, A., Bex, F.: A basic framework for explanations in argumentation. IEEE Intelligent Systems **36**(2), 25–35 (2021)
7. Bouzeghoub, A., Jabbour, S., Ma, Y., Raddaoui, B.: Handling conflicts in uncertain ontologies using deductive argumentation. In: WI'17. pp. 65–72 (2017)
8. Breuker, J., Valente, A., Winkels, R.: Use and reuse of legal ontologies in knowledge engineering and information management. In: Law and the Semantic Web, pp. 36–64 (2005)
9. Budán, M.C., Simari, G.I., Viglizzo, I., Simari, G.R.: An approach to characterize graded entailment of arguments through a label-based framework. International Journal of Approximate Reasoning pp. 242–269 (2017)
10. Casellas, N., Blázquez, M., Kiryakov, A., Casanovas, P., Poblet, M., Benjamins, R.: Opjk into proton: Legal domain ontology integration into an upper-level ontology. In: OTM 2005. pp. 846–855
11. Cayrol, C., Royer, V., Saurel, C.: Management of preferences in assumption-based reasoning. In: IPMU '92. pp. 13–22 (1992)
12. Collins, H.: What the tortoise said to achilles donald mackenzie, mechanizing proof: Computing, risk, and trust. inside technology. cambridge, ma and london: Mit press, 2001. pp. xi+427. isbn 0-262-13393-8.£ 30.95 (hardback). The British Journal for the History of Science **35**(4), 469–474 (2002)
13. Čyras, K., Rago, A., Albini, E., Baroni, P., Toni, F.: Argumentative xai: A survey. In: Proceedings of IJCAI-21. pp. 4392–4399 (2021)
14. Dauphin, J., Cramer, M.: Aspic-end: Structured argumentation with explanations and natural deduction. TAFA 2017
15. Dung, P.M.: On the acceptability of arguments and its fundamental role in nonmonotonic reasoning, logic programming and n-person games. Artificial Intelligence **77**(2), 321 – 357 (1995)
16. Dung, P.M., Kowalski, R.A., Toni, F.: Assumption-based argumentation. In: Simari, G., Rahwan, I. (eds.) Argumentation in Artificial Intelligence. pp. 100–218. Springer US (2009)
17. Fan, X., Toni, F.: On computing explanations in argumentation. In: Twenty-Ninth AAAI Conference on Artificial Intelligence (2015)
18. Fang, J., Huang, Z.: Reasoning with inconsistent ontologies. Tsinghua Science & Technology **15**(6), 687–691 (2010)
19. Gangemi, A., Sagri, M.T., Tiscornia, D.: A constructive framework for legal ontologies. In: Law and the semantic web, pp. 97–124. Springer (2005)

20. García, A.J., Simari, G.R.: Defeasible logic programming: An argumentative approach. Theory and Practice of Logic Programming **4**(2), 95–138 (2004)
21. García, A.J., Simari, G.R.: Defeasible logic programming: Delp-servers, contextual queries, and explanations for answers. Argument & Computation **5**, 63–88 (2014)
22. Getman, A.P., Karasiuk, V.V.: A crowdsourcing approach to building a legal ontology from text. Artificial intelligence and law **22**(3), 313–335 (2014)
23. Gómez, S.A., Chesñevar, C.I., Simari, G.R.: Reasoning with inconsistent ontologies through argumentation. Applied Artificial Intelligence **24**(1&2), 102–148 (2010)
24. Griffo, C., Almeida, J.P.A., Guizzardi, G.: Towards a legal core ontology based on alexy's theory of fundamental rights. In: Multilingual Workshop on ICAIL (2015)
25. Hoekstra, R., Breuker, J., Di Bello, M., Boer, A., et al.: The lkif core ontology of basic legal concepts. LOAIT **321**, 43–63 (2007)
26. Lu, Y., Yu, Z.: Argumentation theory for reasoning with inconsistent ontologies. In: Borgwardt, S., Meyer, T. (eds.) DL 2020. vol. 2663 (2020)
27. Lu, Y., Yu, Z., Lin, Y., Schafer, B., Ireland, A., Urquhart, L.: An argumentation and ontology based legal support system for ai vehicle design. In: Proceedings of the 35th International Conference on Legal Knowledge and Information Systems (JURIX 2022) (2022)
28. Martinez, M.V., Deagustini, C.A.D., Falappa, M.A., Simari, G.R.: Inconsistency-tolerant reasoning in datalog$^{\pm}$ ontologies via an argumentative semantics. In: IBERAMIA 2014. pp. 15–27 (2014)
29. Modgil, S., Prakken, H.: A general account of argumentation with preferences. Artificial Intelligence **195**, 361–397 (2013)
30. Oberle, D.: Ontologies and reasoning in enterprise service ecosystems. Informatik Spektrum **37**(4), 318–328 (2014)
31. Poblet, M., Casellas, N., Torralba, S., Casanovas, P.: Modeling expert knowledge in the mediation domain: a middle-out approach to design odr ontologies. In: LOAIT 2009. No. 3è
32. Prakken, H.: An abstract framework for argumentation with structured arguments. Argument & Computation **1**(2), 93–124 (2010)
33. Prakken, H., Sartor, G.: Argument-based extended logic programming with defeasible priorities. Journal of Applied Non-Classical Logics **7**(1-2), 25–75 (1997)
34. Rosenfeld, M.: Dworkin and the one law principle: A pluralist critique. Revue Internationale de Philosophie **59**(233 (3)), 363–392 (2005)
35. Schlobach, S., Cornet, R., et al.: Non-standard reasoning services for the debugging of description logic terminologies. In: Ijcai. vol. 3, pp. 355–362 (2003)
36. Schmidt-Schauß, M., Smolka, G.: Attributive concept descriptions with complements. Artificial Intelligence **48**(1), 1–26 (1991)
37. Tempich, C., Simperl, E., Luczak, M., Studer, R., Pinto, H.S.: Argumentation-based ontology engineering. IEEE Intelligent Systems **22**(6), 52–59 (2007)
38. Valente, A.: Legal Knowledge Engineering: A Modelling Approach. IOS Press (1995)
39. Valente, A., Breuker, J., Brouwer, B.: Legal modeling and automated reasoning with on-line. International Journal of Human-Computer Studies **51**, 1079–1125 (1999)
40. Van Engers, T., Boer, A., Breuker, J., Valente, A., Winkels, R.: Ontologies in the legal domain. In: Digital Government: E-Government Research, Case Studies, and Implementation. pp. 233–261 (2008)
41. Yu, Z., Lu, Y.: An argumentation-based legal reasoning approach for DL-ontology. arXiv preprint arXiv:2209.03070 (2022), https://arxiv.org/abs/2209.03070
42. Zaouk, A.K., Wills, M., Traube, E., Strassburger, R.: Driver alcohol detection system for safety (dadss)–a status update. In: 24th ESV. Gothenburg, Sweden (2015)
43. Zhang, X., Lin, Z., Wang, K.: Towards a paradoxical description logic for the semantic web. In: FoIKS. pp. 306–325. Springer (2010)

# Semantic Model for the Legal Maintenance: the Case of Semantic Annotation of France Legislative and Regulatory Texts

Julien Breton[1,2], Mokhtar Boumedyen Billami[2][0000−0003−4428−4298], Max Chevalier[1][0000−0001−5402−6255], and Cassia Trojahn[1][0000−0003−2840−005X]

[1] IRIT, Toulouse, France
firstname.lastname@irit.fr
[2] Berger-Levrault, Labège, France
firstname.lastname@berger-levrault.com

**Abstract.** In different domains, compliance with legal documents about industrial maintenance is crucial. Legal industrial maintenance is the legal commitment of a company to control, maintain and repair its equipments. With the evolution of legal texts, companies are increasingly adopting automatic processing of legal texts in order to extract their key elements and to support the task of analysis and compliance. To perform such a task of knowledge extraction, a number of state-of-the-art proposal relies on a semantic model. Based on existing models from both legislative and industrial maintenance domains, we propose a new semantic model for the legal industrial maintenance: SEMLEG (SEmantic Model for the LEGal maintenance). This model results from an analysis of documents extracted from the Légifrance French governmental website.

**Keywords:** Semantic model · legal maintenance · industrial maintenance.

## 1 Introduction

In December 2021, the CNIL (French National Commission for Computing and Liberties)[3] has applied a record penalty against Google of 150 million euros for non-compliance with the law. This example shows that companies are obliged to respect the law and risks, otherwise, severe sanctions can be applied. In order to carry out a legal monitoring, companies usually rely on human expertise to manually analyse legal documents. As highlighted in [17], in France, there are "more than 10,500 laws, 120,000 decrees, 7,400 treaties, 17,000 community's texts, tens of thousands of pages in 62 different codes. Some are constantly being modified: 6 modifications per working day for the 2006 Tax Code". Therefore, a first problem is the quantity of legislative documents and their constant updates. Secondly, the domain-specific vocabulary can make it difficult to understand the legislation. Third, the abundance of cross-references in legal documents makes reading them tedious.

---

[3] https://www.cnil.fr/

In order to address these issues, companies and the scientific community consider the automated processing of legislative documents. With the advantage of helping to process massive information, the processing of such documents aims to extract and to represent legislative rules. In this paper, we focus on structuring the legal maintenance information through a semantic model: SEMLEG (SEmantic Model for the LEGal maintenance). These extraction and structuring objectives have concrete applications. As an example, elevator maintenance technicians can be assisted by automated tools that analyse legal documents and propose to them a synthetic view of the elevator maintenance plan.

The rest of this paper is organized as follows. Section 2 presents a motivating example to illustrate our goals. Section 3 introduces the main related works on knowledge extraction from textual documents, with a focus on legislative semantic models and on semantic models of industrial maintenance. Section 4 presents SEMLEG model, which aims to cover the domain of legal industrial maintenance. Finally, the paper is concluded in Section 5.

## 2   Motivating example

Figure 1 shows an example of document from the French governmental website Légifrance. This document has been translated with Google Translate. It describes the legal regulation for companies working on lifting devices, and more specially on their checks. As in Figure 1, several cross-references are illustrated in Section 1; the text mentions different articles (*R4323-23 to R4323-27, R4535-7, [...]*) and refers the reader to the labour code. Section 2 of the document is dedicated to define what is a lifting device and a lifting accessory. It is worthwhile, with an automated process, to extract and synthesize the relevant elements from the document. In our case, we can expect knowledge extraction about the definitions in Section 2 and the rules in Section 3. First, the definitions allow systems or human to categorize the equipments into a group bound with a set of rules. In this way, when the technician uses a Computerized Maintenance Management System (CMMS) to operate on a lift, it knows the lift as an asset category and can suggest retrieving the manufacturer's instructions from the head of the establishment according to the order of March 1, 2004, Section 3.a.

The information extraction task can be broken down into multiple subtasks. The first one is the extraction of information from legal documents. The second one is the structuring of this information. The next sections are dedicated to answer these questions. Later in this paper, we will illustrate our proposition with the legal document, as in Figure 1.

## 3   Related work

This section presents a state of the art of the different proposals in information extraction from textual documents. The use of semantic models will be studied

**Order of March 1, 2004 relating to verifications of lifting devices and accessories**

*Last update of the data of this text: January 09, 2011*
NOR: SOCT0410464A

**Version in force on July 01, 2022**

The Minister for Social Affairs, Labor and Solidarity and the Minister for Agriculture, Food, Fisheries and Rural Affairs,

Having regard to Directive 98/34/EC of the European Parliament and of the Council of 22 June 1998 providing for an information procedure in the field of technical standards and regulations and rules relating to information society services, and in particular the notification no. 2003/0262/F;

Considering the labor code, and in particular its articles L. 620-6, R. 233-11, R. 233-11-1, R. 233-11-2;

Considering the decree of December 22, 2000 relating to the conditions and methods of approval of organizations for the verification of the state of conformity of work equipment;

Having regard to the opinion of the Higher Council for the Prevention of Occupational Risks, specialized commission no. 3;

Having regard to the opinion of the National Commission for Occupational Hygiene and Safety in Agriculture,

## Section 1. (Articles 1 to 3)

Section 1                                    Modified by Decree n°2008-244 of March 7, 2008 - art. 9 (V)

This Order determines the work equipment used for lifting loads, elevating workstations or elevating people to which the general periodic checks apply, checks during commissioning and checks during the return to service after any dismantling and reassembly operation or modification likely to jeopardize their safety, provided for by articles R4323-23 to R4323-27 , R4535-7 , R4721-11 , R4323-22 and R4323-28 of the labor code, at the expense of the head of the establishment in which this work equipment is put into service or used.

This order defines, for each of these verifications, their content, the conditions for their execution and, where applicable, their periodicity.

Section 2

The work equipment listed below must undergo the checks defined in Article 1:

a) The lifting devices defined below and their supports:

machines, including those moved by directly employed human power, and their equipment, driven by one or more operators who act on the movements by means of service devices over which they retain control, at least one of whose functions is to move a load made up of goods or equipment and, where applicable, of one or more people, with a significant change in the level of this load during its movement, the load not being permanently linked to the device. A change in level corresponding to what is just necessary to move the load by lifting it off the ground is not considered significant and is not likely to create risks in the event of failure of the load support.

In this decree, the term lifting devices also refers to lifting installations meeting the definition given above and specified in the appendix to this decree;

b) Lifting accessories meeting the following definition:

equipment not incorporated into a machine, a tractor or other equipment and placed between the machine, the tractor or any other equipment and the load, such as slings, lifting beams, self-tightening clamps, magnets, suction cups, lifting keys .

Section 3                                    Amended by Ordinance no. 2007-329 of March 12, 2007 - art. 12 (V)

a) The head of the establishment must make the lifting devices and accessories concerned and clearly identified available to the qualified persons in charge of the checks for the time necessary, taking into account the foreseeable duration of the examinations, tests and tests to be carried out.

b) The head of the establishment must make available to the qualified persons in charge of the examinations, tests and tests to be carried out the necessary documents, such as the manufacturer's instructions, the declaration or the certificate of conformity, the reports of the checks previous records and the aircraft maintenance log.

**Fig. 1.** Example of an 'Order' from Légifrance translated in English.

in Section 3.2 with a particular attention on models related to the law (Section 3.2.1) and models related to the industrial maintenance (Section 3.2.2).

### 3.1 Information extraction

Information extraction is a broad domain with multiple proposals in the scientific community. Nowadays, a wave of works has notably adopted neural networks in labelling or classification tasks in order to extract relevant elements. We can cite, for example, the work of David B. et al. [3] who developed a system allowing to anonymize German financial documents. The system makes it possible by finding entities such as first and last names, postal and e-mail addresses, locations, etc. The study was conducted on different architectures such as RNN (Recurrent Neural Networks), LSTM (Long and Short Term Memory) [18] or CRF (Conditional Random Fields) [12]. The Transformers' technology has been introduced [21] and has surpassed many existing models. It has been demonstrated that this architecture can be successful used to extract named entities [24] and to structure texts into knowledge graphs [5].

A significant part of the works proposes an information extraction based on resources containing, in a structured way, the concepts of a domain as well as the relations between them. These resources are semantic models like, for example, ontologies or knowledge graphs. The creation and the use of a semantic model has been the subject of different proposals in the literature. Munira A. et al. [1] proposed an Ontology-Based Information Extraction (OBIE) system with the objective of extracting, from textual documents, the land suitability for residential use. In the domain of industrial maintenance, [6] developed a system that relies on a semantic model and that allows managing the maintenance assets in industry. In the following section, we introduce the semantic models in the domain of the legal maintenance. We will, at least, present the two main domains related to the legal maintenance: the law and the industrial maintenance.

### 3.2 Semantic models

**Semantic models related to the law.** One of the early works in semantic modelling related to the law [11] created the "Frame" model, which aims to structure legal rules. Many of its concepts will be found in the further works. Van E. et al. [20] detail two of these models dedicated to the representation of legislation: FOLaw and LRI-Core. LRI-Core has been used as a high-level ontology for the modelling of German administrative laws. LKIF [8] is an open source ontology alternative to LRI-Core which can be applied on multi-domain representation of the law in order to facilitate its reuse. This ontology contains for example the notion of "Right" which characterizes the permission, obligation or prohibition to perform an act according to the law. LegalRuleML [2], an XML standard for the legal domain, has been inspired by LKIF to represent the legal knowledge and legal reasoning.

The LKIF ontology has been constructed via a supervised approach in order to manually build a semantic model [9]. While most works have considered the construction of the models in a manual manner, several alternative approaches have considered the construction using automatic approaches on large corpus [16].

Beyond structuring knowledge semantically, other works focus on mathematical formalizations using, e.g., deontic logic rules. Propagated in the scientific community by [23], this formalization of philosophical concepts relies on symbolic reasoning with notions of modality (prohibition, permission, obligation).

**Semantic models for industrial maintenance.** Semantic models dedicated to legal maintenance have also been addressed in scientific research. In 2004, Rasovska I. et al. [15] proposed a system composed of a conceptual model allowing to make decisions related to the industrial maintenance. Few years later, the IMAMO (Industrial MAintenance Management Ontology) [6] has been published. IMAMO is an ontological model with the objective of standardizing information exchanges related to maintenance. It aims at ensuring semantic interoperability while generating data to be used as a decision-making support. Many other works reflect this structure life cycle for industrial equipment [7, 19, 14, 13, 10]. We find in these models essential concepts for industrial maintenance. For example, the notion of 'actions' that group different acts specific to maintenance such as assembly, energization, scrapping, etc. We can mention a last open source model produced by the IOF group (Industrial Ontologies Foundry) which tries to "*create a set of reference and open ontologies covering the whole industry domain*"[4].

### 3.3   Positioning

From all semantic representations listed above, none of them is sufficient on its own to represent legal industrial maintenance. On one hand, we have models specially designed to model legal rules, but we can't find key elements of maintenance such as material entities. On the other hand, industrial maintenance models have been proposed by the scientific community, but do not structure the legal notions. The limit of the state of the art arises here with the inability to fully model the domain of legal industrial maintenance.

The reuse of existing ontologies is in fact a practice in ontology construction. In particular for information retrieval and knowledge engineering, in [22], the authors mention two main constraints for the implementation: the first one is the adequacy of ontologies with the information retrieval task. The second constraint is to find ontologies that represent, jointly and accurately, the domain. However, it appears that the domain of legal industrial maintenance, being very specialized, do not comply with the second constraint. The existing models, not covering the specificities of the domain we are dealing with, we decided to create SEMLEG, reusing as much as possible existing ontologies. This model allows linking the Computerized Maintenance Management System (CMMS) with the legal obligations present in the legal texts.

---
[4] https://www.industrialontologies.org/our-mission/

## 4 Semantic model for legal maintenance: SEMLEG

In this section, we detail the different concepts of SEMLEG. A second part is devoted to illustrate the SEMLEG model on an example taken from Légifrance[5].

### 4.1 Reusing existing vocabularies

As introduced in the previous section, there are two main approaches for building an ontology: automatically with a large corpus or manually by experts. In this paper, we have chosen the construction via experts, supported by the reuse of existing semantic models (as further detailed in the following). This choice is motivated by a desire for interpretability between the models, as well as the reuse of existing concepts rather than trying to recreate them [4]. In SEMLEG we use two existing ontologies: LKIF-CORE is dedicated to the legal elements and IOF to the maintenance elements. In this section, we will describe in more detail the motivation for choosing these ontologies. Both of LKIF-CORE[6] and IOF[7] are hosted on GitHub and can be easily accessed.

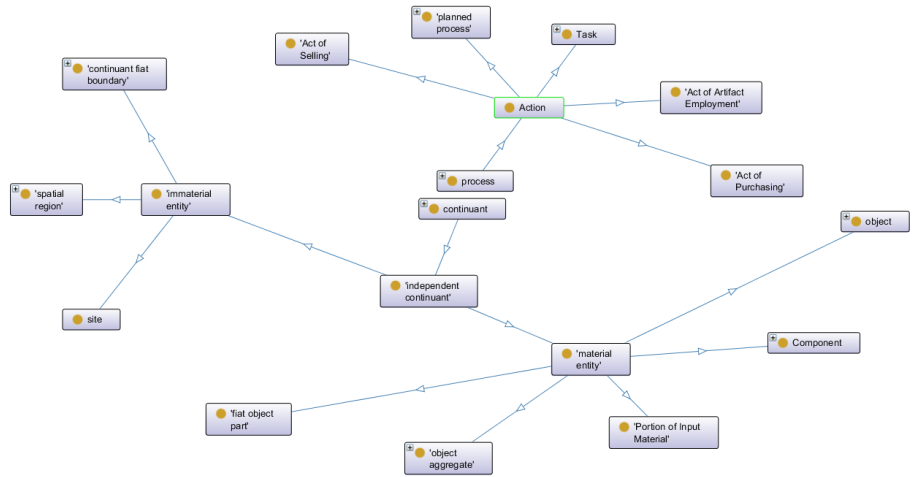

**Fig. 2.** Fragment of the IOF ontology.

Figures 2 and 3 illustrate the fragments of IOF and LIKIF of interest for SEMLEG: *Norm* for LKIF-CORE and *independent continuant*, *action* for IOF. We have the class **lkif-core:Norm** and, with this concept, we can represent the notion of obligation and prohibition as presented before. On the IOF side, we have the class **lkif-core:independent continuant** which allows representing material and immaterial entities. We will find, for example, on the immaterial side: geographical areas. On the material side, we will find: the owner, a buyer, a seller, etc. We will also find more inert elements like: mechanical systems, tools, machines. We also chose to include **iof:Action** which allows representing, for example, maintenance tasks, the action of buying, selling, etc.

---
[5] https://www.legifrance.gouv.fr/
[6] https://github.com/RinkeHoekstra/lkif-core
[7] https://github.com/welschmichel/IOF_Maintenance_Working_Group_Public

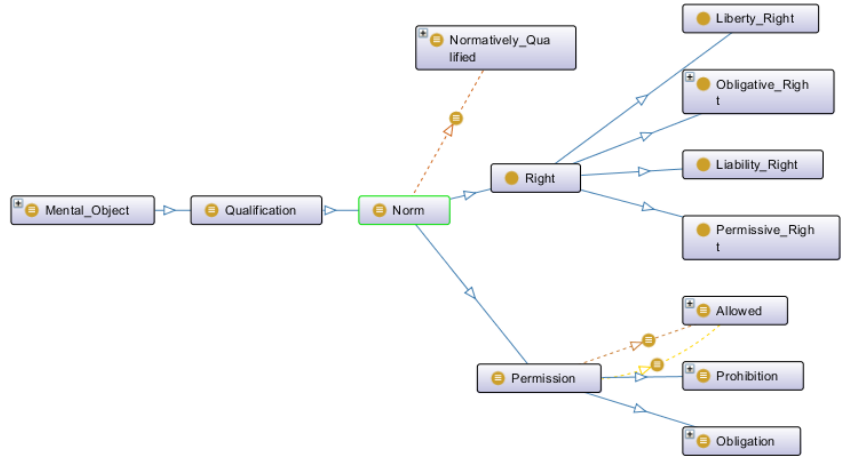

**Fig. 3.** Fragment of the LKIF-CORE ontology.

### 4.2 Construction and explanation of SEMLEG

The process of constructing SEMLEG has been divided in two main steps: (1) the extraction from Légifrance of three representative industrial maintenance decrees in various domains: pressured equipments[8], lifting machines[9], and buried tanks[10]; (2) the identification of the recurrent parts of the rules that compose the orders and the construction of the semantic model by 4 experts, each one coming from a different field of expertise: (a) industrial legal maintenance, (b) ontological knowledge modeling, (c) generalized information systems and (d) automatic natural language processing. This work allowed us to obtain the model illustrated in Figure 4. Its implementation in OWL from the software Protégé[11] is available as open source on GitLab[12].

---

[8] https://www.legifrance.gouv.fr/loda/id/JORFTEXT000036128632/2022-06-01/
[9] https://www.legifrance.gouv.fr/loda/id/JORFTEXT000000439029/2022-07-01/
[10] https://www.legifrance.gouv.fr/loda/id/JORFTEXT000018820571/2022-06-01/
[11] https://protege.stanford.edu/
[12] https://gitlab.irit.fr/semleg/semleg

**Fig. 4.** SEMLEG semantic model.

**Table 1.** SEMLEG concepts and definitions

| Concept | Definition | Example |
| --- | --- | --- |
| Source | Allows, like Russian dolls, to encapsulate a set of rules, definitions or other sources. | Article L. 512-5 |
| Link | Links two sources together. This link can be a repeal, a mention or a modification. | Modified by ORDER of May 11, 2015 |
| Definition | Stores the definition of an act or resource. | A functional test of a lifting device is the test which consists in moving in the most unfavorable positions |
| Rule | Describe the rules and allows aggregating the necessary entities. | The periodic general verification of lifting devices must be done every twelve months. |
| Act | Is the act of the rule. | Verification |
| iof:Action | Represents actions in the field of industrial maintenance. | Verification |
| Resource | Represents the material or immaterial resources of a rule. | Lifting devices |
| Undefined | Represents undefined resources. | |
| iof:independent continuant | Brings together tangible and intangible entities. | Lifting devices |
| Modality | Represents the modality of a rule, which can be an obligation, a prohibition or a permission. | Must (obligation) |
| lkif-core:Norm | Reflects the legal or moral right to do or not to do something or to obtain or not obtain an action, thing or consideration in civil society. | Must (obligation) |
| Operand | Represents the elements on which a logical operation will operate. | |
| Operator | Represents the logical operation operator. | |
| SuccessiveConjunction | Represents performing one action after another. | I do A **and after** B |
| ParallelConjunction | Represents the performance of an action at the same time as another. | I do A **and at the same time** B |
| SimpleDisjunction | Represents the completion of one action or another. | I do A **or** B |
| Condition | Conditions acts, resources and operators. | Lifting devices **used for the transport of persons** |
| TemporalCondition | Conditions via a temporal aspect. | Every twelve months |
| SpatialCondition | Conditions via a spatial aspect. | In the factory area |
| NumericalCondition | Condition via a numeric value. | After 5 cm |

Table 1 presents the list of SEMLEG concepts and the definition attached to each concept. In the rest of this section, we will see in particular how SEMLEG allows us to structure the legislative tree, the rules, the operators between the rules as well as the conditions.

**The legislative tree** is modeled by a successive imbrication of groups, allowing to structure the legal information. For example, we will find a chapter dedicated to an idea which is itself in a section. Chapters, sections, articles, etc. are modeled via the concept **semleg:Source** and their nesting via the relation **semleg:hasSubSource**. To link sources together, we have introduced the concept **semleg:Link**. It allows a source to abrogate, mention or modify another source. At the leaves of this tree, which is the legislative document, there are two types of elements: definitions and rules. Definitions are a legal explanation of what is meant by the use of a term. They aim to characterize objects, acts, actors, etc. Rules, on the other hand, are responsible for carrying the main legislative information.

**The rule** is characterized, in SEMLEG, by: a subject (**semleg:hasSubject**), an act (**semleg:Act**), a modality (**semleg:Modality**) and an object (**semleg:hasObject**).

For the construction of SEMLEG, we have reused existing concepts; this is why we can see on the diagram equivalence relations between SEMLEG classes and IOF or LKIF-CORE classes. So we have equivalence between the classes: **lkif-core:Norm** and **semleg:Modality**, **lkif-core:independent continuant** and **semleg:Resource**, **iof:Action** and **semleg:Act**.

Note that a resource can sometimes be undefined (**semleg:Undefined**). For example, in the sentence: "A draft of the report must be sent at the end of the audit", we do not find any mention of the subject (who must hand in the report). As a result, the rule will therefore be modelled with an undefined subject.

**Sequence of rules.** Like the mathematical operations, we have introduced operators (**semleg:Operator**) and operands (**semleg:Operand**) which are concepts that allow us to reason about the sequence between rules. We have so far identified 3 different operators: simple disjunction, the parallel conjunction and the successive conjunction. The simple disjunction can be translated as: "I do A <u>or</u> B". The parallel conjunction can be translated as: "I do A <u>and at the same time</u> B". The successive conjunction can be translated as: "I do A <u>and then only</u> B". The operators thus make it possible to structure the chains of rules (the operands) in the legislative texts.

**Conditions** are concepts that can be used by acts, resources and operators. For example, conditions can add a notion of time to the performance of an act: "When the installation is shut down [...]". Within the framework of our study, we met mainly three sub-types of conditions: temporal conditions (**semleg:TemporalCondition**), spatial conditions (**semleg:SpatialCondition**) and numerical conditions (**semleg:NumericalCondition**). This list is not exhaustive, and the more particular conditions can be structured via the parent class : **semleg:Condition**.

### 4.3 Extract from Légifrance structured via SEMLEG

In this section, we illustrate the use of SEMLEG with an example (Figure 5) extracted from Légifrance. Consider the following example: "*When the installation is permanently shut down, the tanks and pipes are degassed.*"[13].

**Fig. 5.** Order of 18 April 2008 from Légifrance.

As we can see in Figure 5, the example is depicted in a sequence of nested blocks. In our case, the rule is in Section 5, itself in Title A, itself in the Order of 18 April 2008. This nesting introduces the first concept: **semleg:Source**. This separation allows two sources to be linked together via **semleg:Link**. In our example, "Section 5" is a source connected to "Order of August 9, 2017 - art. 2" by a modification link. It is through this process that we are able to model the evolution over time.

The following sentence available in Section 2 illustrates the concept of definition: "*A tank is said to be buried when it is completely or partially below the level

---

[13] https://www-legifrance-gouv-fr.translate.goog/loda/article_lc/
LEGIARTI000035650389/2022-06-01?_x_tr_sl=fr&_x_tr_tl=en&_x_tr_hl=fr&
_x_tr_pto=wapp

*of the surrounding ground, whether it is directly in the ground or in a pit. Tanks installed in premises are not considered buried, even when the premises are located below the surrounding ground.*"[14]. This sentence is an essential element in understanding and defining what a buried tank is.

In Figure 6, we have labelled our example sentence using the concepts we define in SEMLEG (we do not detail the links with other ontologies). We thus find the following concepts: modality, resources, act and conditions.

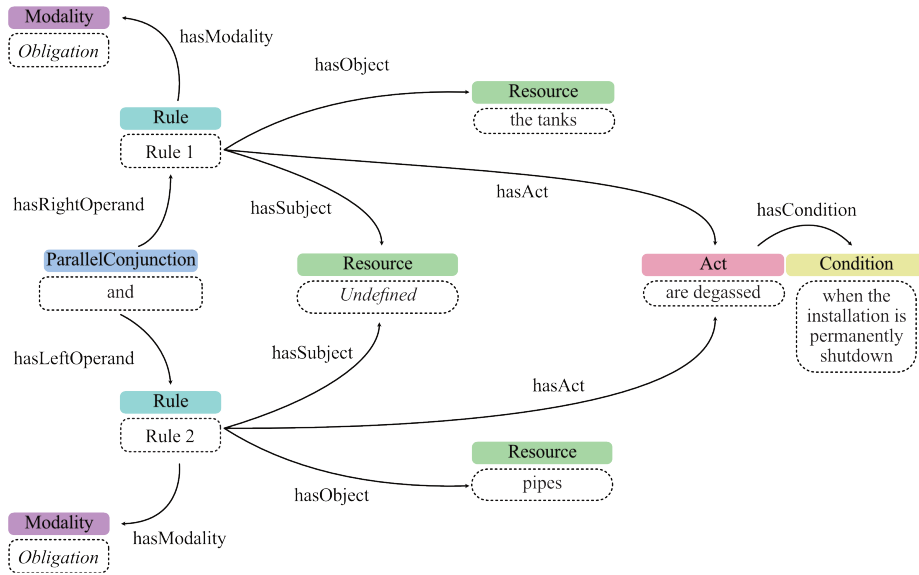

**Fig. 6.** Example of a Légifrance sentence structured with SEMLEG

In our example, the modality is an obligation in implicit form. Indeed, unlike explicit modalities which can have verbs like "must" and "can", implicit modalities do not have these verbs. The second element is the act which is in our sentence: "are degassed". This action is performed by a subject not mentioned in this sentence. Finally, the subject will perform an action on another resource considered as the object of the rule: the tanks and the pipes. This second resource can find a more precise characterization in IOF via the concept: **iof:Maintainable Item**.

Then, the operator is in our case a simple disjunction thanks to the word "or" which allows choosing between 'a competent person' and 'an organization'. The operator acts here as a junction between the rule proposing 'a competent person' and the second one proposing 'an organization'.

---

[14] https://www-legifrance-gouv-fr.translate.goog/loda/article_lc/
LEGIARTI000030706280/2022-06-01?_x_tr_sl=auto&_x_tr_tl=en&_x_tr_hl=
fr&_x_tr_pto=wapp

Finally, we have the conditions. In our example, "when the installation is permanently shutdown" is a condition that qualifies the act.

## 5 Conclusions and future work

In this paper, we have presented a semantic model called SEMLEG which allows structuring textual documents of legal industrial maintenance. This model results from a manual analysis of documents from Légifrance involving domain experts. It combines existing ontologies from legal and industrial maintenance domains. While we have illustrated the use of SEMLEG with a subset of representative documents, the next step of this work will be to extend the evaluation of the adequacy of the proposed model with a larger set of documents within a task of information extraction. This can led to the evolution of the model. We plan also to assess the benefits of SEMLEG in the extraction task, by comparing the task performance without SEMLEG and with SEMLEG.

## References

1. Al-Ageili, M., Mouhoub, M.: An ontology-based information extraction system for residential land use suitability analysis. ArXiv (2022)
2. Athan, T., Boley, H., Governatori, G., Palmirani, M., Paschke, A., Wyner, A.: Oasis legalruleml. In: proceedings of the fourteenth international conference on artificial intelligence and law. pp. 3–12 (2013)
3. Biesner, D., Ramamurthy, R., Stenzel, R., Lübbering, M., Hillebrand, L.P., Ladi, A., Pielka, M., Loitz, R., Bauckhage, C., Sifa, R.: Anonymization of german financial documents using neural network-based language models with contextual word representations. International Journal of Data Science and Analytics pp. 151–161 (2022)
4. Carriero, V.A., Daquino, M., Gangemi, A., Nuzzolese, A.G., Peroni, S., Presutti, V., Tomasi, F.: The landscape of ontology reuse approaches. In: Applications and Practices in Ontology Design, Extraction, and Reasoning (2020)
5. Friedman, S.E., Magnusson, I.H., Sarathy, V.: From unstructured text to causal knowledge graphs: A transformer-based approach. ArXiv (2022)
6. Hedi Karray, M., Chebel-Morello, B., Zerhouni, N.: A Formal Ontology for Industrial Maintenance. In: Terminology & Ontology : Theories and applications, TOTh Conference 2011. pp. 1–20 (May 2011), https://hal.archives-ouvertes.fr/hal-00602182
7. Hodkiewicz, M., Klüwer, J.W., Woods, C., Smoker, T., Low, E.: An ontology for reasoning over engineering textual data stored in FMEA spreadsheet tables. Computers in Industry (Oct 2021). https://doi.org/10.1016/j.compind.2021.103496, https://linkinghub.elsevier.com/retrieve/pii/S0166361521001032
8. Hoekstra, R., Breuker, J., Di Bello, M., Boer, A., et al.: The lkif core ontology of basic legal concepts. LOAIT pp. 43–63 (2007)
9. Kingston, J., Schafer, B., Vandenberghe, W.: Towards a Financial Fraud Ontology: A Legal Modelling Approach. Artificial Intelligence and Law (4), 419–446 (Dec 2004). https://doi.org/10.1007/s10506-005-4163-0, http://link.springer.com/10.1007/s10506-005-4163-0

10. Koukias, A., Nadoveza, D., Kiritsis, D.: Semantic Data Model for Operation and Maintenance of the Engineering Asset. In: Advances in Production Management Systems. Competitive Manufacturing for Innovative Products and Services, pp. 49–55. Berlin, Heidelberg (2013)
11. Kralingen, R.v.: A Conceptual Frame-based Ontology for the Law. In: Proceedings of the First International Workshop on Legal Ontologies. pp. 6–17 (1997)
12. Lafferty, J., McCallum, A., Pereira, F.C.: Conditional random fields: Probabilistic models for segmenting and labeling sequence data (2001)
13. Matsokis, A., Kiritsis, D.: An ontology-based approach for Product Lifecycle Management. Computers in industry pp. 787–797 (2010)
14. Otte, J.N., Kiritsi, D., Ali, M.M., Yang, R., Zhang, B., Rudnicki, R., Rai, R., Smith, B.: An ontological approach to representing the product life cycle. Applied Ontology pp. 179–197 (Apr 2019). https://doi.org/10.3233/AO-190210, https://www.medra.org/servlet/aliasResolver?alias=iospress&doi=10.3233/AO-190210
15. Rasovska, I., Chebel-Morello, B., Zerhouni, N.: A conceptual model of maintenance process in unified modeling language. IFAC Proceedings pp. 497–502 (2004)
16. Saravanan, M., Ravindran, B., Raman, S.: Improving legal information retrieval using an ontological framework. Artificial Intelligence and Law pp. 101–124 (Jun 2009). https://doi.org/10.1007/s10506-009-9075-y, http://link.springer.com/10.1007/s10506-009-9075-y
17. Sassier, P., Lansoy, D.: Ubu loi. Arthème Fayard, France (2008)
18. Sherstinsky, A.: Fundamentals of recurrent neural network (rnn) and long short-term memory (lstm) network. ArXiv (2018)
19. Smith, B., Ameri, F., Cheong, H., Kiritsis, D., Sormaz, D.N., Will, C., Otte, J.N.: A first-order logic formalization of the industrial ontologies foundry signature using basic formal ontology. In: JOWO : Joint Ontology Workshops (2019)
20. Van Engers, T., Boer, A., Breuker, J., Valente, A., Winkels, R.: Ontologies in the Legal Domain, pp. 233–261. Boston, MA (2008). https://doi.org/10.1007/978-0-387-71611-4_13
21. Vaswani, A., Shazeer, N.M., Parmar, N., Uszkoreit, J., Jones, L., Gomez, A.N., Kaiser, L., Polosukhin, I.: Attention is all you need. In: NIPS (2017)
22. Wimalasuriya, D.C., Dou, D.: Using multiple ontologies in information extraction. In: Proceedings of the 18th ACM conference on Information and knowledge management. pp. 235–244 (2009)
23. von Wright, G.H.: Deontic Logic. Mind pp. 1–15 (1951), https://www.jstor.org/stable/2251395
24. Yan, H., Deng, B., Li, X., Qiu, X.: Tener: Adapting transformer encoder for named entity recognition. ArXiv (2019)

# Author Index